\documentclass{Style}
\usepackage[T2A]{fontenc}
\usepackage[cp1251]{inputenc}
\usepackage[english]{babel}
\usepackage{cite,array}

\twocolumn

\rus

\title{Anomalous superconductivity and superfluidity in repulsive fermion
systems}

\rtitle{Anomalous superconductivity and superfluidity in repulsive
fermion systems}

\author{M.\,Yu. Kagan$^{a,b}$, V.\,A. Mitskan$^{c,d}$, M.\,M.~Korovushkin$^{c}$\/}

\address{$^{a)}$Kapitza Institute for Physical Problems,
Russian Academy of Sciences, 119334 Moscow, Russian Federation\\
$^{b)}$National Research University Higher School of Economics,
101000 Moscow, Russian Federation\\
$^{c)}$Kirensky Institute of Physics, Siberian Branch of the Russian Academy of Sciences,
660036 Krasnoyarsk, Russian Federation\\
$^{d)}$Reshetnev Siberian State Aerospace University, 660014
Krasnoyarsk, Russian Federation}

\rauthor{M.\,Yu. Kagan, V.\,A. Mitskan, M.\,M.~Korovushkin}

\abstract{We discuss the mechanisms of unconventional
superconductivity and superfluidity in 3D and 2D fermionic systems
with purely repulsive interaction at low densities. We construct
phase diagrams of these systems and find the areas of the
superconducting state in free space, as well as on the lattice in
the framework of the Fermi-gas model with hard-core repulsion, the
Hubbard model, the Shubin-Vonsovsky model, and the $t-J$ model. We
demonstrate that the critical superconducting temperature can be
greatly increased in the spin-polarized case or in a two-band
situation already at low densities. The proposed theory is based
on the Kohn-Luttinger mechanism or its generalizations and
explains or predicts anomalous $p$-, $d$-, and $f$-wave pairing in
various materials, such as high-temperature superconductors, the
idealized monolayer and bilayer of doped graphene, heavy-fermion
systems, layered organic superconductors, superfluid $^3$He,
spin-polarized $^3$He mixtures in $^4$He, ultracold quantum gases
in magnetic traps, and optical lattices.
}

\begin{document}

\maketitle

\tableofcontents

\section{Introduction}
\label{sec_intro}

The recent discovery of Cooper pairing at a recordly high
temperature of 190 K in metallic hydrogen sulfide
H$_2$S~\cite{Duan14,Drozdov14} under pressure of the order of 1
Mbar raises our hopes to move forward from ‘high-temperature’ to
‘room-temperature’ superconductors. At the same time, there are
many interesting low-temperature superconducting and superfluid
systems with anomalous types of pairing and a nontrivial structure
of the order parameter. In this review, we consider systems with a
low density of fermions in the framework of the nonphonon
mechanisms of superconductivity, such as the famous Kohn-Luttinger
mechanism and its generalizations, and the exchange mechanisms
connected with the antiferromagnetic attraction of spins on
neighboring sites, which are topical, in particular, for the $t-J$
model, in which both low and high critical temperatures of the
superconductive transition and anomalous $p$-, $d$-, and $f$-wave
Cooper pairing can appear.

As is known from textbooks, the conduction electrons in metals,
together with the positively charged ions, form a solid-state
plasma, which determines the combination of their electric,
galvanomagnetic, kinetic, and superconducting properties. The
coupling between the subsystem of massive positive ions and the
subsystem of light fermions leads to the appearance of the
electron-phonon interaction, which affects the properties of the
electron subsystem. In particular, the effective interaction
between electrons in a solid-state plasma can differ significantly
from the Coulomb interaction of electrons in the vacuum and can
even change sign. This most important effect is the basis of the
electron-phonon mechanism of Cooper instability in standard
superconductors~\cite{BCS57}.

It is obvious that the role of the mediator (coupling to which
initiates the renormalization of the Coulomb interaction) can be
played by any other subsystem. It is only necessary that the
interaction of the electron gas with this subsystem lead to
polarization effects that cause the generation of electrons and
holes in the vicinity of the Fermi surface. Notably, in many
theoretical studies on high-temperature superconductors,
collective excitations of the subsystem of localized spins of
copper ions serve as such a mediator. This, in particular,
determines the spin-fluctuation mechanism of Cooper instability,
which leads to the formation of a superconducting phase with
$d-$type symmetry of the order parameter.

Within the formalism of the secondary quantization of fermions,
the operator of Coulomb interaction of electrons contains terms
that in the higher orders of the perturbation theory initiate
polarization contributions to the ground-state energy, which also
leads to the renormalization of the Coulomb interaction of
electrons. Therefore, the effective interaction of electrons in
such a metal can differ significantly from the electron-electron
interaction in the vacuum. This makes topical the problem (first
formulated by Anderson~\cite{Anderson87}) of such a
renormalization of Coulomb interaction under which the effective
electron-electron interaction in a substance would have an
attractive rather than repulsive nature. In other words, the
problem consists in searching for conditions under which the
above-mentioned polarization effects in the electron plasma of a
metal would lead to a change in the sign of the resulting
interaction between the electrons. From the mathematical
standpoint, the problem reduces to calculating the effective
pairwise interaction of electrons with multiparticle effects in
the electron ensemble taken into account. No less important,
according to Anderson, is the problem of explaining the
unconventional properties of the normal state of many strongly
correlated electronic systems at temperatures higher than the
critical temperature, especially in the pseudogap state.

In recent decades, considerable progress has been achieved in
experimental and theoretical studies of superconductive systems
with a nonphonon nature of the Cooper pairing and with a complex,
nontrivial structure of the order parameter. The first
experimentally discovered systems with unconventional triplet
$p-$wave pairing (the total spin of the Cooper pair
$S_{\textrm{tot}}=1$ and the orbital momentum of the relative
motion of the pair $l=1$) were the superfluid A and B phases of
$^3$He with low critical temperatures, $T_c\sim1\,\textrm{mK}$.
Another example of systems in which the $p-$wave pairing is
realized are $^6$Li$_2$ and $^{40}$K$_2$ molecules in magnetic
traps in the regime of the $p-$wave Feshbach resonance with
ultralow critical temperatures:
$T_c\sim10^{-6}-10^{-7}\,\textrm{K}$~\cite{Regal03,Schunck05}. It
is assumed that unconventional $p-$wave pairing with critical
temperatures $T_c\sim0.5-1\,\textrm{K}$ is realized in some
heavy-fermion intermetallic compounds, such as
U$_{1-x}$Th$_x$Be$_{13}$ and UNl$_2$Al$_3$, with large effective
masses $m^*\sim100-200m_e$~\cite{Ott84,Kromer98}. Frequently, the
$p-$wave pairing is mentioned in connection with organic
superconductors, such as $\alpha-$(BEDT-TTF)$_2$I$_3$ with
$T_c\sim5\,\textrm{K}$~\cite{Kuroki06}. Finally, $p-$wave pairing
with $T_c\sim1\,\textrm{K}$ is apparently realized in the
ruthenates Sr$_2$RuO$_4$~\cite{Maeno01,Rice95}, and it cannot be
excluded in the layered dichalcogenides CuS$_2$--CuSe$_2$ or the
semimetals and semimetallic superlattices
InAs--GaSb,PbTe--SnTe~\cite{Murase86}, The heavy-fermionic
intermetallic compound UPt$_3$ with $m^*\sim200m_e$ and
$T_c\sim0.5\,\textrm{K}$, as well as a large class of
high-temperature cuprate superconductors with critical
temperatures from $T_c=36\,\textrm{K}$ (for the lanthanum-based
compounds) to $T_c=160\,\textrm{K}$ (obtained in mercury-based
superconductors under pressure), are related to unconventional
superconductors with the singlet $d-$wave pairing
($S_{\textrm{tot}}=0,\,l=2$). Finally, in connection with the
problems of applied superconductivity, it is also necessary to
mention new multiband superconductors with a more conventional
$s-$wave pairing, such as MgB$_2$~\cite{Nagamatsu01}, and the
recently discovered superconductors based on iron arsenide
~\cite{Kamihara08} and the H$_2$S and PoH$_2$ metallic compounds
already noted above~\cite{Liu15}.

Along with the problems of Cooper pairing in the above-mentioned
electron systems, also of significant interest is the search for
fermionic superfluidity in three-dimensional (3D) and
two-dimensional (2D) (thin films, submonolayers) solutions of
$^3$He in $^4$He ~\cite{Vollhardt90,Volovik92,Volovik03} and for
superconductivity in doped graphene ~\cite{Novoselov04}, which are
problems that have still not been solved experimentally. These
systems are among the most promising ones from the standpoint of
the experimental and theoretical description of a wide class of
physical phenomena and of the nature of multiparticle correlations
in them.

Notably, submonolayers of $^3$He adsorbed on different substrates,
such as a solid substrate or the free surface of superfluid
$^4$He, with the variation of the particle density in wide ranges,
make it possible to realize different regimes in the system --
from the ultrararefied Fermi gas to strongly correlated Fermi
systems~\cite{Kagan94_1}. This makes the solutions ideal objects
for the development and verification of different methods of the
Fermi liquid theory. The unbalanced (spin-polarized) ultracold
Fermi gases in 3D and especially in 2D magnetic traps are also
very promising~\cite{Ong15,Ries14}.

Of significant interest from both the fundamental and applied
standpoints is graphene, because of its unique electronic
properties~\cite{Lozovik08,Kotov12}. Near the Fermi level, the
electrons in graphene have a linear dispersion, and the energy gap
between the valence band and the conduction band is absent;
therefore, the electrons in graphene can be described by a 2D
Dirac equation for massless charged
quasiparticles~\cite{Wallace47}. The properties of these
quasiparticles, such as their two-dimensionality, the spinor
nature of their spectrum, the zero mass, and the absence of the
gap in the spectrum, lead to a number of phenomena that have no
counterparts in other physical systems~\cite{Castro09}.

The above-mentioned studies have stimulated an intensive search
for alternative mechanisms of pairing based on strong correlations
in the Fermi liquid. The most promising in this respect are the
Kohn-Luttinger mechanism~\cite{Kohn65}, proposed in 1965, and its
generalizations (see, e.g.,~\cite{Miyake07}). The Kohn-Luttinger
mechanism assumes the transformation of the pure repulsive
interaction of two particles in the vacuum in the presence of a
fermionic background into an effective attraction in the substance
in the channel with a nonzero orbital angular momentum of the
pair.

This review is devoted to the description of basic results
obtained in recent decades concerning Kohn-Luttinger
superconductivity in repulsive Fermi systems and its
generalizations as well as the exchange mechanisms of
superconductivity in the generalized $t-J$ model.

\section{Superconductivity in the model of a Fermi gas with repulsion}
\label{sec_gas}

The basic model for studying the nonphonon mechanisms of
superconductivity in low-density electron systems is the model of
a Fermi gas. In the case of a Fermi gas with attraction, the
scattering length $a$ is negative ($a < 0$), which results in a
traditional $s$-wave pairing (total spin $S=0$, orbital angular
momentum $l=0$) with the critical temperature
\begin{equation}\label{Tcs}
T^{s}_c\approx0.28\,\varepsilon_F\exp\biggl(-\frac{\pi}{2|a|p_F}\biggr).
\end{equation}
where $\varepsilon_F$ is the Fermi energy and $p_F$ is the Fermi
momentum.

This result was obtained in Ref.~\cite{Gor'kov61} soon after the
appearance of the Bardeen-Cooper-Schrieffer (BCS)
theory~\cite{BCS57}. Result (\ref{Tcs}) differs from the classical
formula given in~\cite{BCS57} by the presence of a quantity
$0.28\,\varepsilon_F$ in the preexponential factor instead of the
Debye frequency $\omega_D$ typical for the phonon models in
conventional superconductors.
\begin{figure}[t]
\begin{center}
\includegraphics[width=0.48\textwidth]{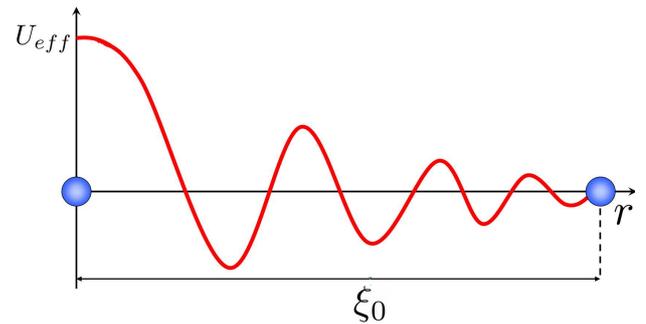}
\caption{Fig.~1. Friedel oscillations in the effective interaction
of two particles as a result of the polarization of the fermionic
background, where $\xi_0$ is the coherence length of the Cooper
pair~\cite{Kagan14d}.} \label{friedel}
\end{center}
\end{figure}

In the model of a Fermi gas with repulsion, the scattering length
$a$ is positive ($a>0$) and the superconductivity corresponds to
the Kohn-Luttinger mechanism in the low-temperature region. The
physical reason for this consists in the effective interaction of
quasiparticles, which arises as a result of the polarization of
the fermionic background. Due to a sharp boundary existing in the
momentum space which is equal to the diameter of the Fermi sphere
$2p_F$ and separates the occupied states from the empty ones, the
effective interaction of quasiparticles that are located on the
Fermi level does not decrease exponentially, but has an
oscillating form (Friedel oscillations~\cite{Freidel54,Freidel58})
\begin{equation}\label{oscillations}
U_{\textrm{eff}}(r)\sim\frac{\cos(2p_Fr)}{(2p_Fr)^3}.
\end{equation}
If the distance between two electrons in a Cooper pair is
relatively large, effective interaction (\ref{oscillations}) in
the coordinate space has a large number of maxima and minima
(Fig.~\ref{friedel}). Then the integral effect determined by the
averaging over the potential relief of Friedel oscillations can,
in principle, lead to an effective attraction and the appearance
of superconductivity in the system.

\begin{figure}[t]
\begin{center}
\includegraphics[width=0.46\textwidth]{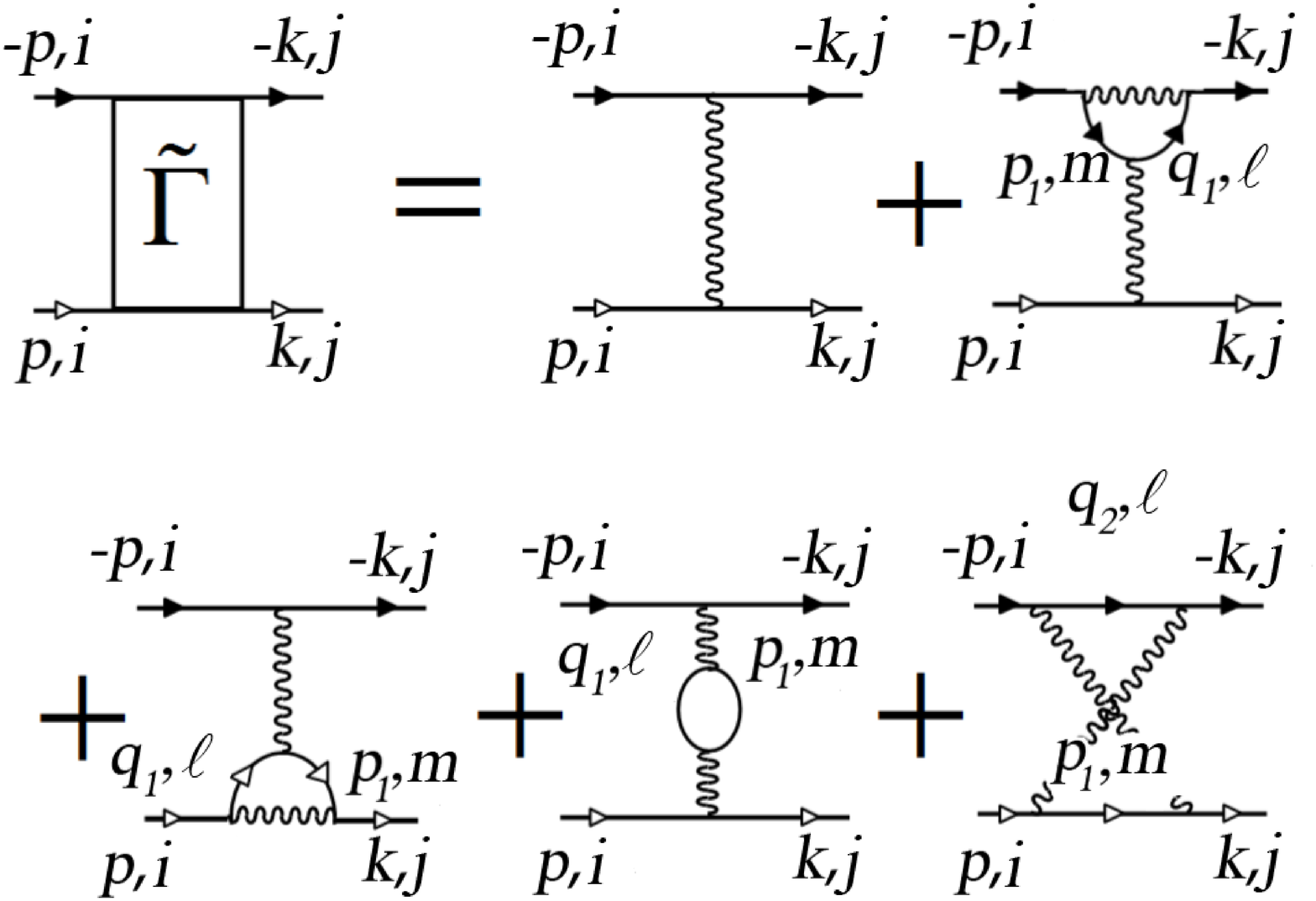}
\caption{Fig.~2. Diagrams of the first and second order for the
effective interaction of electrons $U_{\textrm{eff}}$. The solid
lines with light (dark) arrows correspond to the Green functions
of electrons with the spin projections $+{\textstyle{1 \over
2}}$~($-{\textstyle{1 \over 2}}$);
$\textbf{q}_1=\textbf{p}_1+\textbf{p}-\textbf{k}$ и $\textbf{q}_2
=\textbf{p}_1-\textbf{p}-\textbf{k}$. The wavy lines correspond to
the bare interaction. In the case of a Fermi gas (see
Section~\ref{sec_gas}~), the Hubbard model (see
Section~\ref{sec_Hubbard}~), and the Shubin-Vonsovsky model (see
Section~\ref{sec_SV}~), the indices in the diagrams are
$i=j=l=m=1$. In the case of a graphene monolayer (see
Section~\ref{sec_monolayer}~) $i=j=1$, and $l,m=1,2$. In the case
of a graphene bilayer (see Section~\ref{sec_bilayer}~) $i,j=1,2$
and $l,m=1,2,3,4$~\cite{Kagan14}.}\label{diagrams_alpha}
\end{center}
\end{figure}

The first to advocate this mechanism of superconductivity were
Kohn and Luttinger~\cite{Kohn65}, who considered 3D repulsive
Fermi systems. They showed that the effective interaction in the
first two orders of the perturbation theory in the gas parameter
(more precisely, in the scattering length $a$) is described by the
sum of the five diagrams shown in Fig.~\ref{diagrams_alpha}. The
first diagram corresponds to the bare interaction of two electrons
in the Cooper channel. The next four diagrams (Kohn-Luttinger
diagrams) are due to second-order processes and take into account
the polarization effects of the filled Fermi sphere. In the case
of a short-range potential, the contribution to the effective
interaction is determined only by the fourth exchange diagram, and
in the first two orders of the perturbation theory, the expression
for $U_{\textrm{eff}}$ can be written as
\begin{equation}
U_{\textrm{eff}}(\textbf{p},\textbf{k})=\frac{4\pi
a}{m}+\biggl(\frac{4\pi a}{m}\biggr)^2\Pi(\textbf{p}+\textbf{k}),
\end{equation}
where $4\pi a/m$ is the pseudopotential, which corresponds to the
wavy line in Fig.~\ref{diagrams_alpha}, and
$\Pi(\textbf{p}+\textbf{k})$ is the static polarization operator,
which is described by the standard Lindhard
function~\cite{Lindhard54,Ashcroft79}
\begin{equation}\label{Lindhard}
\Pi(\textbf{p}+\textbf{k})=\frac1N\sum_{\textbf{p}_1}\frac{n_F(\varepsilon_{\textbf{p}_1-\textbf{p}-\textbf{k}})-
n_F(\varepsilon_{\textbf{p}_1})} {\varepsilon_{\textbf{p}_1}-
\varepsilon_{\textbf{p}_1-\textbf{p}-\textbf{k}}}.
\end{equation}
This operator is responsible for the charge screening in the
electron plasma in metals. The plus sign in the argument of the
polarization operator is due to the so-called crossing, which, in
the case of short-range repulsion, distinguishes the exchange
diagram from the true polarization loop, which contains the
argument $\textbf{p}-\textbf{k}$. In the absence of a lattice,
$\varepsilon_{\textbf{p}}=\displaystyle\frac{\textbf{p}^2}{2m}$ is
the energy spectrum, $$n_F(x)=(\exp(\frac{x-\mu}{T})+1)^{-1}$$ is
the Fermi-Dirac distribution function, and $\mu$ is the chemical
potential.

It was noted in the early work of Migdal~\cite{Migdal58} and
Kohn~\cite{Kohn59} that at low temperatures ($T\ll\varepsilon_F$),
the polarization operator contains, apart from a regular part, a
singular part -- the so-called Kohn anomaly, which in the 3D case
has the form
\begin{equation}
\Pi_{\textrm{sing}}(\widetilde{q})\sim(\widetilde{q}-2p_F)\ln|\widetilde{q}-2p_F|,
\end{equation}
where we have $\widetilde{q}=|\textbf{p}+\textbf{k}|$ in the
cross-channel. In the coordinate space, the singular part of
$\Pi_{\textrm{sing}}$ leads to Friedel oscillations
(\ref{oscillations}) in the effective interaction (see
Fig.~\ref{friedel}).

Thus, the purely repulsive short-range potential between two
particles in vacuum induces an effective interaction in the
electron gas in a metal with the competition between repulsion and
attraction. It turns out that the singular part in
$U_{\textrm{eff}}$ favors attraction, ensuring a contribution that
always exceeds the repulsive contribution caused by the regular
part of $U_{\textrm{eff}}$. At large orbital momenta $l\gg1$, this
leads to a superconducting instability with the critical
temperature $T_{cl}\sim\varepsilon_F\exp(-l^4)$. In this case, the
conventional singlet pairing in the $s$-wave channel
($S_{\textrm{tot}}=0,\,l=0$) is suppressed by the short-range
Coulomb repulsion caused by the main maximum in $U_{\textrm{eff}}$
(see Fig.~\ref{friedel}), and superconductivity is realized at
large orbital momenta, $l\gg1$. We note that at $l\neq0$, the role
of the main maximum is weakened by the centrifugal potential,
which improves the conditions for the appearance of
superconductivity in channels with anomalous pairing.

From Ref.~\cite{Kohn65}, a nontrivial conclusion followed that no
Fermi systems exist in the normal state at a zero temperature; any
3D electronic system with a purely repulsive interaction between
particles is unstable with respect to the transition to the
superconducting state with a large orbital angular momentum of the
relative motion of a Cooper pair ($l\gg1$). However, the estimates
carried out in~\cite{Kohn65} for the critical temperature in
electronic systems in metals with realistic parameters and for
superfluid helium at $l=2$ gave very low values of the critical
temperature: $T_c\sim10^{-16} \textrm{K}$ for $^3$He and
$T_c\sim10^{-11} \textrm{K}$ for metallic plasma. The low value of
$T_c$ was one of the reasons why the Kohn-Luttinger mechanism was
not popular among researchers for a sufficiently long period and
was unjustly forgotten.

Later on, in Refs~\cite{Fay68,Kagan88}, it was shown that the
temperature of the superconducting transition in~\cite{Kohn65} was
underestimated because of the utilization of an asymptotic
expression for large values of the orbital angular momentum,
$l\gg1$. In reality, at $l=1$, an exact analytic calculation shows
that the contributions to $U_{\textrm{eff}}$ that correspond to
the attraction of quasiparticles dominate over the repulsive
contributions. As a result, the repulsive 3D Fermi gas is unstable
with respect to the superconducting transition with the triplet
$p$-wave pairing at the critical
temperature~\cite{Fay68,Kagan88,Baranov92b,Baranov96}
\begin{eqnarray}\label{Tcp}
T_{c1}&\approx&\varepsilon_F\exp\biggl(-\frac{5\pi^2}{4(2\ln2-1)(ap_F)^2}\biggr)\nonumber\\
&=&\varepsilon_F\exp\biggl(-\frac{13}{\lambda^2} \biggr),
\end{eqnarray}
where $\lambda=\displaystyle\frac{2ap_F}{\pi}$ is the effective 3D
Galitskii gas parameter~\cite{Galitskii58}, We note that for
$l=1$, the contribution from the Kohn anomaly only increases the
value of $T_{c1}$, but does not play a decisive role in the
appearance of the triplet superconductivity itself.

It was demonstrated in Ref.~\cite{Kagan89} the critical
temperature of a superfluid transition can be substantially
increased already at low fermionic densities by placing the system
of neutral Fermi particles into a magnetic field or by creating
spin polarization ($n_{\uparrow}>n_{\downarrow}$). This occurs
because the paramagnetic suppression of the superconductivity
(which takes place for $s$-wave pairing) is absent in the $p$-wave
channel for the so-called $A_1$ phase and the increase in $T_c$ is
possible due to the enhancement of the effective interaction and
the changes in the character of the Kohn anomaly. The highest
critical temperatures then correspond to the $A_1$ phase, where
the Cooper pair is formed by two spins up, and the effective
interaction for them is prepared by two spins down. In this case,
$T_c$ is a function of the ratio of the density of spin-up
particles to the density of spin-down particles,
$~n_{\uparrow}/n_{\downarrow}$, or more precisely of the spin
polarization
$\alpha=\displaystyle\frac{n_{\uparrow}-n_{\downarrow}}{n_{\uparrow}+n_{\downarrow}}$.

In the case of a repulsive 2D electron gas, in the first two
orders of the perturbation theory in the gas parameter, the
effective interaction takes the form~\cite{Chubukov93,Efremov00b}
\begin{equation}
U_{\textrm{eff}}(\textbf{p},\textbf{k})=\frac{4\pi}{m}f_0+
\Bigl(\frac{4\pi}{m}f_0\Bigr)^2\Pi(\textbf{p}+\textbf{k}),
\end{equation}
where $f_0=\displaystyle\frac{1}{2\ln(p_Fr_0)}$ is the Bloom 2D
gas parameter~\cite{Bloom75} and $\Pi(\textbf{p}+\textbf{k})$ is
the 2D polarization operator in the cross channel.

In the 2D situation, the effective interaction in the coordinate
space also contains Friedel oscillations:
\begin{equation}\label{oscillations2D}
U_{\textrm{eff}}(r)\sim f_0^2\frac{\cos(2p_Fr)}{(2p_Fr)^2},
\end{equation}
which are even much stronger than in the 3D case. But in the
momentum space, the 2D Kohn anomaly has one-sided
character~\cite{Afanasiev62},
\begin{equation}\label{one_sided}
U^{\textrm{sing}}_{\textrm{eff}}(\widetilde{q})\sim
f_0^2\textrm{Re}\sqrt{\widetilde{q}-2p_F}=0
\end{equation}
for $\widetilde{q}=|\textbf{p}+\textbf{k}|\leq2p_F$, and,
therefore, is ineffective for the problem of superconductivity (in
which $\widetilde{q}\leq2p_F$). Thus, the 2D Fermi gas with
repulsion remains in the normal state at least in the first two
orders of the perturbation theory in the gas parameter $f_0$.
Nevertheless, it was shown in~\cite{Chubukov93} that the
superconducting $p$-wave pairing appears in the next (third) order
of the perturbation theory in $f_0$, in which, for the singular
contribution to the effective interaction, the expression under
the square root in (\ref{one_sided}) reverses sign:
\begin{equation}
U^{\textrm{sing}}_{\textrm{eff}}(\widetilde{q})\sim
f_0^3\textrm{Re}\sqrt{2p_F-\widetilde{q}}.
\end{equation}
In this case, an exact calculation~\cite{Efremov00a} of the
critical temperature taking all irreducible third-order diagrams
into account, yields
\begin{equation}\label{Tcp3order}
T_c\sim\varepsilon_F\biggl(-\frac{1}{6.1f_0^3} \biggr).
\end{equation}
For the submonolayers of $^3$He on the surface of superfluid films
of $^4$He~\cite{Oh94}, the temperature of the superconducting
transition is estimated as $10^{-4}$K~\cite{Chubukov93,Efremov00a}
for the maximal densities at which the Fermi-gas description is
still applicable, and this estimate is quite reasonable for
experimental observation.

\section{Superconductivity in the 3D and 2D Hubbard model with repulsion}
\label{sec_Hubbard}

In connection with the discovery of high-temperature
superconductivity~\cite{Bednorz86}, the Hubbard
model~\cite{Hubbard63} acquired substantial popularity as one of
the basic models for describing the anomalous properties of
cuprates. The Hubbard model is a particular case of the general
model of interacting electrons, whose band structure is described
within the strong-coupling approximation and is the minimal model
that accounts for band electron motion in the metal as well as the
strong electron-electron
interaction~\cite{Izyumov94,Izyumov95,Georges96,Tasaki98,VVVSGO01}.
This model is of special importance in the description of
narrow-band metals~\cite{Efremov00a}. The Hamiltonian of the
Hubbard model on a lattice has the form
\begin{eqnarray}\label{HubbardHamiltonian}
\hat{H'}&=&\sum\limits_{f\sigma}(\varepsilon-\mu)
n_{f\sigma}+\sum\limits_{fm\sigma}t_{fm}c_{f\sigma}^\dag
c_{m\sigma} +U\sum\limits_f
n_{f\uparrow}n_{f\downarrow},\nonumber\\
\end{eqnarray}
where  $c_{f\sigma}^\dag (c_{f\sigma})$  is the operator of
creation (annihilation) of an electron with a spin projection
$\sigma=\pm1/2$ at a site $f$; $\varepsilon$ is the on-site energy
of the electron; $\mu$ is the chemical potential of the system;
$n_f=\sum\limits_\sigma n_{f\sigma}=\sum\limits_\sigma
c_{f\sigma}^\dag c_{f\sigma}$ is the operator of the number of
particles on the site $f$; the matrix element $t_{fm}$ stands for
electron hoppings from site $f$ to site $m$; and $U$ is the
parameter of the Coulomb interaction of two electrons that are
located on the same site and have opposite spin projections
(Hubbard repulsion).
\begin{figure}[t]
\begin{center}
\includegraphics[width=0.49\textwidth]{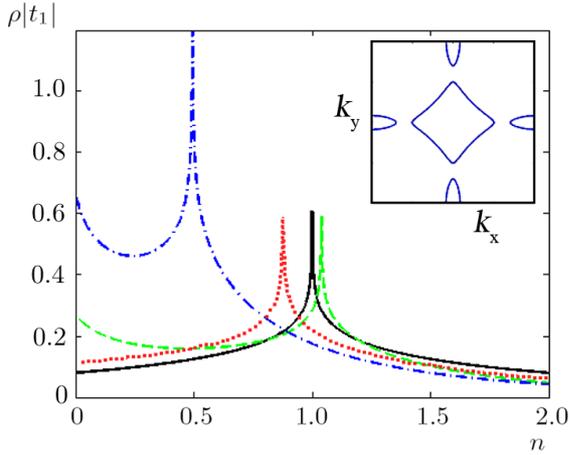}
\caption{Fig.~3. Modification of the density of electronic states
and the shift of the Van Hove singularity in the Hubbard model on
a square lattice upon a change in the hopping integrals:
$t_2=t_3=0$ (solid curve), $t_2=0.15,~t_3=0$ (dotted curve),
$t_2=0.15,~t_3=0.1$ (dashed curve), $t_2=0.44,~t_3=0$
(dashed-dotted curve). The inset shows the formation of a
multisheet Fermi contour at $t_2=0.44,~t_3=-0.1,~\mu=2$ (all the
parameters are given in the units of $|t_1|$)~\cite{Kagan14d}.}
\label{DOS_Hubbard}
\end{center}
\end{figure}

Since extensive experimental data have indicated that the basic
dynamics of Fermi excitations in cuprates is developed in the
CuO$_2$ planes, the nonphonon mechanisms of superconductivity were
mainly based on the 2D Hubbard model on a simple square lattice.
In the momentum space, the Hamiltonian of the model has the form
\begin{eqnarray}\label{Hubbard_momentum}
\hat{H'}
&=&\sum\limits_{\textbf{p}\sigma}(\varepsilon_{\textbf{p}}-\mu)
c^{\dagger}_{\textbf{p}\sigma}c_{\textbf{p}\sigma} +
U\sum_{\textbf{p}\textbf{p'}\textbf{q}}c^{\dagger}_{\textbf{p}\uparrow}
c^{\dagger}_{\textbf{p'}+\textbf{q}{\downarrow}}c_{\textbf{p}+\textbf{q}{\downarrow}}
c_{\textbf{p'}{\uparrow}},\nonumber\\
\end{eqnarray}
where the energy of an electron, including distant hoppings, which
are determined by the parameters $t_2$ and $t_3$, is given by
\begin{eqnarray}
\varepsilon_{\textbf{p}}&=&2t_1(\textrm{cos}\,p_xa+\textrm{cos}\,p_ya)+
4t_2\textrm{cos}\,p_xa\,\textrm{cos}\,p_ya \nonumber\\
&+&2t_3(\textrm{cos}\,2p_xa+\textrm{cos}\,2p_ya).\label{Hubbard_spectra}
\end{eqnarray}
where $a$ is the lattice constant (intersite distance).

We note that when simulating electron spectrum
(\ref{Hubbard_spectra}) and constructing the phase diagram of the
superconducting state in the Hubbard model, going beyond the
framework of the nearest-neighbor approximation
($t_2\neq0,\,t_3\neq0$) becomes essential. This is because the
leading contribution to the effective coupling constant comes from
the interaction of electrons that are located near the Fermi
surface, whose geometry depends on the structure of the energy
spectrum. An important role is also played by the fact that
account for distant hoppings shifts the Van Hove singularity in
the density of electronic states from the position at half-filling
($n=1$) into the region of smaller or higher electronic densities
(Fig.~\ref{DOS_Hubbard}). We note that the introduction of
hoppings to the third coordination sphere of the square lattice,
$t_3\neq0$, can lead to a qualitative change in the geometry of
the Fermi surface, which connected with the formation of a
multisheet Fermi contour (see the inset in
Fig.~\ref{DOS_Hubbard}).

Thus, an account for distant hoppings can lead to a modification of
the phase diagram that determines the regions of the realization
of the superconducting states with different types of the order
parameter symmetry.

In the Hubbard model, the perturbation theory can be constructed
in two limiting cases: (1) the Born weak-coupling approximation,
$U\ll W$ ($W=2zt$ is the bandwidth; $z$ is the number of nearest
neighbors) and an arbitrary electron density, $0<n<1$; and (2) the
strong-coupling approximation $U\gg W$ at low electron density,
$n\ll1$. The utilization of the weak-coupling approximation $U\ll
W$ in the analysis of the feasibility of Kohn-Luttinger
superconducting pairing allows us to calculate $U_{\textrm{eff}}$
for the Cooper channel at all the densities $0<n\leq1$ by
restricting ourselves to the second order diagrams in the
interaction $U$ (see Fig.~\ref{diagrams_alpha}). In the opposite
limit of strong coupling, $U\gg W$, the restriction to first- and
second-order diagrams is justified only in the region of low
electron density $n<<1$, where the Galitskii-Bloom Fermi-gas
expansion is valid~\cite{Galitskii58,Bloom75}.

In one of the first studies~\cite{Baranov92a}, the authors
analyzed the conditions for the realization of the Kohn-Luttinger
superconductivity in the 2D Hubbard model with Hamiltonian
(\ref{Hubbard_momentum}) in the weak-coupling limit $U\ll W$, in
the nearest-neighbor approximation ($t_2=t_3=0$) at low electron
densities ($p_Fa\ll1$). In this case, the following expansion is
valid for the electron spectrum:
\begin{eqnarray}
\varepsilon_{\textbf{p}}-\mu&=&2t_1(\textrm{cos}\,p_xa+\textrm{cos}\,p_ya)-\mu\nonumber\\
&\approx&\frac{p^2-p^2_F}{2m}-\frac{(p_x^4+p_y^4)a^2}{24m}+\frac{(p_x^6+p_y^6)a^4}{720m}.
\end{eqnarray}
where $m=1/2t_1a^2$ is the band mass. It can be seen that in the
chosen approximation, the bare spectrum of electrons at $p_Fa\ll1$
almost coincides with the spectrum of a free Fermi gas, and the
Hubbard Hamiltonian is equivalent to the Hamiltonian of a weakly
nonideal Fermi gas with short-range repulsion between the
particles~\cite{Baranov91}. To determine the possibility of the
superconducting pairing in this approximation, the effective bare
vertex for the Cooper channel was calculated in first two orders
of perturbation theory:
\begin{eqnarray}\label{Gamma_waveU}
&&U_{\textrm{eff}}(\textbf{p},\textbf{k})=U+U^2\Pi(\textbf{p}+\textbf{k}),
\end{eqnarray}
where  $\Pi(\textbf{p}+\textbf{k})$ is polarization operator
(\ref{Lindhard}). To solve the problem of superconducting pairing,
the function $U_{\textrm{eff}}(\textbf{p},\textbf{k})$ was
expanded in a series with eigenfunctions of the irreducible
representations of the $C_{4v}$ symmetry group of the square
lattice (see Section~\ref{sec_SV}~), and then the sign of the
expressions for $U_{\textrm{eff}}^{\gamma}$ was analyzed for each
representation $\gamma$. As a result, it was shown that the 2D
electron system described by the Hubbard model for small filling
and $U\ll W$ is unstable towards the superconducting pairing with
the $d_{xy}$-type symmetry of the order parameter
$\Delta(\phi)\sim\sin(4m+2)\phi$, where the integer $m$ satisfies
the condition $m\in[0,\infty)$.
\begin{figure}[t]
\begin{center}
\includegraphics[width=0.41\textwidth]{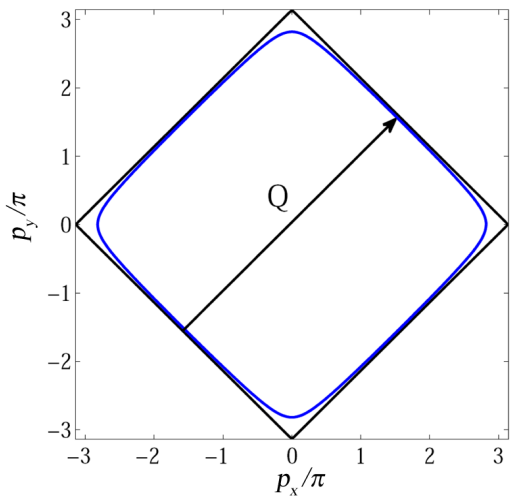}
\caption{Fig.~4. Fermi surface for a nearly half filling
($n\approx1$) in the 2D Hubbard model on a square lattice:
$\textbf{Q}=(\pi/a,\pi/a)$ is the nesting vector~\cite{Kagan14d}.}
\label{nesting}
\end{center}
\end{figure}

The weak-coupling limit $U\ll W$ in the 3D and 2D Hubbard model
near the half-filling, $n\sim1$, was analyzed
in~\cite{Scalapino86a,Scalapino86b,Kozlov89}. In the 2D
case~\cite{Kozlov89}, in the nearest-neighbor approximation at
$n\approx1$, the electron spectrum becomes quasihyperbolic
\begin{equation}
\varepsilon_{\textbf{p}}\approx\pm\frac{p_x^2-p_y^2}{2m}
\end{equation}
near the corner points $(0,\pm \pi)$ and $(\pm \pi,0)$ at which
the Fermi surface almost touches the Brillouin zone
(Fig.~\ref{nesting}). As it is well known, the density of
electronic states has a logarithmic singularity in these regions
near the Van Hove filling, namely
$\rho(E)\displaystyle\sim\ln\frac{t}{|\mu|}$, where $|\mu|\ll t$
is the modulus of the chemical potential near the half-filling. It
can be seen from Fig.~\ref{nesting} that there are almost flat
regions of the Fermi surface, which satisfy the condition for
ideal nesting in the exactly half-filled case ($n=1$):
\begin{equation}\label{epQ}
\varepsilon_{\textbf{p}+\textbf{Q}}=-\varepsilon_{\textbf{p}}.
\end{equation}
where $\textbf{Q}=(\pi/a,\pi/a)$ is the nesting vector for the 2D
square lattice. In these regions, the polarization operator reads
$\Pi(\textbf{Q})\displaystyle\sim\ln^2\frac{t}{|\mu|}$~\cite{Kozlov89,Dzyaloshinskii88b},
where one logarithmic factor comes from the density of states and
the other one is due to the Kohn anomaly. The following quantity
serves here as the parameter of the perturbation theory in the 2D
weak-coupling limit:
\begin{equation}
f_0=\frac{U}{8\pi t}\ll1,
\end{equation}
and in the second order of the perturbation theory in $f_0$, the
expression for the effective interaction takes the form
\begin{equation}
U_{\textrm{eff}}\sim f_0+f_0^2\ln^2\frac{t}{|\mu|}.
\end{equation}
Since the expression for the Cooper loop $L$ at $n\approx1$, apart
from the usual Cooper logarithm, also contains a logarithm due to
the Van Hove singularity, we have
\begin{eqnarray}
L(\xi_{\textbf{p}})=\displaystyle
\frac1N\sum_{\textbf{p}}\frac{\tanh(\xi_{\textbf{p}}/2T)}{2\xi_{\textbf{p}}}\sim\ln\frac{\mu}{T}
\ln\frac{t}{|\mu|},
\end{eqnarray}
where $\xi_{\textbf{p}}=\varepsilon_{\textbf{p}}-\mu$. Therefore,
the expression for the critical temperature with the order
parameter of the typical for cuprates $d_{x^2-y^2}$-wave symmetry
obtained in Ref.~\cite{Kozlov89} in the leading logarithmic
approximation has the form
\begin{equation}
f_0^2\ln^3\frac{t}{|\mu|}\ln\frac{\mu}{T_c}\sim1,
\end{equation}
or
\begin{equation}\label{Tcdx2y2}
T^{d_{x^2-y^2}}_c\sim\mu\exp\Biggl(-\frac{1}{f_0^2\ln^3\frac{t}{|\mu|}}\Biggr).
\end{equation}
It can be seen from expression (\ref{Tcdx2y2}) that the
denominator in the right-hand side, in spite of the low value of
$f_0^2$ at $f_0\,\ll\,1$, increases substantially due to the large
value of $\displaystyle\ln^3\frac{t}{|\mu|}\gg1$.

The results in~\cite{Baranov92a} on the realization of
$d_{xy}$-wave pairing at $n\lesssim0.6$ and on the
$d_{x^2-y^2}$-wave pairing at
$n\sim1$~\cite{Scalapino86a,Scalapino86b,Kozlov89} in the
weak-coupling limit were subsequently confirmed by other authors.
In~\cite{Hlubina99}, a phase diagram of the superconducting state
of the 2D Hubbard model was constructed at small and intermediate
occupation numbers, which shows how the results of the competition
of different types of the order parameter symmetry to depend on
the value of the parameter $t_2$ of the electron hoppings to the
next-nearest-neighbor sites. The phase diagram obtained in the
second order of the perturbation theory shows that at $t_2=0$ in
the region of the low electron densities, $0<n<0.52$,
superconductivity with the $d_{xy}$-type symmetry of the order
parameter is realized. In the range of $0.52<n<0.58$, the ground
state corresponds to the phase with the $p$-wave pairing, and at
$n>0.58$ the $d_{x^2-y^2}$-wave pairing appears. Similar results
were obtained in~\cite{Zanchi96} in the framework of the
renormalization group approach.

In vicinity of the half-filling, $0.95<n<1$, where strong
competition between superconductivity and antiferromagnetism is
observed, the problem of Cooper instability was examined in
Refs~\cite{Dzyaloshinskii88a,Dzyaloshinskii88b,Zheleznyak97}. In
these studies, the so-called parquet diagrams were summed up and
at $\mu\sim T_c$ the relation
\begin{equation}
f_0^2\ln^4\frac{t}{|\mu|}\sim f_0^2\ln^4\frac{t}{T_c}\sim1.
\end{equation}
was obtained. This yields an elegant estimate for the maximal
critical temperature:
\begin{equation}\label{Tcdx2y2_parquet}
T^{d_{x^2-y^2}}_c\sim
t\exp\Biggl(-\frac{\textrm{const}}{\sqrt{f_0}}\Biggr).
\end{equation}

The maximal critical temperature of the superconducting transition
in the 2D Hubbard model was obtained in Ref.~\cite{Raghu10} in the
regime $U/W\sim1$ at optimal electron concentrations
$n\sim0.8-0.9$. According to the estimate obtained
in~\cite{Raghu10}, the critical temperature can reach the values
$T^{d_{x^2-y^2}}_c\approx100\,\textrm{K}$, which is quite
reasonable for optimally doped cuprate superconductors.

\section{Enhancement of the critical temperature in the two-band Hubbard
model and in a spin-polarized Fermi gas}
\label{sec_Hubbard_two_bands}

Along with the possibility to enhance the anomalous
superconductivity by applying a magnetic field to a system of
neutral Fermi particles as described in Section~\ref{sec_gas}~,
there is also another possibility to increase $T_c$ significantly
already at a low electron density. It is related to the two-band
situation~\cite{Kagan91} or a multilayer system with geometrically
separated layers. In this case, the role of spins-up is played by
the electrons of the first band (or layer), and the role of
spins-down by the electrons of the second band (or layer). The
coupling between electrons of two bands is achieved by means of
the interband Coulomb interaction $U_{12}n_1n_2$. As a result, the
following exciton-type mechanism of superconductivity becomes
possible: the electrons of one band form a Cooper pair via
polarization of the electrons of the other
band~\cite{Kagan91,KaganValkov11a,KaganValkov11b}. In this case,
the role of spin polarization is played by the relative filling of
two bands $n_1/n_2$.

We now examine the two-band Hubbard model with one broad band and
one narrow band~\cite{KaganValkov11a,KaganValkov11b}, which
accordingly contains ‘heavy’ ($n_1=n_H$) and ‘light’ ($n_2=n_L$)
electrons. This model is a natural generalization of the
well-known Falikov-Kimball model~\cite{Falicov69} for systems with
a mixed valence; however, it exhibits a richer physics in view of
the presence of a finite bandwidth of heavy electrons instead of a
localized level. In the Hubbard model with one narrow band, the
effective interaction, as it was shown in
Refs~\cite{KaganValkov11a,KaganValkov11b,Kagan86,Kagan87}, is
determined mainly by the interband Coulomb repulsion of heavy and
light electrons  $U_{12}=U_{HL}$. The corresponding critical
temperature of the superconducting transition depends
nonmonotonically on the relative filling of the bands $n_H/n_L$
and exhibits a wide and clearly pronounced maximum at $n_H/n_L=4$
in the 2D case. The maximal critical temperature can be expressed
as
\begin{equation}\label{Tcpolaron}
T_{c\,\textrm{max}}=T_{c1}\biggl(\frac{n_H}{n_L}\approx4\biggr)=\varepsilon_F\exp\biggl(-\frac{1}{2f_0^2}\biggr),
\end{equation}
and corresponds to the triplet $p$-wave pairing of heavy particles
via the polarization of light particles. In the Born weak-coupling
approximation, the effective gas parameter depends linearly on the
interband Coulomb repulsion and on the square root of the product
of the masses~\cite{Kagan91}:
$$f_0=\frac{\sqrt{m_Hm_L}}{2\pi}U_{HL}.$$
In the opposite limit of strong
coupling~\cite{KaganValkov11a,KaganValkov11b},
$$f_0=\sqrt{\frac{m_H}{m_L}}\frac{1}{\ln[1/(p_Fa)^2]}.$$
In the so-called unitary limit of the strongly screened Coulomb
interaction ($f_0\rightarrow1/2$) and of the strong
electron-polaron effect, $m^*_H/m_L\sim(m_H/m_L)^2$.
Correspondingly, the maximal critical temperature in this case
yields~\cite{KaganValkov11a,KaganValkov11b}
\begin{equation}
T_{c1}\sim\varepsilon^*_{FH}\exp\biggl(-\frac{1}{2f_0^2}\biggr)\sim\varepsilon^*_{FH}\exp(-2),
\end{equation}
where $\varepsilon^*_{FH}$ is the renormalized (strongly narrowed)
Fermi energy of heavy particles
$$\varepsilon^*_{FH}=\frac{p^2_{FH}}{2m^*_H}\sim30-50\,\textrm{K}.$$
Let us stress that in the unitary limit the sharp increase in the
effective mass of heavy particles up to $m^*_H\sim100m_e$ is
caused by the many-body electron-polaron
effect~\cite{Kagan86,Kagan87}. As a result, quite reasonable
critical temperatures are obtained for the superconducting
$p$-wave pairing: $T_c\sim5\,\textrm{K}$, typical for
uranium-based heavy-fermion compounds, mentioned in the
Introduction, U$_{1-x}$Th$_x$Be$_{13}$ and UNl$_2$Al$_3$ with
large effective masses $m^*\sim100-200m_e$~\cite{Ott84,Kromer98}
and for organic superconductors.

We note that the electron-polaron effect, which leads to a
significant increase in the effective mass in the model, is
connected with the nonadiabatic part of the wave function that
describes a heavy electron surrounded by a cloud of virtual
electron-hole pairs that belong to the band of light
electrons~\cite{Kagan86,Kagan87}.

The discussed mechanism of superconductivity possibly can be
realized in the bismuth- and thallium-based cuprate
superconductors, as well as in the $\textrm{PbTe-SnTe}$
superlattices~\cite{Murase86} and dichalcogenides CuS$_2$ and
CuSe$_2$ with geometrically separated layers. We note that, in
general, the two bands can belong to the same layer or to
different layers. It is also reasonable to assume that this
mechanism can be fulfilled in the ruthenates
Sr$_2$RuO$_4$~\cite{Maeno01} and in the ultracold Fermi gas of
$^6$Li atoms in magnetic traps with a strong imbalance of the
hyperfine components (see Section~\ref{sec_perspectives}~).

We note again that in the presence of the band of heavy and light
electrons with strongly different masses, $m_H\gg m_L$, and
different densities, $n_H>n_L$, the critical temperature $T_c$ is
determined mainly by the pairing of the heavy electrons via the
polarization of the light electrons. However, taking into account
even an infinitely small Geilikman-Moskalenko-Suhl
term~\cite{Suhl59,Geilikman65,Geilikman66,Geilikman73,Baranov92c}
$K\displaystyle\sum_{\textbf{p}\textbf{p}'}a^{\dag}_{\textbf{p}}a^{\dag}_{\textbf{-p}}b_{\textbf{p}'}b_{\textbf{-p}'}$
(where $K$ is the parameter of interaction corresponding to the
rescattering of the Cooper pairs between the ‘heavy’ and ‘light’
bands), leads to the opening of superconducting gaps in both bands
at the same critical temperature close to one in
(\ref{Tcpolaron}).

Let us consider the application of this theory to low-temperature
physics in more details. We emphasize that the ultracold quantum
gases in magneto-dipole traps, as well as with the spin-polarized
solutions of $^3$He in $^4$He, especially in the
quasi-two-dimensional situation (in which the largest increase in
the temperature of the triplet $p$-wave pairing occurs), are the
excellent systems for verifying the theoretical predictions
of~\cite{Kagan94_1,Baranov96,Kagan89,Kagan91,KaganValkov11a,KaganValkov11b,Baranov93b}.
Good experimental opportunities for the realization of the
‘high-temperature’ superfluidity in spin-polarized (imbalanced)
Fermi gases in quasi-two-dimensional magnetic traps, in
particular, has G E Thomas’s group in North Carolina~\cite{Ong15}.
We also note that in the 1990s G Frossati’s group in Leyden
experimentally obtained a 20\% increase (from 2.5 up to 3.14 mK)
in the critical temperature of the superfluid transition
$T_{c1}^{\uparrow\uparrow }$ in the $A_1$ phase of the superfluid
$^3$He in magnetic fields up to $H=15$ T (at a spin polarization
of 7\%)~\cite{Frossati90,Wiegers90}. In this case, at the maximum
for the critical temperature of the triplet $p-$wave pairing
(reached at the spin polarization
$\alpha=\displaystyle\frac{n_{\uparrow}-n_{\downarrow}}{n_{\uparrow}+n_{\downarrow}}=48$\%),
the theory of~\cite{Kagan94_1,Kagan89} predicts an increase in
$T_{c1}^{\uparrow\uparrow }$ by a factor of 6.4 in comparison with
the nonpolarized case. A similar value for
$T_{c1}^{\uparrow\uparrow }$ at the maximum (with the maximum
$T_{c1}^{\uparrow\uparrow }=5.6 T_{c1}$), but for a 35\% spin
polarization, was also predicted in the metamagnetic model in
Refs~\cite{Bedell86,Frossati86} in the framework of the so-called
$s-p$ approximation for the Landau Fermi liquid
theory~\cite{Dy69}. We note, however, that the approach
in~\cite{Kagan94_1,Kagan89} in the framework of the enhanced
Kohn-Luttinger mechanism of superfluidity is characterized by the
only one fitting parameter, the gas parameter $ap_F$, and
therefore this approach is more model-independent than the one
in~\cite{Bedell86}, which uses two fitting parameters that are not
connected with each other, namely, the $s$ and $p$ harmonics of
the scattering amplitude of quasiparticles on the Fermi surface.

We also note that for solutions of $^3$He in $^4$He, the theory
of~\cite{Kagan94_1} and the results in~\cite{Bashkin81,Ostgaard92}
predict a phase diagram for the fermionic superfluidity of $^3$He
in the 3D case with the regions of $s$-wave pairing at small
concentration of $^3$He in the solution ($x=1-2\%$) and the
regions of $p-$wave pairing at larger concentrations of $^3$He
($x>2-4\%$). The critical temperatures of the $s$-wave pairing in
the solutions are maximal at a zero magnetic fields and at
$x\sim1\%$. According to the estimates
in~\cite{Bashkin81,Ostgaard92}, the $s$-wave critical temperature
is in the range $10^{-4}-10^{-5}\,\textrm{K}$.

The temperature of the triplet $p$-wave pairing grows sharply in a
magnetic field, and at the maximal possible concentration
$x_{max}=9.5\%$ of $^3$He in the solution in the field  $H\sim15$
T, we have $T_{c1}^{\uparrow\uparrow
}\sim10^{-5}-10^{-6}\,\textrm{K}$. In 2D solutions of $^3$He in
$^4$He, that is for the submonolayers of $^3$He at the Andreev
levels~\cite{Andreev66,Zinovieva69} (appearing similarly to Tamm
levels on the free surfaces of thin films of superfluid
$^4$He~\cite{Alikacem91,Higley89}) or on the
grafoils~\cite{Saunders90,Lusher91}, the phase diagram of the
solution also contains the regions of $s$- and $p$-wave pairings.

We emphasize that for $s$-wave pairing in 2D systems, usually two
phenomena are realized simultaneously: the pairing of two
particles in the real space (with the formation of a molecule or a
dimer) and the Cooper pairing in the momentum space. The maximal
$T_c$ for the $s$-wave pairing, according
to~\cite{Miyake83,Randeria89,Schmitt89}, is in the range of
$10^{-3}-10^{-4}\textrm{K}$ at the 2D density $n_3\sim0.01$ of a
monolayer~\cite{Bashkin81,Ostgaard92}. At the same time, the
temperature of the triplet $p$-wave pairing can be increased in
magnetic fields, and at $H\sim15$ T and the 2D density
$n_3\sim0.05$ of a monolayer, it can become quite accessible
experimentally ($T_{c1}^{\uparrow\uparrow }\sim1 \textrm{mK}$,
according to~\cite{Kagan94_1}). The experimental observation of
fermionic superfluidity in the 3D solutions and submonolayers of
$^3$He remains a challenge for the ‘low-temperature
community’~\cite{Kagan14e}. A similar phase diagram with the
regions of $s$- and $p$-wave pairings was theoretically predicted
in~\cite{Baranov96} and in~\cite{Stoof96} for the fermionic
isotope of lithium $^6$Li on the Cooper (BCS) side of the
crossover with a region of the Bose-Einstein condensation (BEC) in
the regime of the $s$-wave Feshbach resonance.

Let us recall that in the fermionic $^6$Li, the quasiresonance
scattering length, which is very large in absolute value
($a=-2.3\cdot10^3 \textrm{{\AA}}$), corresponds to attraction. As
a result, in the balanced case ($n_1=n_2$), a singlet $s$-wave
pairing with the critical temperature determined by formula
(\ref{Tcs}) is realized for the two hyperfine components of the
nuclear spin $I$ that are captured in a magnetic trap. The maximal
temperature $T_{c0}$ in the 3D case, according to~\cite{Stoof96},
is of the order of $10^{-6}\textrm{K}$ at
$\varepsilon_F\sim10^{-5}\,\textrm{K}$. However, if the imbalance
between the hyperfine components is sufficiently large, such that
$(n_1-n_2)/(n_1+n_2)>T_{c0}/\varepsilon_F$, then, according to the
Landau criterion of superfluidity~\cite{Baranov96}, the $s$-wave
pairing is totally suppressed. Nevertheless, in this case a
$p$-wave pairing can arise if the Cooper pairs (as is the case of
the $A_1$ phase of superfluid $^3$He) are formed by Fermi
particles of one hyperfine component while the effective
interaction for them is prepared by Fermi particles of another (or
others) hyperfine component. In this situation, the maximal
critical temperature
$T_{c1}^{\uparrow\uparrow}\sim\varepsilon_F\exp(-7/\lambda ^2)$ of
the $p$-wave pairing for the optimal ratio of the densities of the
hyperfine components, according to~\cite{Baranov96}, can reach
$10^{-7}\,\textrm{K}$ at $\varepsilon_F\sim10^{-5}\,\textrm{K}$
and $\lambda\sim1$.

The effect of $T_c$ enhancement in total analogy with the
solutions of $^3$He in $^4$He, for the $p$-wave pairing in the
imbalanced gases manifests itself much more strongly and clearly
in the quasi-two-dimensional situation~\cite{Kagan91}. Therefore,
the experimental achievements obtained in~\cite{Ong15}, which make
it possible to create the quasi-two-dimensional traps and to
control their essential parameters (such as the density,
temperature, and number of particles on layer-by-layer basis) are
very important.

We finally consider one more promising prediction of this theory.
It was shown in Ref.~\cite{Baranov93b}, that in
quasi-two-dimensional (layered) materials in a magnetic field that
is strictly parallel to the layer, the appearing vector potential
$A_y=Hz$ ($\textbf{H}=H \textbf{e}_x$, with $x$ and $y$ being the
coordinates in the layer) does not change the motion of the
electrons and Cooper pairs in the plane of the layer. Therefore,
the diamagnetic Meissner effect is completely suppressed. As a
result, the electronic monolayer (or the layered system) becomes
equivalent to an uncharged (neutral) fermionic layer of $^3$He.

Thus, for low-density quasi-two-dimensional systems, the phase
diagram of the superconducting state in a magnetic field parallel
to the electronic layer can contain again the regions of
conventional $s$-wave pairing in the absence of a magnetic field
and the regions of triplet $p-$wave pairing (similar to the $A_1$
phase of the superfluid $^3$He) in strong magnetic fields, when
the $s$-wave pairing is totally suppressed paramagnetically.
Moreover, in the magnetic fields $H\sim15$ T and at low Fermi
energies ($\varepsilon_F\sim30\,\textrm{K}$) for sufficiently
large degrees of spin polarization of electrons ($\alpha\geq10\%$)
the reasonable critical temperatures ($T_{c1}^{\uparrow\uparrow
}\sim0.5\,\textrm{K}$) can be obtained. Of course, in this case,
as in the case of graphene (see Sections 8 and 9), the
experimentalists should be very careful analyzing the role of the
structural disorder and nonmagnetic impurities, which lead to the
isotropization of the order parameter and therefore suppress the
nonspherical $p$-wave pairing~\cite{Abrikosov60,Larkin70}.
Furthermore, it is necessary to ensure a high degree of the
parallelism of the magnetic field to the plane of the layer,
because the presence of even a relatively small perpendicular
component would lead to the diamagnetic suppression of
superconductivity~\cite{Kagan14e}. Nevertheless, the proposed
mechanism is very interesting for the possible realization of
superconductivity in very pure heterostructures (see
Section~\ref{sec_perspectives}~).

\section{Nontrivial corrections to the Landau Fermi liquid theory in 2D low-density systems}
\label{sec_UHB}

It is well known that a high temperature of the superconducting
transition in cuprate superconductors is connected with very
unusual properties of these systems in the normal
(nonsuperconducting) state. Among the unconventional properties of
the normal phase, the small jump in the distribution function of
the interacting particles on the Fermi surface and a linear
temperature dependence of the resistivity at temperatures much
lower than Debye temperature in optimally doped cuprate
superconductors are of interest. To explain these facts, Anderson
advanced a concept of a Luttinger-type Fermi liquid with a zero
jump of the distribution function on the Fermi
surface~\cite{Anderson90}. A similar idea of a marginal Fermi
liquid, which is a special case of a Luttinger liquid, was
proposed by the authors of~\cite{Varma89} based on the analysis of
the experimental data.

Later on, Anderson advanced an even more nontrivial idea that not
only a high-density strongly interacting 2D Fermi system but even
a weakly interacting low-density 2D Fermi gas should also
described by a Luttinger Fermi liquid~\cite{Anderson91} rather
than by the Landau Fermi-liquid theory with a finite jump of the
distribution function. In Refs~\cite{Anderson90,Anderson91},
Anderson formulated three important points, which led to his
doubts regarding the applicability of the standard Galitskii-Bloom
Fermi-gas approach~\cite{Bloom75,Galitskii58} in the 2D case.
These are, firstly, the problem of the finite scattering
phase-shift for the particles with almost parallel spins, which
leads to the vanishing of the $Z$-factor (Migdal jump) on the
Fermi surface; secondly, the problem (connected with the first
one) concerning the essential role of the upper Hubbard band in
the lattice models already in the case of a low electron density;
and, finally, the problem of the possible existence of a strong
singularity in the Landau $f$-function for the quasiparticles
interaction, which arises in a 2D Fermi gas even in the absence of
a lattice.

Many theorists participated in the discussion developed after the
publication of Anderson’s work; most of
them~\cite{Engelbrecht90,Fukuyama91,Fabrizio92,Prokofiev93,Baranov93}
supported the Fermi-gas ideology and attempted to prove its
consistency in the 2D case using ladder and parquet approximations
in terms of the diagrammatic technique. Anderson continued to
insist on his point of view, assuming that the diagrammatic
technique is inapplicable to the 2D systems even at the level of
summing up an infinite series of parquet diagrams. In fact, the
dispute considered the problem of the choice of a correct ground
state, which would allow us to construct a regular procedure of
successive approximations in the interaction (or, to be more
precise, in its part that was not taken into account in choosing
the ground state). According to Anderson’s qualitative
considerations, the Landau function $f(\textbf{p},\textbf{p}')$ of
the interaction of quasiparticles with almost parallel momenta
$\textbf{p}$ and $\textbf{p}'$ and opposite spins of the colliding
particles in the 2D case, when there is a small deviation from the
Fermi surface for $\textbf{p}$ and $\textbf{p}'$, contains a
singular part of the form
\begin{equation}
f_{\textrm{sing}}(\textbf{p},\textbf{p}')\sim\frac{1}{|\textbf{p}-\textbf{p}'|}.
\end{equation}
The existence of such a strong singularity leads to a logarithmic
divergence of all Landau harmonics $f_0,f_1,\ldots$, and, thus, to
the complete crush of the Fermi-liquid theory. The accurate
calculation of the Landau quasiparticles interaction function
$f(\textbf{p},\textbf{p}')$ carried out in the second order of the
perturbation theory in Ref.~\cite{Prokofiev93} and, independently,
in~\cite{Baranov93}, leads to a significantly weaker singularity
in $f$ of the form $|\textbf{p}-\textbf{p}'|^{-1/2}$, which, in
addition to that, exists only in a small window of angles
$\phi\propto|\textbf{p}-\textbf{p}'|^{3/2}$ near the parallel
orientation of momenta $\textbf{p}$ and $\textbf{p}'$. As a
result, this singularity leads only to nontrivial temperature
corrections to the $f$-function rather than to the destruction of
the Fermi-liquid picture as whole.

Concerning the second point of the discussion raised by Anderson,
the authors of~\cite{Woelfle11} examined the 2D Hubbard model in
the limit of strong coupling ($U\gg W$) and low electron density
($n\ll1$) in the Kanamori $T-$matrix
approximation~\cite{Kanamori63}. In the low-energy region
$\varepsilon\leq\varepsilon_F$, in the framework of this
description, the 2D Hubbard model becomes equivalent to a 2D
electron gas with a quadratic spectrum and short-range repulsion
(see Section~\ref{sec_gas}~). This model can be characterized by
the 2D Bloom gas parameter
$f_0\approx\displaystyle\frac{1}{\textrm{ln}(1/na^2)}$~\cite{Bloom75},
which allows to conduct a controlled diagrammatical expansion
(here, $n=p^2_F/2\pi$ the electron density in the 2D case for both
spin projections). For the first iteration of the self-consistent
$T-$matrix approximation, the authors of~\cite{Woelfle11} found
the contribution from the $T-$matrix pole corresponding to the
upper Hubbard band. As a result, a dressed one-particle Green’s
function was obtained with a double-pole
structure~\cite{Woelfle11}, which resembles the Green function in
the Hubbard-I approximation~\cite{Hubbard63}:
\begin{eqnarray}\label{Green}
G(\omega,\textbf{k})\approx\frac{\displaystyle\bigl(1-na^2/2\bigr)}
{\displaystyle\omega-\xi_\textbf{k}\bigl(1-na^2/2\bigr)+io}\nonumber\\
+\frac{na^2/2}
{\displaystyle\omega-U\bigl(1-na^2/2\bigr)-\xi_\textbf{k}na^2/2+io}.
\end{eqnarray}
where $o$ is the notation for infinitesimally small imaginary part
The first term in the right-hand side of (\ref{Green}) corresponds
to the contribution from the lower Hubbard band, and the second
term corresponds to the contribution from the upper Hubbard band.
We note that the second iteration of the self-consistent
$T$-matrix approximation does not change the principal properties
of expression (\ref{Green}). Thus, the presence of the upper
Hubbard band leads to nontrivial corrections to the Landau Fermi
liquid picture at low electron densities without total destruction
of this picture in the 2D case. More specifically, they produce
only small Hartree-Fock corrections to the thermodynamic
potential.

We note that all the results concerning superconductivity in the
Hubbard model obtained in the single-pole approximation for the
one-particle Green function remain valid at $U\gg W$ and low
electron density (up to small corrections $\sim W/U$, where $W$ is
the bandwidth), when the second pole is taken into account. Thus,
this result concerning the two-pole structure of the Green
function plays the role of a very interesting bridge connecting
the exact results of Galitskii and the Hubbard-I approximation
(the Gutzwiller approximation) in the Hubbard model. At the same
time, it does not affect the type of pairing or the estimate of
the critical temperature at low electron densities.

\section{Superconductivity in the Shubin-Vonsovsky model}
\label{sec_SV}

The authors of~\cite{Alexandrov11} raised an important problem of
the role of full (not reduced) Coulomb interaction in the
nonphonon superconductivity mechanisms, which in real metals does
not limited to the short-range Hubbard repulsion. The authors
of~\cite{Alexandrov11} examined the 3D jelly model with realistic
values of the electron density, when the Wigner-Seitz correlation
radius is not very large $r_S\leq20$,
\begin{equation}\label{rS}
r_S=\frac{1.92}{p_Fa_B},
\end{equation}
where $a_B=\displaystyle\frac{\varepsilon_0}{me^2}$ is the Bohr
radius of the electron ($\hbar=1$). In calculation of the
effective interaction, the contributions from the first and second
order caused by all diagrams presented in
Fig.~\ref{diagrams_alpha} were taken into account. The authors
of~\cite{Alexandrov11} noted that the previous studies of
Kohn-Luttinger superconductivity were mainly limited to the
calculation of only the short-range Hubbard interaction of
electrons $U$, in view of the computational difficulties connected
with taking into account the first and higher orders of the
Fourier transform of the long-range Coulomb repulsion
$V_{\textbf{q}}$ (depending on the wave vector ${\textbf{q}}$) in
the diagrams. As a result, the strong long-range Coulomb repulsion
in the first order of the perturbation theory (the first diagram
in Fig.~\ref{diagrams_alpha}) was ignored, and the contribution of
the electrons to the effective interaction in the Cooper channel,
which was determined only by the last second-order (exchange)
diagram in Fig.~\ref{diagrams_alpha}, had an attractive nature and
ensured $p$-wave pairing in the 3D case~\cite{Fay68,Kagan88} and
$d$-wave pairing in the 2D case~\cite{Baranov92b,Raghu10}.

In Ref.~\cite{Alexandrov11}, the long-range Coulomb interaction
$V_{\textbf{q}}$ was chosen in the form of the Fourier transform
of the Yukawa potential
$V(r)=\displaystyle\frac{e^2}{r}\exp(-\kappa r)$, which in the 3D
case takes the standard form
\begin{equation}\label{screening}
V_{\textbf{q}}=\frac{4\pi e^2}{q^2+\kappa^2},
\end{equation}
where $\kappa$ is the reciprocal Debye screening length. The
authors of~\cite{Alexandrov11} concluded, based on the results of
calculations, that the low and intermediate values of the Hubbard
repulsion $U$ in the presence of the long-range part of Coulomb
interaction (\ref{screening}) do not lead to the Cooper
instability both in 3D and 2D Fermi systems in the $p$-wave and
$d$-wave channels, irrespective of how small the screening length
is. The pairing appearing at large orbital momenta ($l\geq3$)
leads to the almost zero values of the critical temperature at any
reasonable value of the Fermi energy. According to the authors
of~\cite{Alexandrov11}, the anomalous pairing caused by strong
Coulomb repulsion cannot be measured experimentally in practice,
since the corresponding condensation energy (if it exists) is
several times lower than the condensation energy caused by the
electron-phonon interaction.

The growth of interest in the problem of account for the
long-range part of Coulomb correlations in the description of the
phase diagram of high-temperature superconductors raised the
popularity of the extended Hubbard model that includes the
interaction between the electrons located on different sites of
the crystal lattice (in the Russian literature, this model is
often called the Shubin-Vonsovsky
model~\cite{Shubin34,Shubin35,Shubin36}).

In the historical aspect, the Shubin-Vonsovsky model, which was
formulated almost immediately after the creation of quantum
mechanics, is a predecessor of some important models in the theory
of strongly correlated electronic systems, in particular, the
$s-d(f)$ model and the Hubbard model itself. The Shubin-Vonsovsky
model was actively used in studies of polar states in
solids~\cite{Vonsovsky79a,Vonsovsky79b}, for describing the
metal-insulator transition~\cite{Zaitsev80}, and also in the study
of the influence of the intersite Coulomb repulsion on the
effective band structure and superconducting properties of
strongly correlated systems~\cite{Zaitsev88,Zaitsev89,Valkov11}.

In the Wannier representation, the Hamiltonian of the
Shubin-Vonsovsky model can be written as
\begin{eqnarray}\label{SVHamiltonian}
\hat{H'}
&=&\sum\limits_{f\sigma}(\varepsilon-\mu)c^{\dagger}_{f\sigma}c_{f\sigma}
+
\sum\limits_{fm\sigma}t_{fm}c^{\dagger}_{f\sigma}c_{m\sigma}\nonumber\\
&&+U\sum_f\hat{n}_{f\uparrow}\hat{n}_{f\downarrow}+\frac{1}{
2}\sum_{fm\sigma\sigma'}V_{fm}\hat{n}_{f\sigma}\hat{n}_{m\sigma'},
\end{eqnarray}
where the last term in the right-hand side corresponds to the
energy $V_{fm}$ of the Coulomb interaction of electrons that are
located on different sites of the crystalline lattice. The last
three terms together in the right-hand side of
(\ref{SVHamiltonian}) reflect the fact that the screening radius
in the systems in question is equal to several lattice
spacings~\cite{Zaitsev80}. This determines the efficiency of the
Shubin-Vonsovsky model, in which the intersite Coulomb interaction
is taken into account within several coordination spheres. In the
momentum representation, the Hamiltonian (\ref{SVHamiltonian})
takes the form
\begin{eqnarray}
\hat{H'}
&=&\sum\limits_{\textbf{p}\sigma}(\varepsilon_{\textbf{p}}-\mu)
c^{\dagger}_{\textbf{p}\sigma}c_{\textbf{p}\sigma} +
U\sum_{\textbf{p}\textbf{p'}\textbf{q}}c^{\dagger}_{\textbf{p}\uparrow}
c^{\dagger}_{\textbf{p'}+\textbf{q}{\downarrow}}c_{\textbf{p}+\textbf{q}{\downarrow}}
c_{\textbf{p'}{\uparrow}}\nonumber\\
&+&\frac12\sum_{\textbf{p}\textbf{p'}\textbf{q}\sigma\sigma'}V_{\textbf{p}-\textbf{p'}}\,c^{\dagger}_{\textbf{p}\sigma}
c^{\dagger}_{\textbf{p'}+\textbf{q}{\sigma'}}c_{\textbf{p}+\textbf{q}{\sigma'}}c_{\textbf{p'}{\sigma}},
\end{eqnarray}
where the Fourier transform of the Coulomb interaction between the
electrons located on the nearest-neighbor sites, $V_1$, and on the
next-nearest sites, $V_2$, of the square lattice in the 2D case is
written as
\begin{equation}\label{Vq}
V_{\textbf{q}}=2V_1(\textrm{cos}\,q_xa+\textrm{cos}\,q_ya)+4V_2
\textrm{cos}\,q_xa~\textrm{cos}\,q_ya.
\end{equation}

The authors of~\cite{Kagan11} made a contribution to the
discussion in~\cite{Raghu10,Alexandrov11} by investigating the
conditions of the appearance of the Kohn-Luttinger superconducting
pairing in the 3D and 2D Shubin-Vonsovsky models with a Coulomb
repulsion of the electrons located on neighboring sites
($V_1\neq0,\,V_2=0$). As for the interaction, they considered the
maximally strong Coulomb repulsion on both the same and
neighboring sites: $U\gg V_1\gg W$ ($W$ is the bandwidth; $W=12t$
for the 3D cubic lattice and $W=8t$ for the 2D square lattice).

On the cubic lattice in the 3D case, we have the following
expressions for the bare interaction of electrons in vacuum in the
$s$-wave and $p$-wave channels:
\begin{eqnarray}\label{UsUp}
U^s_{\textrm{vac}}=U+6V+o(p^2a^2),\qquad
U^p_{\textrm{vac}}=2V\textbf{p}\textbf{p}'a^2.
\end{eqnarray}
In this case, the $T$-matrices in the appropriate channels in the
strong-coupling limit are determined as
\begin{eqnarray}\label{TsTp}
T_s=\frac{4\pi}{m}a_s,\qquad
T_p=\frac{4\pi}{m}2a_p\textbf{p}\textbf{p}'a^2,
\end{eqnarray}
where $a_s\sim a$ and $a_p\sim a$ for the scattering lengths in
the $s$-wave and $p$-wave channels. As a result, the dimensionless
gas parameter in the $s$-wave channel takes the form
$\lambda_s=\lambda=\displaystyle\frac{2ap_F}{\pi}$, just as in the
Hubbard model, whereas the bare gas parameter $\lambda_p$ in the
$p$-wave channel is proportional to $(p_Fa)^3$, in accordance with
the general quantum-mechanical results for slow particles
($p_Fa<1$) in vacuum~\cite{Landau89}.

Thus, even in the maximally repulsive 3D Shubin-Vonsovsky model,
which is the most unfavorable for the appearance of effective
attraction and superconductivity, the normal state in the
strong-coupling regime with low electron density is unstable with
respect to the triplet $p$-wave pairing. Notably, the effective
interaction of electrons at $l=1$ in the substance takes the
form~\cite{Kagan11}
\begin{equation}\label{Utriplet3D}
\rho_{3D}U^{l=1}_{\textrm{eff}}=\lambda_p-\frac{\lambda^2_s}{13},
\end{equation}
where $\rho_{3D} = mp_F/(2\pi^2)$ is the density of states in the
3D Fermi gas. As was mentioned above, the contribution from the
$p$-wave harmonic of the polarization operator $\Pi_{l=1}$ in
substance, $-\lambda^2_s/13<0$, favors attraction, and it cannot
be compensated by the contribution from the intersite Coulomb
repulsion $V_1$ in the $p$-wave channel, which is proportional to
$(p_Fa)^3$.

Similarly, in the 2D case, in the regime of strong coupling and
low electron density, the dimensionless gas parameter in the
$s$-wave channel is
$f_s=f_0\sim\displaystyle\frac{1}{\textrm{ln}(1/na^2)}\sim\frac{1}{\textrm{ln}[1/(p_Fa)^2]}$,
just as in the 2D Hubbard model, whereas the dimensionless gas
parameter in the $p$-wave channel is $f_p\sim p_F^2a^2$, again in
agreement with the results for slow particles in vacuum. The
effective interaction in the 2D case in substance takes the
form~\cite{Kagan11}
\begin{equation}\label{Utriplet2D}
\rho_{2D}U^{l=1}_{\textrm{eff}}=-6.1f^3_s+2p_F^2a^2.
\end{equation}
where $\rho_{2D} = m/(2\pi)$ is the density of states of a 2D
Fermi gas. Since $f^3_s\gg p_F^2a^2$ for $p_Fa\ll1$, we obtain
$U^{l=1}_{\textrm{eff}}\approx-6.1f^3_s$, as in the case $V_1=0$
(see Section~\ref{sec_gas}~).

Thus, the previous results concerning the realization of
superconducting $p$-wave pairing in both 3D and 2D repulsive
Hubbard models at strong coupling ($U\gg W$) with low electron
density remain valid even when we take the strong Coulomb
repulsion $V_1\gg W$ of electrons at the nearest sites into
account in the framework of the Shubin-Vonsovsky model. As a
result, the same expressions for the main exponent (which
determines the critical temperature of $p$-wave pairing
(\ref{Tcp}) and (\ref{Tcp3order})), are obtained just as in the
absence of the intersite Coulomb repulsion ($V_1=0$) in both
three-dimensional and two-dimensional cases. Account for $V_1$
changes only the preexponential factor~\cite{Efremov00a}, which
means that the superconducting $p$-wave pairing can be developed
in Fermi systems with pure repulsion~\cite{Kagan11} (in the
absence of electron-phonon interaction) even in the presence of
the long-range Coulomb repulsion.

The authors of~\cite{Raghu12} carried out a similar analysis for
the extended Hubbard model in the Born weak-coupling approximation
and came to the same conclusions as the authors of~\cite{Kagan11}.
Moreover, it was noted in~\cite{Raghu12} that in the weak-coupling
regime $W>U>V$, the effect of the long-range Coulomb interactions
is also suppressed, and does not impair the conditions for the
development of Cooper instability. This is explained by the fact
that the long-range interactions in the lattice models usually
contribute only to some specific channels of pairing and do not
affect the other channels. At the same time, the polarization
contributions that are described by the diagrams shown in
Fig.~\ref{diagrams_alpha} have components in all the channels and
usually more than one of them favors attraction. In this
situation, it turns out that the long-range interactions either do
not influence the principal components of the effective
interaction which lead to the pairing or suppress the main
components but do not affect the secondary ones [see the
discussion after expression (\ref{Delta_s})].

In this connection, a phase diagram was constructed in
Ref.~\cite{Raghu12} based on the extended Hubbard model in the
framework of the Kohn-Luttinger mechanism, which clearly reflects
the result of the competition of the superconducting phases with
different types of the symmetry of order parameter. In the
calculations of the effective coupling constant, an expression for
the renormalized scattering amplitude in the Cooper channel was
used in the form
\begin{eqnarray}\label{Gamma_wave_1storder}
U_{\textrm{eff}}(\textbf{p},\textbf{q})=U+V_{\textbf{p}-\textbf{q}}+U^2\Pi(\textbf{p}+\textbf{q}),
\end{eqnarray}
where $V_{\textbf{p}-\textbf{q}}$ is the Fourier transform of the
intersite Coulomb repulsion of electrons, Eqn (\ref{Vq}), and
$\Pi(\textbf{p}+\textbf{q})$ is the Lindhard function
(\ref{Lindhard}). Thus, the intersite Coulomb interaction $V$ was
taken into account only in the first order of the perturbation
theory, and the polarization contributions were determined only by
the term of the order of $U^2$. It was shown in~\cite{Raghu12}
that although the long-range interactions have a tendency to
suppress the anomalous pairing in some channels, the
Kohn-Luttinger superconductivity survives in the entire region of
electron concentrations $0<n<1$ and for all relations of the model
parameters.

It was noted in Refs~\cite{Kagan13a,Kagan13b} that the effective
interaction $U_{\textrm{eff}}({\bf q})$ is characterized by a
dependence that is quadratic in the quasimomentum only in the
region of $\textbf{q}\textbf{a}\ll1$. Outside this region, the
dependence of $V_{\textbf{q}}$ on the momentum is determined by
periodic functions. As a result, the behavior of
$U_{\textrm{eff}}({\bf q})$ is modified significantly in
comparison with the behavior of the momentum dependence of the
Fourier transform of the Yukawa potential. These factors
substantially affect the conditions of the realization of Cooper
instability at large electron densities, when the Fermi surfaces
do not have the spherical symmetry. Therefore, it can be expected
that the conditions for the realization of superconducting pairing
in the framework of the Kohn-Luttinger mechanism are determined
not only by the dynamic effects caused by the Coulomb interactions
but also by the effects related to the Brillouin zone.

The authors of~\cite{Kagan13a,Kagan13b} discussed the influence of
the Coulomb interaction of electrons located in the first and
second coordination spheres on the development of Cooper
instability in the Born weak-coupling approximation, $W>U>V$.
Accordingly, they used the effective interaction
$U_{\textrm{eff}}(\textbf{p},\textbf{k})$, which is determined in
the graphic form by the sum of five diagrams (see
Fig.~\ref{diagrams_alpha}) and for the Shubin-Vonsovsky model has
the following analytic form
\begin{eqnarray}\label{Gamma_wave_SV}
&&U_{\textrm{eff}}(\textbf{p},\textbf{k})=U+V_{\textbf{p}-\textbf{k}}+
\delta U(\textbf{p},\textbf{k}),
\\&&\delta U(\textbf{p},\textbf{k})=\frac1N\sum_{\textbf{p}_1}(U+V_{\textbf{p}-\textbf{k}})
(2V_{\textbf{p}-\textbf{k}}-V_{\textbf{p}_1+\textbf{p}}-V_{\textbf{p}_1-\textbf{k}})\nonumber\\
&&\qquad\qquad\times\frac{n_F(\varepsilon_{\textbf{p}_1})-n_F(\varepsilon_{\textbf{p}_1+\textbf{p}-\textbf{k}})}
{\varepsilon_{\textbf{p}_1}-\varepsilon_{\textbf{p}_1+\textbf{p}-\textbf{k}}}+\nonumber\\
&&\qquad\qquad+\frac1N\sum_{\textbf{p}_1}(U+V_{\textbf{p}_1-\textbf{p}})(U+V_{\textbf{p}_1-\textbf{k}})\nonumber\\
&&\qquad\qquad\times\frac{n_F(\varepsilon_{\textbf{p}_1})-n_F(\varepsilon_{\textbf{p}_1-\textbf{p}-\textbf{k}})}
{\varepsilon_{\textbf{p}_1-\textbf{p}-\textbf{k}}-\varepsilon_{\textbf{p}_1}}.
\label{Gamma_wave_b}
\end{eqnarray}
The presence of the renormalized expression for the effective
interaction allows us to analyze the conditions for the
realization of the Cooper instability. Taking into account the
fact that the leading contribution to the total scattering
amplitude of two electrons with opposite momenta and spin
projections $\Gamma$ (the total amplitude in the Cooper channel)
is determined by electron scattering near the Fermi surface, the
dependence of $\Gamma$ on the Matsubara frequency can be neglected
in the Bethe-Salpeter integral equation. As a result, this
equation is simplified taking the form
\begin{eqnarray}
\label{IntegralEq} \Gamma(\textbf{p}\,|\textbf{k})=
U_{\textrm{eff}}(\textbf{p},\textbf{k})-\frac
1N\sum_{\textbf{q}}U_{\textrm{eff}}(\textbf{p},\textbf{q})
L(\xi_{\textbf{q}}) \Gamma(\textbf{q}\,|\textbf{k}),
\end{eqnarray}
where
$L(\xi_{\textbf{q}})=\tanh(\xi_{\textbf{q}}/2T)/2\xi_{\textbf{q}}$
is the standard expression for the kernel of the Cooper loop.
\begin{figure}[t]
\begin{center}
\includegraphics[width=0.50\textwidth]{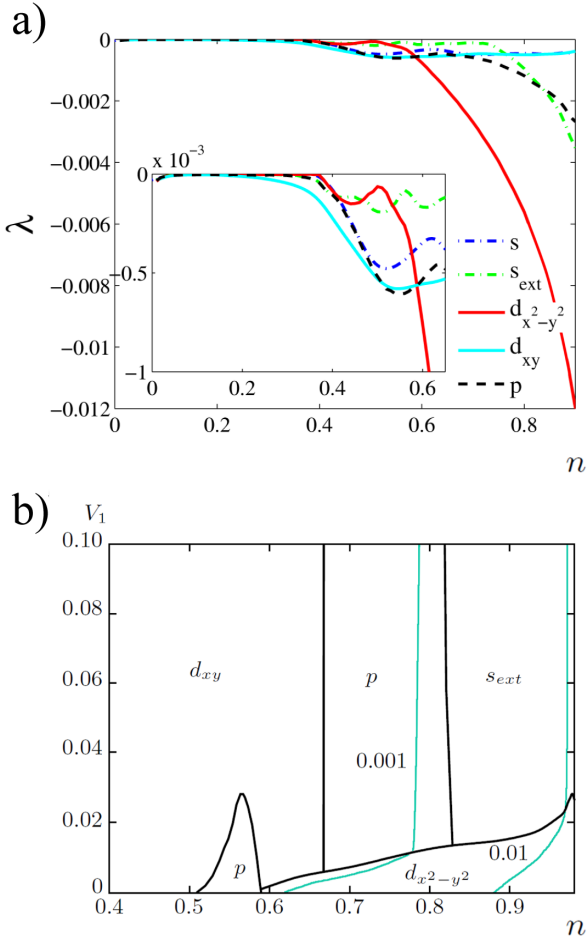}
\caption{Fig.~5. (a) Dependencies of $\lambda$ on the electron
concentration $n$ at $t_2=t_3=0,~U=|t_1|$ and $V_1=V_2=0$; and (b)
the "$n-V_1$" phase diagram of the Shubin-Vonsovsky model on a
square lattice at $t_2=t_3=0,~U=|t_1|$ and $V_2/V_1=0$. The
intersite Coulomb repulsion is taken into account only in the
first order of the perturbation theory~\cite{Kagan13a,Kagan13b}.
For all points that belong to one thin line in (b), the indicated
value of $|\lambda|$ is constant.} \label{PD1}
\end{center}
\end{figure}

It is known~\cite{Gor'kov61} that the appearance of Cooper
instability can be extracted from an analysis of the homogeneous
part of Eqn (\ref{IntegralEq}). In this case, the dependence of
$\Gamma$ on the momentum $\textbf{k}$ is factorized and can be
neglected. As a result, we proceed from (\ref{IntegralEq}) an
integral Gor’kov equation for the superconducting gap
$\Delta(\textbf{p})$. Passing to the integration with respect to
the constant-energy contours (in the 2D case), we obtain that the
possibility of the Cooper pairing is determined by the
characteristics of the energy spectrum in the vicinity of the
Fermi level and by the effective interaction
$U_{\textrm{eff}}(\textbf{p},\textbf{k})$ of the electrons that
are located near the Fermi surface
$\varepsilon_{\textbf{q}}=\mu$~\cite{Scalapino86a,Scalapino86b}.
As a result, the analysis of the Cooper instability reduces to the
solution of the eigenvalue problem
\begin{equation}
\label{IntegralEqPhi}
\frac{1}{(2\pi)^2}\oint\limits_{\varepsilon_{\textbf{q}}=\mu}
\frac{d\hat{\textbf{q}}} {v_F(\hat{\textbf{q}})}
U_{\textrm{eff}}(\hat{\textbf{p}}\,,\hat{\textbf{q}})
\Delta(\hat{\textbf{q}})=\lambda\Delta(\hat{\textbf{p}}),
\end{equation}
in which the superconducting order parameter
$\Delta(\hat{\textbf{q}})$ plays the role of an eigenvector, and
we have the eigenvalues of $\lambda^{-1}\simeq\ln(T_c/W)$. In this
case, the quasimomenta $\hat{\textbf{p}}$ and $\hat{\textbf{q}}$
lie on the Fermi surface, and $v_F(\hat{\textbf{q}})$ is the Fermi
velocity. The feasible solutions of Eqn (\ref{IntegralEqPhi}) with
$\lambda<0$ are determined not only by the effective interaction
$U_{\textrm{eff}}(\bf{p},\bf{q})$ but also by the shape of
isoenergetic contours. As far as the concrete structure of these
contours is closely connected with the energy spectrum, it is
obvious that if we are not limited to the nearest-neighbor
approximation and take into account the distant hoppings, we can
substantially influence the conditions for the realization of the
Cooper instability and significantly modify the structure of the
phase diagram of the superconducting state.

To solve Eqn (\ref{IntegralEqPhi}), we represent its kernel in the
form of a superposition of the functions which belong to one of
the irreducible representations of the symmetry group $C_{4v}$ of
the square lattice. As it is well known, this group has five
irreducible representations~\cite{Landau89}, and for each of them
Eqn (\ref{IntegralEqPhi}) has its solution with an appropriate
effective coupling constant $\lambda$. Subsequently, the following
classification is used for the symmetry of the order parameter:
the representation $A_1$ corresponds to $s$-wave type symmetry;
$A_2$ to the extended $s$-wave type symmetry ($s_{ext}$); $B_1$ to
the $d_{xy}$-wave type symmetry; $B_2$ to the $d_{x^2-y^2}$-wave
type symmetry; and $E$ to the $p$-wave type symmetry.

For the irreducible representations $\gamma = A_1, A_2, B_1, B_2,
E$, the solution of Eqn (\ref{IntegralEqPhi}) is searched in the
form
\begin{equation}\label{solution1}
\Delta^{(\gamma)}(\phi)=\sum\limits_{m}\Delta_{m}^{(\gamma)}g_{m}^{(\gamma)}(\phi),
\end{equation}
where $m$ is the number of the basis function of the
representation $\gamma$, and $\phi$ is the angle that
characterizes the direction of the momentum $\hat{\textbf{p}}$
with respect to the axis $p_x$. The explicit form of
$g_{m}^{(\gamma)}(\phi)$ is determined by the following
expressions:
\begin{eqnarray}\label{harmon}
&&A_1\rightarrow~g_{m}^{(s)}(\phi)=\frac{1}{\sqrt{(1+\delta_{m0})\pi}}\,
\textrm{cos}\,4m\phi,~~m\in[\,0,\infty),\label{invariants_s}\nonumber\\
&&A_2\rightarrow~g_{m}^{(s_{ext})}(\phi)=\frac{1}{\sqrt{\pi}}\,\textrm{sin}\,
4(m+1)\phi,\label{invariants_s1}\nonumber\\
&&B_1\rightarrow~g_{m}^{(d_{xy})}(\phi)=\frac{1}{\sqrt{\pi}}\,
\textrm{sin}\,(4m+2)\phi,\label{invariants_dxy}\\
&&B_2\rightarrow~g_{m}^{(d_{x^2-y^2})}(\phi)=\frac{1}{\sqrt{\pi}}\,
\textrm{cos}\,(4m+2)\phi,\label{invariants_dx2y2}\nonumber\\
&&E~\rightarrow~g_{m}^{(p)}(\phi)=\frac{1}{\sqrt{\pi}}\,(A\,\textrm{sin}\,
(2m+1)\phi\nonumber\\
&&\qquad\qquad\qquad\qquad+B\,\textrm{cos}\,(2m+1)\phi)\label{invariants_p}\nonumber.
\end{eqnarray}
The basis functions satisfy the orthonormalization conditions
\begin{equation}\label{norma}
\int\limits_0^{2\pi}d\phi\,g_{m}^{(\gamma)}(\phi)g_{
n}^{(\beta)}(\phi)=\delta_{\gamma\beta}\delta_{mn}.
\end{equation}

Substituting (\ref{solution1}) in Eqn (\ref{IntegralEqPhi}),
integrating with respect to the angles, and using the
orthonormalization condition for the functions
$g_{m}^{(\gamma)}(\phi)$, we obtain
\begin{equation}\label{EqDelta}
\sum_n\Lambda^{(\gamma)}_{mn}\Delta_{n}^{(\gamma)}=\lambda_{\gamma}\Delta_{m}^{(\gamma)},
\end{equation}
where
\begin{eqnarray}
\label{matrix}
\Lambda^{(\gamma)}_{mn}&=&\frac{1}{(2\pi)^2}\oint\limits_0^{2\pi}d\phi_{\textbf{p}}
\oint\limits_0^{2\pi}d\phi_{\textbf{q}}\frac{d\hat{\textbf{q}}}
{d\phi_{\textbf{q}}v_F(\hat{\textbf{q}})}
U_{\textrm{eff}}(\hat{\textbf{p}}\,,\hat{\textbf{q}})\nonumber\\
&\times&g_{m}^{(\gamma)}(\phi_{\textbf{p}}) g_{
n}^{(\gamma)}(\phi_{\textbf{q}}).
\end{eqnarray}
Since $T_c\sim W\exp\bigl(1/\lambda \bigr)$, each negative
eigenvalue $\lambda_{\gamma }$ corresponds to a superconducting
phase with the $\gamma$-wave symmetry of the order parameter. The
expansion of the order parameter $\Delta^{(\gamma)}(\phi)$ in the
basis functions includes many harmonics in general, but the
leading contribution is determined by only several first terms.
The largest value of the critical temperature is associated with
the largest value of $|\lambda_{\gamma}|$.
\begin{figure}[t]
\begin{center}
\includegraphics[width=0.45\textwidth]{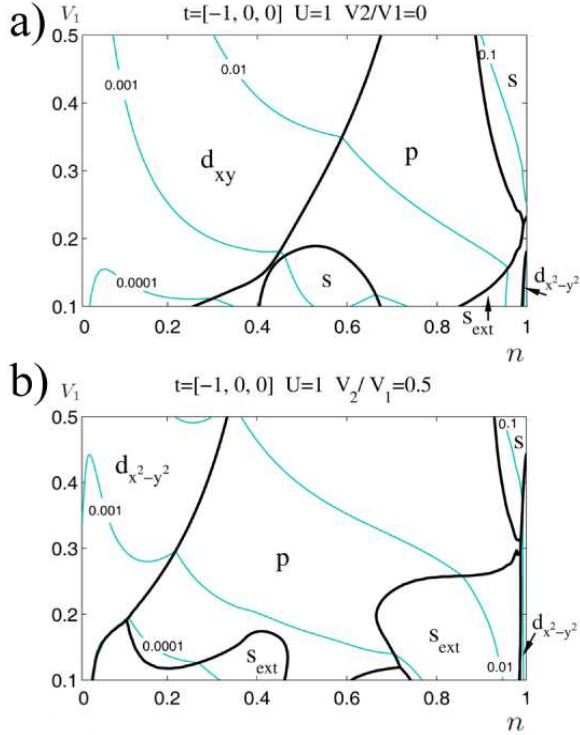}
\caption{Fig.~6. The "$n-V_1$" phase diagram of the
Shubin-Vonsovsky model on a square lattice constructed taking into
account the contributions of the second order in $V$ at the
parameters $t_2=t_3=0,~U=|t_1|$ and the ratios (a) $V_2/V_1=0$ and
(b) $V_2/V_1=0.5$. The thin lines correspond to the lines of
constant $|\lambda|$~\cite{Kagan13b}.} \label{PD2}
\end{center}
\end{figure}

If the intersite Coulomb interaction is taken into account only
for electrons that are located on the nearest sites ($V_1\neq0,
V_2=0$ in (\ref{Vq})) and the excitation spectrum is described by
only one hopping parameter ($t_1\neq0,\,t_2=t_3=0$), then the
phase diagram of the superconducting state at $U=|t_1|$
(Fig.~\ref{PD1}) contains five regions. Figure~\ref{PD1}a displays
the dependence of the effective coupling constants $\lambda$ for
different types of the symmetry of the superconducting order
parameter on the electron density $n$ obtained at $V_1=V_2=0$.
Based on the $\lambda(n)$ dependencies, a phase diagram for
different values of the intersite Coulomb repulsion $V_1$, which
reflects the competition between the superconducting phases with
the different symmetry types of the order parameter can be
constructed (Fig.~\ref{PD1}b). The case depicted in
Fig.~\ref{PD1}a corresponds to the abscissa axis in
Fig.~\ref{PD1}b. To construct this phase diagram, Eqn
(\ref{Gamma_wave_1storder}) for the effective interaction of
electrons in the Cooper channel was used, which takes
contributions of only the first order in $V$ into account and
ignores the contributions proportional to $UV$ and $V^2$. The
region of the phase diagram that lie on the abscissa axis
($V_1=0$) are in a good agreement with the phase diagrams regions
obtained in Refs~\cite{Hlubina99,Deng14}. In the region of low and
intermediate densities of electrons, $n=0-0.52$, in the two first
orders of the perturbation theory, superconductivity with the
$d_{xy}$ symmetry type of the order
parameter~\cite{Baranov92a,Baranov92b} is realized; in the range
of $n=0.52-0.58$, the ground state corresponds to the phase with
$p-$wave pairing, but in this case $|\lambda_p|$ insignificantly
exceeds $|\lambda_{d_{xy}}|$ (see the inset in Fig.~\ref{PD1}a).
We note that according to the calculations of the authors
of~\cite{Raghu12}, $d_{xy}$-wave pairing is realized instead of
$p-$wave pairing in this interval of electron densities. At
$n>0.58$, the $d_{x^2-y^2}$-wave type of superconductivity
appears, which is relevant for cuprate superconductors.

Note that with an account of the Coulomb repulsion $V_1$ on
neighboring sites in the first order of the perturbation theory,
the absolute value of $\lambda$ decreases for all types of
symmetry. In this case, the superconducting $d_{x^2-y^2}$-wave
phase is suppressed most strongly and, with increasing $V_1$, the
phases with the order parameters that have different symmetry
types are realized at these concentrations.
\begin{figure*}[t]
\begin{center}
\includegraphics[width=0.75\textwidth]{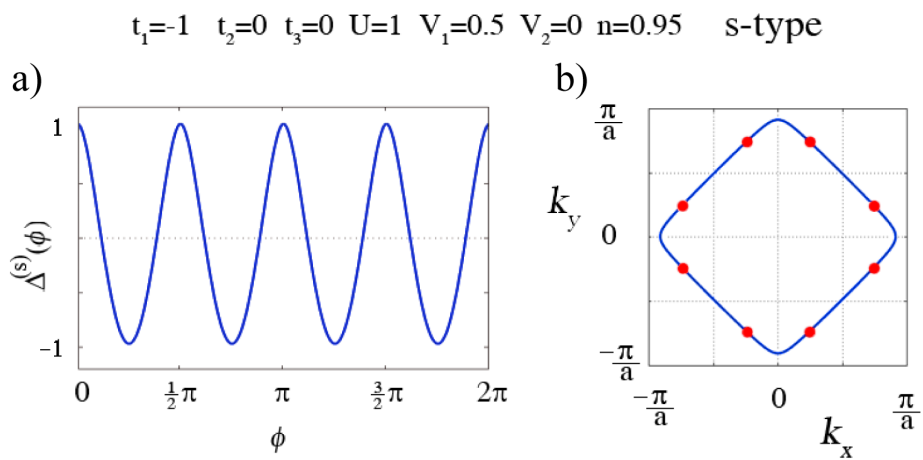}
\caption{Fig.~7. (a) Angular dependence of the superconducting
order parameter $\Delta^{(s)}(\phi)$ and (b) the arrangement of
the nodal points (zeroes of the gap $\Delta^{(s)}(\phi)$) on the
Fermi contour calculated at the parameters
$t_2=t_3=0,~U=|t_1|,~V_1=0.5|t_1|,~V_2=0,$ and
$n=0.95$~\cite{Kagan13b}.} \label{Delta_theta}
\end{center}
\end{figure*}

The first order of the perturbation theory in the intersite
Coulomb repulsion always has a tendency to suppress the
superconducting pairing. Hence, the possibility of realization of
the Cooper instability based on the Kohn-Luttinger mechanism is
connected with the appearance in the second order of the
perturbation theory of the attractive and sufficiently strong
contributions to the effective interaction matrix
(\ref{Gamma_wave_b}). Thus, in order to take into account the
polarization effects for the intersite Coulomb interaction, we
should use the full expression (\ref{Gamma_wave_SV}) for
$U_{\textrm{eff}}(\textbf{p},\textbf{q})$ but not reduced one
(\ref{Gamma_wave_1storder}). In this case, the polarization
effects proportional to $UV$ and $V^2$, even at small values of
$V_1$, substantially change and complicate the structure of the
phase diagram (Fig.~\ref{PD2}a). With an increase of the intersite
Coulomb interaction parameter $V_1$, an increase of $|\lambda|$
occurs for $T_c\sim W \exp(-1/|\lambda|)$. In this case, only
three phases are stabilized, which correspond to the $d_{xy}$-,
$p$-, and $s$-wave symmetry types of the superconducting order
parameter.

We note that in the region of high electron concentrations and at
$0.25<V_1/|t_1|<0.5$, the Kohn-Luttinger polarization effects lead
to the appearance of a superconducting $s$-wave pairing. This
qualitative effect clearly demonstrates the importance of the
second-order processes in calculating the effective interaction of
electrons in the Cooper channel and in constructing the phase
diagram presented in Fig.~\ref{PD2}. A quantitative comparison of
the different partial contributions to the total effective
interaction showed that the realization of $s$-wave pairing was
due to the polarization contributions proportional to $V^2$. In
this case, the leading contribution is determined by the angular
harmonic $g_{1}^{(s)}(\phi)=\displaystyle\frac{1}{\sqrt{\pi}}\cos
4\phi$ rather than by the constant (as in the case of the usual
$s$-wave pairing in the isotropic situation and in the absence of
a lattice).

The above-mentioned scenario of the realization of superconducting
$s$-wave pairing due to the higher angular harmonics correlates
well with the experimental data recently obtained in
Ref.~\cite{Okazaki12}, which presents the results of the
experimental studies of a superconductor based on the iron
arsenide KFe$_2$As$_2$ by angle-resolved photoemission
spectroscopy (ARPES). These studies showed that this compound is a
nodal superconductor (containing zeroes of the gaps) with an
$s$-wave type symmetry of the order parameter, which has eight
points at which the gap vanishes.

Figure~\ref{Delta_theta}a shows the angular dependence of the
superconducting order parameter
\begin{eqnarray}
\Delta^{(s)}(\phi) &=& \frac{\Delta^{(s)}_0}{\sqrt{2}} +
\Delta^{(s)}_1\cos 4\phi + \Delta^{(s)}_2\cos 8\phi \nonumber\\
&+&\Delta^{(s)}_3\cos 12\phi + \Delta^{(s)}_4\cos
16\phi,\label{Delta_s}
\end{eqnarray}
calculated in Ref.~\cite{Kagan13b} for the point of the phase
diagram at which the $s$-wave pairing is realized at large
electron densities. This dependence demonstrates the presence of
eight nodal points at which the gap vanishes. Their arrangement on
the Fermi contour (Fig.~\ref{Delta_theta}b) according to the
results of calculations~\cite{Kagan13b} is in a qualitative
agreement with the experimental picture presented
in~\cite{Okazaki12}.
\begin{figure}[b]
\begin{center}
\includegraphics[width=0.48\textwidth]{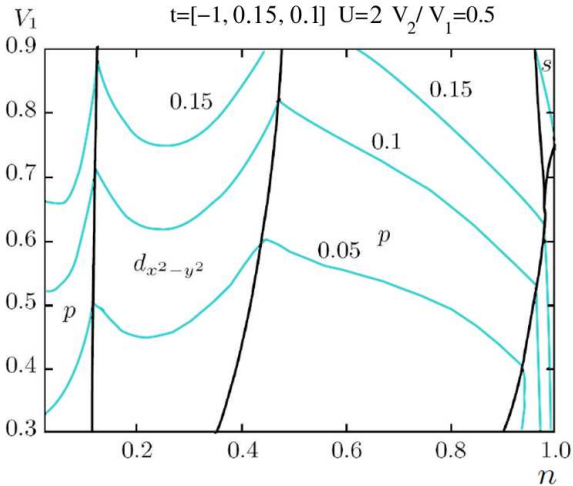}
\caption{Fig.~8. The "$n-V_1$" phase diagram of the
Shubin-Vonsovsky model on a square lattice, calculated at the
parameters $t_2=0.15|t_1|,~t_3=0.1|t_1|,~U=2|t_1|,~V_2/V_1=0.5$.
The thin lines are lines of constant $|\lambda|$~\cite{Kagan13b}.}
\label{PD5}
\end{center}
\end{figure}

A similar scenario for the realization of the superconductivity is
also observed in the $p$-wave channel. Here, the pairing obtained
by taking into account the second order of perturbation theory in
the Coulomb interaction is suppressed by the bare repulsion only
for the first harmonic
$g_0^{(p)}(\phi)=\displaystyle\frac{1}{\sqrt{\pi}}\,(A\,\textrm{sin}\,
\phi+B\,\textrm{cos}\,\phi)$. In this case, the leading
contribution to $\Delta^{(p)}(\hat{\textbf{p}})$ comes from the
next $p$-wave pairing harmonic on the lattice,
$g_1^{(p)}(\phi)=\displaystyle\frac{1}{\sqrt{\pi}}\,(A\,\textrm{sin}(3\phi)+B\,\textrm{cos}(3\phi))$.

The authors of~\cite{Kagan13b} have also analyzed the influence of
the Coulomb repulsion between next-to-nearest neighbors
($V_2\neq0$) and distant electron hoppings ($t_2\neq0,\,t_3\neq0$)
on the phase diagram of the superconducting state of the
Shubin-Vonsovsky model. Figure~\ref{PD5} shows a modification of
the phase diagram of the Shubin-Vonsovsky model that takes place
with an increase in the Hubbard repulsion parameter. It can be
seen that in the region of low electron densities and in the
region of densities close to the Van Hove singularity, we get a
superconducting phase with a $d_{x^2-y^2}$-wave of the order
parameter and with the sufficiently large values of
$|\lambda|\sim0.1-0.2$. This result can be interesting for the
possibility of the implementation of the Kohn-Luttinger mechanism
to cuprate superconductors. Note that at $|\lambda|\sim0.2$ the
critical temperatures of the superconducting transition can reach
values $T_c\sim 100\,\textrm{K}$ which are quite reasonable for
the cuprates.

\section{Superconductivity in the 2D \textit{t-J} model}
\label{sec_tJ}

\begin{figure}[t]
\begin{center}
\includegraphics[width=0.47\textwidth]{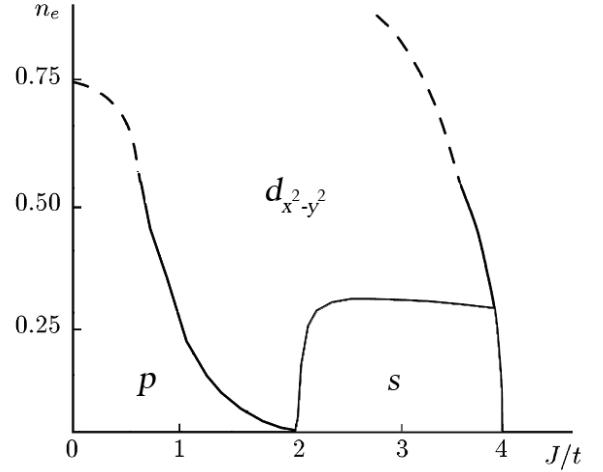}
\caption{Fig.~9. Phase diagram of the superconducting state of the
2D $t-J$ model at small and intermediate electron
densities~\cite{Kagan94_2,Kagan14e}.} \label{tJ_PD}
\end{center}
\end{figure}

After Anderson advanced an idea~\cite{Anderson87} that the
electronic properties of cuprate superconductors can be described
by the Hubbard model in the strong-coupling limit $U\gg W$, the
so-called $t-J$ model acquired great popularity among the
researchers (see
reviews~\cite{Izyumov91,Brenig95,Izyumov97,Plakida02}). This model
was initially derived by a canonical transformation from the
Hubbard model near the half-filling, $n\rightarrow1$, in the limit
$t/U<<1$~\cite{Bulaevskii68,Chao77}. Later on, for cuprates, a
generalized $t-J$ model was suggested~\cite{Hybertsen89,Unger93},
The Hamiltonian of the generalized 2D $t-J$ model with a weakened
constraint and an arbitrary ratio $J/t$ derived from the
three-band Emery model~\cite{Emery87,Varma87} is written
as~\cite{Kagan94_2,Hybertsen89,Unger93}
\begin{eqnarray}\label{tJHamiltonian}
\hat{H}
&=&\sum\limits_{f\sigma}(\varepsilon-\mu)c^{\dagger}_{f\sigma}c_{f\sigma}
+
t\sum\limits_{\langle fm\rangle\sigma}c^{\dagger}_{f\sigma}c_{m\sigma}\\
&&+U\sum_f\hat{n}_{f\uparrow}\hat{n}_{f\downarrow}+ J\sum_{\langle
fm\rangle}\biggl(\textbf{S}_f\textbf{S}_m-\frac{\hat{n}_f\hat{n}_m}{4}\biggr).\nonumber
\end{eqnarray}
In fact, this is a model with a strong Hubbard repulsion between
electrons on one site and weak antiferromagnetic attraction $J>0$
on neighboring sites. The hierarchy of the parameters of the model
is $U\gg\{J,t\}$. The phase diagram of the $t-J$ model constructed
in~\cite{Kagan94_2} is presented in Fig.~\ref{tJ_PD}.

For the parameters that are realistic for optimally doped cuprate
superconductors, $J/t\sim0.5$ and $n=2\varepsilon_F/W=0.85$, the
critical temperature of the superconducting transition has been
estimated as
\begin{equation}\label{Tcdx2y2_tJ}
T^{d_{x^2-y^2}}_c\sim\varepsilon_F\exp\biggl(-\frac{\pi
t}{2Jn^2}\biggr)\sim10^2\,\textrm{K}.
\end{equation}
We note that a similar estimate for the critical temperature of
the $d_{x^2-y^2}$-wave pairing has been obtained in the framework
of a more rigorous spin-polaron theory
in~\cite{Plakida01,Plakida03} with the use of the Hubbard
operators~\cite{Hubbard65}.

The authors of~\cite{Plakida01,Plakida03} also used the
generalized $t-J$ model derived from the Emery model in the limit
of a small number of holes by constructing the Zhang-Rice
singlets~\cite{Zhang88} at $J<t$ and neglecting the Coulomb
repulsion between the charge carriers from the energy bands of
copper and oxygen ($V_{pd}=0$). In this approximation, the
weakened constraint is also not very important, and we can neglect
the kinematic interaction~\cite{Zaitsev87,Zaitsev04}.

We note that a very interesting physics appears in the so-called
‘difficult’ comer of the phase diagram of the generalized $t-J$
model, namely, in the case of small $J$ ($J/t<<1$) and low doping
$\delta=(1-n_e)<<1$ (this region is frequently called the
pseudogap). For this part of the phase diagram, in accordance with
the ideas of Laughlin~\cite{Laughlin88,Fetter89} on the
spin-charge confinement of spin and charge in two-dimensional and
three-dimensional strongly correlated electron systems (see
also~\cite{Kagan04,Kagan06}), a concept  of a strongly interacting
Fermi-Bose mixture of spinons and holons has been proposed. The
spinons and holons in the mixture form the composite holes in the
confinement potential of an antiferromagnetic (AFM)
string~\cite{Bulaevskii68,Brinkman70}). In this case, according to the assumption made
in~\cite{Kagan14e,Kagan06}, the phase diagram of cuprate
superconductors in the region of low doping can be considered in
the framework of the scenario of the BCS-BEC crossover between the
local and extended pairs for pairing of two composite holes (two
spin polarons or two AFM strings) in the $d_{x^2-y^2}$-wave
channel. Certainly, the transition from the region of optimal
doping with a large Fermi surface and extended Cooper pairs to the
region of low doping with local pairs and hole-type pockets (small
Fermi surface) can be realized in a very nontrivial way and can
contain a singularity in the middle, such as a quantum critical
point (QCP) (see, e.g.,~\cite{Sachdev10,Lee13,Castellani95}), or
even certain intermediate phases.

Returning to the region of the extended Cooper pairing and optimal
doping $n_e\geq0.85$, we emphasize that the development and
utilization of the Kohn-Luttinger ideas for the strong-coupling
regime would be one of the most challenging directions in the
modern theory of superconductivity in strongly correlated systems.
However, the solution of this problem requires an account of the
strong on-site correlations in all the orders of perturbation
theory. Moreover, the intersite correlations must be described
taking into account at least the second-order contributions. One
of the possibilities to develop the theory in this direction is to
use the atomic representation~\cite{Hubbard65} and the diagram
technique for the Hubbard operators~\cite{Zaitsev75,Zaitsev76}.
The relevant models that can be used for the investigation of the
Kohn-Luttinger renormalization are the generalized $t-J-V$
model~\cite{Eremin01,Eremin12,Plakida13,Plakida14} and the
$t-J^*-V$ model (which takes three-center interactions into
account), whose important role in describing the superconducting
state was studied
in~\cite{Hirsch89,Yushankhai90,VVV02,VVV05,Korshunov04,VVAG08,VVV08,VVV11}.
These models, which can be derived from the Shubin-Vonsovsky model
in a certain range of parameters, effectively represent its
low-energy versions.

Concluding this section, we note that ultracold quantum gases in
optical lattices also provide an excellent experimental
opportunity to simulate strongly correlated systems on a lattice,
in particular, to study the phase diagram of the $t-J$ model and
even the structure of the AFM string and spin polarons in a
situation with well-controlled and easily tunable parameters $t$,
$J$, and $n_{e}$~\cite{Bermudez15}.

\section{Superconductivity in an idealized graphene monolayer}
\label{sec_monolayer}

Nowadays, the popularity of the Kohn-Luttinger mechanism continues
to grow due to the possibility of its utilization in other
physical systems. For example, the conditions of its appearance in
topological superfluid liquids~\cite{Marienko12}, as well as in
the idealized monolayer and bilayer of graphene (where the effect
of impurities and the van der Waals potential of the substrate are
ignored) are being discussed actively.
\begin{figure}[t]
\begin{center}
\includegraphics[width=0.42\textwidth]{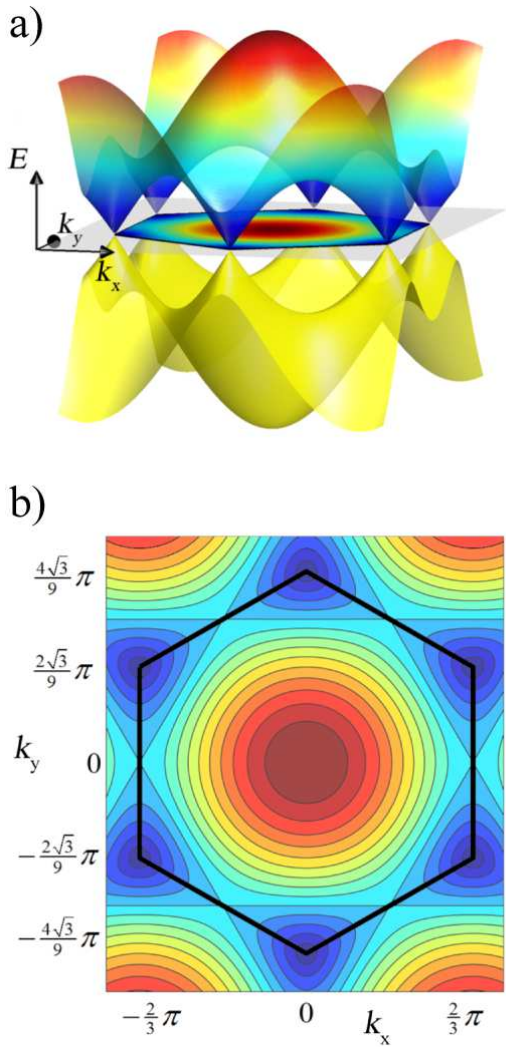}
\caption{Fig.~10. (a) Energy structure of the graphene monolayer
and (b) the energy contours around saddle points in the conduction
band of graphene, obtained in the tight-binding model in the
nearest-neighbors approximation.} \label{monolayer_energy}
\end{center}
\end{figure}

At present, graphene is of significant interest from both the
fundamental and applied points of view because of its
unconventional transport, pseudo-relativistic, and
quantum-electrodynamic
properties~\cite{Lozovik08,Castro09,Kotov12}. These properties of
graphene are caused, first of all, by its unique gapless energy
band structure with the cone-shaped valence and conduction bands
(Fig.~\ref{monolayer_energy}), touching each other at the corners
of the first Brillouin zone at the Dirac points~\cite{Wallace47}.
It has been established that near the Dirac points, the electrons
propagating in graphene are similar to massless fermions with
linear dispersion~\cite{Novoselov05}, minimal conductivity at
vanishing concentration of carriers~\cite{Novoselov05,Tan07}, high
mobility~\cite{Morozov08,Bolotin08,Garcia08}, Klein
tunneling~\cite{Geim06,Young09}, oscillating motion
(\textit{Zitterbewegung})~\cite{Katsnelson06,Rusin09}, universal
absorption of light~\cite{Nair08}, and many other properties that
have no analogs in other physical systems.

Placed in contact with superconductors, graphene manifests
unconventional superconducting properties~\cite{Munoz12}. The
authors of~\cite{Heersche07} experimentally studied the Josephson
effect~\cite{Josephson62} in mesoscopic junctions consisting of a
short sample of graphene monolayer placed between two closely
spaced superconducting electrodes. By cooling this device below
the critical temperature of the electrodes
($T_c\approx1.3\,\textrm{K}$), the authors of~\cite{Heersche07}
observed a supercurrent in the graphene monolayer (a similar
result was obtained in~\cite{Shailos07}). By changing the voltage
of the electric field at the gate electrode, the researchers could
shift the Fermi level from the valence band to the conduction band
and thus control the density of charge carriers in the graphene
monolayer. Irrespective of the position of the Fermi level in the
system, a Josephson current was observed, which indicates that
this device works as a bipolar supercurrent transistor. Namely,
the supercurrent was transferred by $p$-wave type Cooper pairs
when the Fermi level was located in the valence band, and by
electron Cooper pairs when the Fermi level was located in the
conduction band. More important is the fact that the supercurrent
could flow in the graphene monolayer even when the Fermi level was
located precisely at the Dirac point, i.e., at the zero density of
carriers. This behavior was explained within the theory of
ballistic transport from graphene to the Josephson
junctions~\cite{Titov06}; however, later
experiments~\cite{Du08,Ojeda09,Tomori10} have demonstrated that
the transport in the superconductor-graphene-superconductor
junctions is, rather, of a diffusive nature.

Although so far no confirmation has been found that the Cooper
instability can be developed in graphene itself (possibly because
of permanently present structural disorder), the results of the
above-mentioned experiments indicate that the Cooper pairs can
propagate in graphene coherently. In this connection, a question
arises as to whether it is possible to modify graphene
structurally (for example, by introducing twinning planes and
grain boundaries) or chemically, such that it would become a
magnet~\cite{Peres05} or even a true
superconductor~\cite{Esquinazi13}.

The theoretical analysis in~\cite{Marino06} has shown that the
model with a conical dispersion requires a minimal strength of
pairing interaction for the development of Cooper instability in
the undoped system. Furthermore, several attempts were undertaken
to analyze the possibility of realization of the superconducting
state in doped graphene monolayer. In~\cite{Gonzalez01}, the role
of topological defects in the realization of Cooper pairing in
this material was studied. The superconducting phase was also
investigated and the symmetry of the order parameter was
determined on a hexagonal lattice of graphene. In the case of the
attractive Hubbard interaction $U$ between electrons, as it was
shown in~\cite{Zhao06}, the usual singlet pairing with an $s$-wave
type symmetry of the order parameter is realized.
In~\cite{Uchoa07}, a phase diagram of the superconducting state
was constructed (in the mean-field approximation) for a graphene
monolayer in the extended Hubbard model with attraction, and the
plasmon mechanism of superconductivity was investigated, which led
to low critical temperatures in the $s-$wave channel at realistic
values of the electron concentration. Furthermore, it was
demonstrated that in the presence of the attractive interaction
$V$ at the nearest sites, the realization of an exotic combination
of the $s$-wave and the $p$-wave pairings becomes
possible~\cite{Uchoa07}.

At present, along with the frequently investigated problem of
implementation of the superconducting state in a graphene
monolayer within the electron-phonon pairing
mechanism~\cite{Kopnin08,Basko08,Lozovik10,Einenkel11,Classen14},
the possibility of the development of Cooper instability in
graphene as a result of the electron-electron interactions is
being studied actively. In~\cite{Black07}, in the $t-J$ model
within the renormalized mean-field theory, the possibility of the
realization of the superconducting pairing in a graphene monolayer
was studied. Both superconductivity with $s$-wave symmetry of the
order parameter and a chiral superconductivity with the $d$-wave
symmetry (which is described by a two-dimensional representation
and breaks the symmetry with respect to time reversal), were shown
to be possible. In this case, a significant predominance of
$d$-wave pairing over $s$-wave pairing was demonstrated
in~\cite{Black07}.

When we discuss chirality with respect to the superconducting
state, we understand that this state is characterized both by
spontaneous time-reversal symmetry breaking and by parity
violation (see~\cite{Black14a} and also review~\cite{Black14b}).
In other words, this type of superconductivity necessarily
includes a linear combination of the two order parameters that
belong to a unified higher-dimensional representation of the point
symmetry group of the crystal. The chiral superconducting state
with the $d$-wave symmetry of the order parameter in graphene is
the spin-singlet $d_{x^2-y^2}\pm id_{xy}$-wave state. Since the
hexagonal lattice belongs to the symmetry group $C_{6v}$, the two
$d$-wave states make similar contributions but have a phase shift
$\pi/2$ relative to each other. This causes the appearance of a
superconducting state in graphene at any finite doping
level~\cite{Black14a,Black14b}.

The appearance of chiral superconductivity with the
$d_{x^2-y^2}\pm id_{xy}$-wave symmetry of the order parameter was
investigated earlier in cuprates in the presence of magnetic
fields~\cite{Krishana97,Elhalel07} and magnetic
impurities~\cite{Balatsky98}, and also in 2D superfluid
$^3$He~\cite{Volovik88}. We note that the realization of
$d+id$-wave chiral superconductivity was observed experimentally
in the pnictide SrPtAs near the Van Hove filling
$n_{VH}$~\cite{Biswas13}. Note that this compound with
$T_c=2.4\,\textrm{K}$~\cite{Nishikubo11} consists of a set of
weakly bound layers of PtAs that form a hexagonal lattice.

In~\cite{Honerkamp08}, in the framework of the $t-J$
model~\cite{Daul00} with the Coulomb interaction $V$ between the
fermions at neighboring carbon atoms of the hexagonal lattice of
graphene (investigated by the method of the functional
renormalization group), a triplet $f$-wave pairing and a singlet
$d + id$-wave pairing far from the half-filling were detected. The
possibility to realize $d + id$-wave chiral superconductivity due
to the spin-fluctuation mechanism was also confirmed by quantum
Monte Carlo calculations~\cite{Ma11,Chen13}.

The situation where the Fermi level is near one of the Van Hove
singularities in the density of states of graphene monolayer was
considered in~\cite{Gonzalez08}. It is known that these
singularities can enhance magnetic and superconducting
fluctuations~\cite{Markiewicz97}, According to the scenario
described in~\cite{Gonzalez08}, the appearance of Cooper
instability is caused by the strong anisotropy of the Fermi
contour at the Van Hove filling $n_{VH}$, which is in fact related
to the Kohn-Luttinger mechanism. It is emphasized
in~\cite{Gonzalez08} that the realization of this mechanism is
possible in graphene, since the electron-electron scattering
becomes strongly anisotropic and therefore a channel can arise in
which the scattering amplitude has an attractive component with a
nontrivial angular dependence on the Fermi surface. According
to~\cite{Gonzalez08}, such a Cooper instability in the idealized
graphene monolayer is developed predominantly in the $d+id$-wave
channel, and it can lead to critical temperatures up to
$T_c\sim10\,\textrm{K}$, depending on the possibility to tune the
level of the chemical potential maximally close to the Van Hove
singularity. In~\cite{Valenzuela08}, the possible coexistence and
competition between the Pomeranchuk and Kohn-Luttinger
instabilities in graphene monolayer were discussed.

The authors of~\cite{McChesney10} obtained experimentally a
heavily doped monolayer of graphene by a chemical method
(different combinations of K and Ca were used
in~\cite{McChesney10} as dopants), and investigated the prepared
sample by ARPES. It was found that the many-body interactions
significantly deform the Fermi surface, leading to the so-called
extended Van Hove singularity at the $M$ point of the hexagonal
Brillouin zone and inducing a topological transition in the
electron system.

Note that the extended Van Hove singularity~\cite{Gofron94} leads
to the divergence in the density of electron states which appears
when the energy band of a system is almost flat (up to 1 meV) in
one of the directions of the Brillouin zone. In this case, a set
of simple saddle points appears forming a critical line or the
so-called extended saddle point~\cite{Gofron94}. Such an extended
saddle point induces a stronger square root-type of the Van Hove
divergence in the density of electron states, in contrast to the
ordinary saddle point in the energy band, which leads to the usual
logarithmic divergence (in the 2D case)~\cite{VanHove53}. This
square root-type divergence, in turn, can favor a significant
increase in the superconducting critical temperature of the
transition~\cite{Gofron94}.

Besides the experimental investigation of a heavily doped graphene
monolayer, the authors of~\cite{McChesney10} studied the ground
state theoretically and analyzed the competition between the
ferromagnetic and superconducting instabilities. The analysis
showed that the tendency to superconductivity prevails in this
competition, as a result of the strong modulation of the effective
interaction along the Fermi contour, i.e., due to
electron-electron interactions only. The superconducting
instability is then predominantly developed in the $f$-wave
channel~\cite{McChesney10}.

The authors of~\cite{Nandkishore12a,Nandkishore12b} used the
method of the functional renormalization
group~\cite{Dzyaloshinskii87,Schulz87,Shankar94} to analyze the
competition between the superconductivity caused by
electron-electron interaction and the phases of spin and
charge-density waves at the Van Hove filling in graphene
monolayer~\cite{Makogon11,Li12,Wang12}. The analysis showed that
three Van Hove saddle points with an ideal nesting lead to the
domination of the superconducting pairings. The renormalization
group analysis has indicated that under these conditions, a
spin-singlet superconducting state with the $d+id$-wave symmetry
type of the order parameter is realized, which spontaneously
breaks the symmetry with respect to the time reversal and leads to
the chiral Andreev states at the boundaries of the sample.
In~\cite{Nandkishore12c}, it was stressed that upon a small shift
of the Fermi level from the Van Hove singularity, a transition to
a spin-density-wave (SDW) phase occurs, and hence the region of
the coexistence of superconductivity and the antiferromagnetic
ordering in the doped graphene is absent.

It was noted in~\cite{Kiesel12} that the long-range Coulomb
interactions can substantially influence the competition between
the superconducting phases with different symmetry types of the
order parameter in doped graphene. In particular, it was shown in
the extended Hubbard model for graphene that far away from the Van
Hove singularity, where the $d+id$-wave spin-singlet pairing is
realized, the SDW phase experiences strong competition with the
charge-density-wave (CDW) phase enhanced by the long-range Coulomb
interactions, which can favor the realization of the triplet
$f-$wave pairing~\cite{Raghu10} (see below).

The importance of the correct account of the long-range part of
the Coulomb interaction when we derive an effective many-particle
model for graphene and graphite from \textit{ab initio}
calculations was emphasized in~\cite{Wehling11}. in fact, from
these calculations we can properly determine the values of the
partially screened frequency-dependent Coulomb interaction. The
Hubbard repulsion in graphene was found to be $U=9.3\,\textrm{eV}$
in agreement with the estimation given in~\cite{Levin74}, but
contradicting the intuitive expectations of a small $U$ and weak
coupling $U<W$; it is known~\cite{Reich02} that
$t_1=2.8\,\textrm{eV}$ in graphene. The authors
of~\cite{Wehling11} also calculated the Coulomb repulsion
parameters for electrons located on the nearest and
next-to-nearest carbon atoms in a graphene monolayer and get
$V_1=5.5\,\textrm{eV}$ and $V_2=4.1\,\textrm{eV}$, respectively.
We note that other researchers (see, e.g.,~\cite{Perfetto07})
assume that these parameters are much smaller.

The competition of superconducting phases with different symmetry
types in a wide range of the concentrations of the charge carriers
$1<n\leq n_{VH}$ in the idealized monolayer of doped graphene was
investigated in~\cite{Kagan14,Nandkishore14}. It was shown that at
the intermediate electron densities, the distant Coulomb
interactions stimulate superconductivity with the $f$-wave
symmetry of the order parameter and upon approaching the Van Hove
singularity, the superconducting pairing with the $d+id$-wave
symmetry type is realized~\cite{Kagan14,Nandkishore14}.
\begin{figure}[t]
\begin{center}
\includegraphics[width=0.45\textwidth]{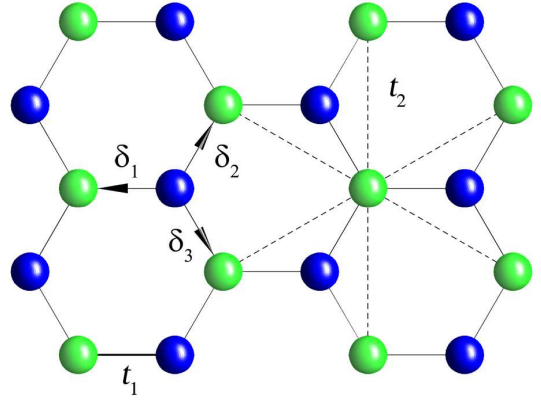}
\caption{Fig.~11. Crystal structure of a graphene monolayer. The
carbon atoms belonging to different sublattices are given by
different colors; $\delta_1$, $\delta_2$ and $\delta_3$ are the
vectors in the directions of the nearest neighbors; $t_1$ and
$t_2$ are the hopping integrals between the nearest and
next-nearest neighbors.} \label{monolayer_lattice}
\end{center}
\end{figure}

In the hexagonal lattice of graphene, each unit cell contains two
carbon atoms; the lattice can therefore is divided into two
sublattices, $A$ and $B$ (Fig.~\ref{monolayer_lattice}). The
Hamiltonian of the Shubin-Vonsovsky model for a monolayer of
graphene, with the electron hoppings between the nearest and the
next-to-nearest atoms and the Coulomb repulsion of electrons
located on the same or different atoms taken into account, in the
Wannier representation is given by
\begin{eqnarray}\label{grapheneHamiltonian}
\hat{H'}&=&\hat{H'}_0+\hat{H}_{\textrm{int}},\\
\hat{H'}_0&=&-\mu\Biggl(\sum_{f\sigma}\hat{n}^A_{f\sigma}+\sum_{g\sigma}
\hat{n}^B_{g\sigma}\Biggr)\nonumber\\
&-&t_1\sum_{
f\delta\sigma}(a^{\dag}_{f\sigma}b_{f+\delta,\sigma}+\textrm{h.c.})\nonumber\\
&-&t_2\Biggl(\sum_{\langle\langle
fm\rangle\rangle}a^{\dag}_{f\sigma}a_{m\sigma}+\sum_{\langle\langle
gn\rangle\rangle}b^{\dag}_{g\sigma}b_{n\sigma}+
\textrm{h.c.}\Biggr),\nonumber
\end{eqnarray}
\begin{eqnarray}\label{monolayerHint}
\hat{H}_{\textrm{int}}&=&U\Biggl(\sum_f
\hat{n}^{A}_{f\uparrow}\hat{n}^{A}_{f\downarrow}+\sum_g
\hat{n}^{B}_{g\uparrow}\hat{n}^{B}_{g\downarrow}\Biggr)\nonumber\\
&+&V_1\sum_{ f\delta\sigma\sigma'}
\hat{n}^{A}_{f\sigma}\hat{n}^{B}_{f+\delta,\sigma'}\nonumber\\
&+&\frac{V_2}{2}\Biggl(\sum_{\langle\langle
fm\rangle\rangle\sigma\sigma'}\hat{n}^{A}_{f\sigma}\hat{n}^{A}_{m\sigma'}+\sum_{\langle\langle
gn\rangle\rangle\sigma\sigma'}\hat{n}^{B}_{g\sigma}\hat{n}^{B}_{n\sigma'}\Biggr).\nonumber
\end{eqnarray}
Here, $a^{\dag}_{f\sigma}(a_{f\sigma})$ are the creation
(annihilation) operators of an electron with the spin projection
$\sigma=\pm1/2$ on the site $f$ of the sublattice $A$;
$\displaystyle\hat{n}^{A}_{f\sigma}=a^{\dag}_{f\sigma}a_{f\sigma}$
is the operator of the number of fermions on the site $f$ of the
sublattice $A$ (and similarly for the sublattice $B$); the vector
$\delta$ connects the nearest-neighbor atoms of the hexagonal
lattice; the double angular brackets mean that the summation is
carried out only over the next-to-nearest neighbors; $t_1$ is the
hopping integral between the neighboring atoms (hoppings between
the different sublattices); $t_2$ is the hopping integral between
next-to-nearest neighbors (hoppings over one sublattice); $U$ is
the parameter of the Hubbard repulsion; and $V_1$ and $V_2$ are
the respective Coulomb repulsions of electrons located on the
nearest and next-to-nearest carbon atoms. It is assumed that the
position of the chemical potential $\mu$ and the number of current
carriers $n$ in the graphene monolayer can be controlled by the
electric field of the gate electrode.

After passing to the momentum space and carrying out the $u-v$
Bogoliubov transformation,
\begin{eqnarray}\label{uv}
&&a_{\textbf{k}\sigma}=w_{11\textbf{k}}\alpha_{1\textbf{k}\sigma}+w_{12\textbf{k}}\alpha_{2\textbf{k}\sigma},\nonumber\\
&&b_{\textbf{k}\sigma}=w_{21\textbf{k}}\alpha_{1\textbf{k}\sigma}+w_{22\textbf{k}}\alpha_{2\textbf{k}\sigma},
\end{eqnarray}
where $\alpha_{1\textbf{k}\sigma}$ and
$\alpha_{2\textbf{k}\sigma}$ are the operators that describe the
respective dynamics of electrons in the upper and lower bands of
the graphene, the Hamiltonian $\hat{H'}_0$ is diagonalized and, as
a result, a well-known expression~\cite{Wallace47} for the
two-band energy spectrum is obtained (see
Fig.~\ref{monolayer_energy}):
\begin{eqnarray}\label{monolayerspectra}
E_{1\textbf{k}}=t_1|u_{\textbf{k}}|-t_2f_{\textbf{k}},\qquad
E_{2\textbf{k}}=-t_1|u_{\textbf{k}}|-t_2f_{\textbf{k}},
\end{eqnarray}
where
\begin{eqnarray}\label{f_k}
&&f_{\textbf{k}}=2\cos(\sqrt{3}k_ya)+
4\cos\biggl(\frac{\sqrt{3}}{2}k_ya\biggr)\cos\biggl(\frac{3}{2}k_xa\biggr),\\
&&u_{\textbf{k}}=\displaystyle\sum_{\delta}e^{i
\textbf{k}\delta}=e^{-ik_xa}+
2e^{\frac{i}{2}k_xa}\cos\biggl(\frac{\sqrt{3}}{2}k_ya\biggr),\label{u_k}\\
&&|u_{\textbf{k}}|=\sqrt{3+f_{\textbf{k}}}.
\end{eqnarray}
The coefficients of the Bogoliubov transformation are
\begin{eqnarray}\label{wz}
&&{w_{1,1}}(\textbf{k}) = w_{22}^*(\textbf{k}) =
\frac{1}{\sqrt{2}}\frac{u^*_\textbf{k}}{|u_\textbf{k}|},\\
&&{w_{12}}(\textbf{k}) = -w_{21}(\textbf{k})
=-\frac{1}{\sqrt{2}}.\nonumber
\end{eqnarray}

In Bogoliubov representation (\ref{uv}), the operator of the
interaction in (\ref{grapheneHamiltonian}) is determined by an
expression that includes the operators
$\alpha_{1\textbf{k}\sigma}$ and $\alpha_{2\textbf{k}\sigma}$,
\begin{eqnarray}\label{Hint_ab}
\hat H_{\textrm{int}} &=& \frac{1}{N}\sum\limits_{ijlm\atop
\textbf{k}\textbf{p}\textbf{q}\textbf{s}\sigma}
\Gamma_{ij;lm}^{||}(\textbf{k},\textbf{p}|\textbf{q},\textbf{s})
\alpha_{i\textbf{k}\sigma}^\dag \alpha_{j\textbf{p}\sigma}^\dag
\alpha_{l\textbf{q}\sigma}\alpha_{m\textbf{s}\sigma}
\nonumber\\
&\times&\delta (\textbf{k}+\textbf{p}-\textbf{q}-\textbf{s}) \\
&+& \frac{1}{N}\sum\limits_{ijlm\atop
\textbf{k}\textbf{p}\textbf{q}\textbf{s}}\Gamma_{ij;lm}^{\bot}(\textbf{k},\textbf{p}|\textbf{q},\textbf{s})
\alpha_{i\textbf{k}\uparrow}^\dag
\alpha_{j\textbf{p}\downarrow}^\dag \alpha_{l\textbf{q}\downarrow}
\alpha_{m\textbf{s}\uparrow}\nonumber\\
&\times&\delta
(\textbf{k}+\textbf{p}-\textbf{q}-\textbf{s}),\nonumber
\end{eqnarray}
where
$\Gamma_{ij;lm}^{||}(\textbf{k},\textbf{p}|\textbf{q},\textbf{s})$
and
$\Gamma_{ij;lm}^{\bot}(\textbf{k},\textbf{p}|\textbf{q},\textbf{s})$
are bare amplitudes, whose form is given below, and $\delta$ is
the Dirac delta-function.

The scattering amplitude in the Cooper channel can be calculated using
the weak-coupling Born approximation with the hierarchy of the
model parameters
\begin{equation}
W>U>V_1>V_2,
\end{equation}
where $W$ is the bandwidth (at $t_2=0$ in Eqn
(\ref{monolayerspectra})) for the upper and lower branches of the
energy spectrum of graphene. We restrict again the consideration
to the second-order diagrams in the effective interaction of two
electrons with opposite values of momentum and spin, using the
quantity $U_{\textrm{eff}}(\textbf{p}, \textbf{k})$. Graphically,
this quantity is the sum of the diagrams represented in
Fig.~\ref{diagrams_alpha}. Assuming that the chemical potential of
doped graphene falls into the upper energy band $E_{1\textbf{k}}$
and analyzing the conditions of the appearance of the
Kohn-Luttinger superconductivity, we can examine the situation
where both the initial and the final momenta belong to the upper
band.

Analytically, the effective interaction
$U_{\textrm{eff}}(\textbf{p}, \textbf{k})$ is given by
\begin{eqnarray}\label{Gamma_wave}
&&U_{\textrm{eff}}(\textbf{p},\textbf{k})=
\widetilde{\Gamma}_0(\textbf{p},\textbf{k})
+\delta U_{\textrm{eff}}(\textbf{p},\textbf{k}),\\
&&\widetilde{\Gamma}_0(\textbf{p},\textbf{k})=\Gamma^{\bot}_{ii;jj}(\textbf{p},
-\textbf{p}| -\textbf{k},
\textbf{k}),\label{Gamma_wave_0}\\
&&\delta U_{\textrm{eff}} (\textbf{p},\textbf{k})=
\frac{1}{N}\sum\limits_{l,m, \textbf{p}_1}
\Gamma^{\bot}_{il;jm}(\textbf{p}, \textbf{q}_2| -\textbf{k},
\textbf{p}_1)\label{Gamma_wave_delta}\nonumber\\
&&\times\Gamma^{\bot}_{mi;lj}(\textbf{p}_1,
-\textbf{p}|\textbf{q}_2,\textbf{k})
\chi_{l,m}(\textbf{q}_2, \textbf{p}_1)\nonumber\\
&&+\frac{2}{N}\sum\limits_{l,m, \textbf{p}_1}  \Bigl\{
\Gamma^{\bot}_{im;lj}( \textbf{p}, \textbf{p}_1| \textbf{q}_1,
\textbf{k})\nonumber\\
&&\times\left[\Gamma^{||}_{li;mj}( \textbf{q}_1, -\textbf{p}|
\textbf{p}_1, -\textbf{k}) -
\Gamma^{||}_{li;jm}( \textbf{q}_1, -\textbf{p}| -\textbf{k}, \textbf{p}_1) \right] \Bigr.\nonumber\\
&&+\Bigl.\Gamma^{\bot}_{li;jm}(\textbf{q}_1, -\textbf{p}| -\textbf{k}, \textbf{p}_1)\nonumber\\
&&\times\left[\Gamma^{||}_{im;jl}( \textbf{p}, \textbf{p}_1|
\textbf{k}, \textbf{q}_1) - \Gamma^{||}_{im;lj}( \textbf{p},
\textbf{p}_1| \textbf{q}_1, \textbf{k})\right]
\Bigr\}\chi_{l,m}(\textbf{q}_1,\textbf{p}_1).\nonumber
\end{eqnarray}
where the expressions
\begin{eqnarray}
\Gamma_{ij;lm}^{||}(\textbf{k},\textbf{p}|\textbf{q},\textbf{s})
&=& \frac
12\Bigl(V_{ij;lm}(\textbf{k},\textbf{p}|\textbf{q},\textbf{s})\nonumber\\
&+&V_{ji;ml}(\textbf{p},\textbf{k}|\textbf{s},\textbf{q})\Bigr),\\
V_{ij;lm}(\textbf{k},\textbf{p}|\textbf{q},\textbf{s})&=&V_1
{u_{\textbf{q}-\textbf{p}}} w_{i1}(\textbf{k}) w_{j2}(\textbf{p})
w^*_{l2}(\textbf{q}) w^*_{m1}(\textbf{s})\nonumber\\
&+&\frac{V_2}{2}f_{\textbf{q}-\textbf{p}}\Bigl( w_{i1}(\textbf{k})
w_{j1}(\textbf{p}) w^*_{l1}(\textbf{q}) w^*_{m1}(\textbf{s})\nonumber\\
&+&w_{i2}(\textbf{k}) w_{j2}(\textbf{p}) w^*_{l2}(\textbf{q})
w^*_{m2}(\textbf{s}) \Bigr)
\end{eqnarray}
describe the strength of the interaction of fermions with parallel
projections of spin, and the expressions
\begin{eqnarray}\label{gammaijlm_monolayer}
&&\Gamma_{ij;lm}^{\bot}(\textbf{k},\textbf{p}|\textbf{q},\textbf{s})
=U_{ij;lm}(\textbf{k},\textbf{p}|\textbf{q},\textbf{s})
\nonumber\\
&&\qquad\qquad+V_{ij;lm}(\textbf{k},\textbf{p}|\textbf{q},\textbf{s})+
V_{ji;ml}(\textbf{p},\textbf{k}|\textbf{s},\textbf{q}),\\
&&U_{ij;lm}(\textbf{k},\textbf{p}|\textbf{q},\textbf{s}) =
U\Bigl( w_{i1}(\textbf{k}) w_{j1}(\textbf{p}) w^*_{l1}(\textbf{q}) w^*_{m1}(\textbf{s}) \nonumber\\
&&\qquad\qquad+w_{i2}(\textbf{k}) w_{j2}(\textbf{p})
w^*_{l2}(\textbf{q}) w^*_{m2}(\textbf{s}) \Bigr)\nonumber
\end{eqnarray}
correspond to the interaction of Fermi quasiparticles with
antiparallel projections of spin. In expressions
(\ref{Gamma_wave})-(\ref{gammaijlm_monolayer}), the indices
$i=j=1$, and the indices $l$ and $m$ can take the values 1 or 2
for upper and lower bands. We introduce the generalized
susceptibilities
\begin{equation}
\chi_{l,m}(\textbf{k},\textbf{p}) = \frac{f(E_{l\textbf{k}}) -
f(E_{m\textbf{p}})} {E_{m\textbf{p}} - E_{l\textbf{k}}},
\end{equation}
where $n_F(x)=\{\exp[(x-\mu)/T]+1\}^{-1}$ is the Fermi-Dirac
distribution function, and the energies $E_{i\textbf{k}}$ are
defined in (\ref{monolayerspectra}). For compactness, the
following notation is introduced in Eqn (\ref{Gamma_wave}) for the
combinations of the momenta:
\begin{equation}\label{q1q2}
\textbf{q}_1 =  \textbf{p}_1 + \textbf{p} - \textbf{k},\qquad
\textbf{q}_2 = \textbf{p}_1-\textbf{p}-\textbf{k}.
\end{equation}

Just as in the case of the Shubin-Vonsovsky model, on the square
lattice (see Section~\ref{sec_SV}~), the problem of the Cooper
instability in graphene monolayer can be reduced to the eigenvalue
problem
\begin{equation}
\label{IntegralEqPhigraphene}
\frac{3\sqrt{3}}{8\pi^2}\oint\limits_{\varepsilon_{\textbf{q}}=\mu}
\frac{d\hat{\textbf{q}}} {v_F(\hat{\textbf{q}})}
U_{\textrm{eff}}(\hat{\textbf{\textbf{p}}},\hat{\textbf{q}})
\Delta(\hat{\textbf{q}})=\lambda\Delta(\hat{\textbf{p}}),
\end{equation}
where the integration is carried out over the contour shown in
Fig.~\ref{monolayer_energy}b.

To solve Eqn (\ref{IntegralEqPhigraphene}), we represent its
kernel as a superposition of the eigenfunctions, each belonging to
one of the irreducible representations of the symmetry group
$C_{6v}$ of the hexagonal lattice. As it is known, this symmetry
group has six irreducible representations~\cite{Landau89}: four
one-dimensional and two two-dimensional. For each representation,
Eqn (\ref{IntegralEqPhigraphene}) has a solution with its own
effective coupling constant $\lambda$. We use the following
notation for the classification of the symmetries of the order
parameter, namely, representation $A_1$ corresponds to $s$-wave
symmetry; $A_2$ to extended $s$-wave pairing; $B_1$ and $B_2$ to
$f$-wave symmetry; $E_1$ to $p+ip$-wave symmetry, and $E_2$ to
$d+id$-wave symmetry.

For each irreducible representation $\nu = A_1, A_2, B_1, B_2,
E_1, E_2$, we search the solution to Eqn (\ref{IntegralEqPhi}) in
the form
\begin{equation}\label{solution}
\Delta^{(\nu)}(\phi)=\sum\limits_{m}\Delta_{m}^{(\nu)}g_{m}^{(\nu)}(\phi),
\end{equation}
where $m$ is the number of an eigenfunction of the representation
$\nu$, and $\phi$ is the angle that determines the direction of
the momentum \textbf{p} relative to the axis $p_x$. Explicitly,
the orthonormalized functions $g_{m}^{(\nu)}(\phi)$ are given by
\begin{eqnarray}\label{harmon}
A_1&&\rightarrow~g_{m}^{(s)}(\phi)=\frac{1}{\sqrt{(1+\delta_{m0})\pi}}\,
\textrm{cos}\,6m\phi,~~m\in[\,0,\infty),\label{invariants_s}\nonumber\\
A_2&&\rightarrow~g_{m}^{(A_2)}(\phi)=\frac{1}{\sqrt{\pi}}\,\textrm{sin}\,
(6m+6)\phi,\label{invariants_s1}\nonumber\\
B_1&&\rightarrow~g_{m}^{(f_1)}(\phi)=\frac{1}{\sqrt{\pi}}\,
\textrm{sin}\,(6m+3)\phi,\label{invariants_dxy}\\
B_2&&\rightarrow~g_{m}^{(f_2)}(\phi)=\frac{1}{\sqrt{\pi}}\,
\textrm{cos}\,(6m+3)\phi,\label{invariants_dx2y2}\nonumber\\
E_1&&\rightarrow~g_{m}^{(p+ip)}(\phi)=\frac{1}{\sqrt{\pi}}\,(A\,\textrm{sin}\,
(2m+1)\phi+\nonumber\\
&&\qquad+B\,\textrm{cos}\,(2m+1)\phi),\label{invariants_p}\nonumber\\
E_2&&~\rightarrow~g_{m}^{(d+id)}(\phi)=\frac{1}{\sqrt{\pi}}\,(A\,\textrm{sin}\,
(2m+2)\phi+\nonumber\\
&&\qquad+B\,\textrm{cos}\,(2m+2)\phi)\label{invariants_p}\nonumber.
\end{eqnarray}
Here, for the two-dimensional representations $E_1$ and $E_2$, the
indices $m$ range over to values such that the coefficients $2m +
1$ and $2m + 2$ are not multiples of 3.
\begin{figure}[t]
\begin{center}
\includegraphics[width=0.47\textwidth]{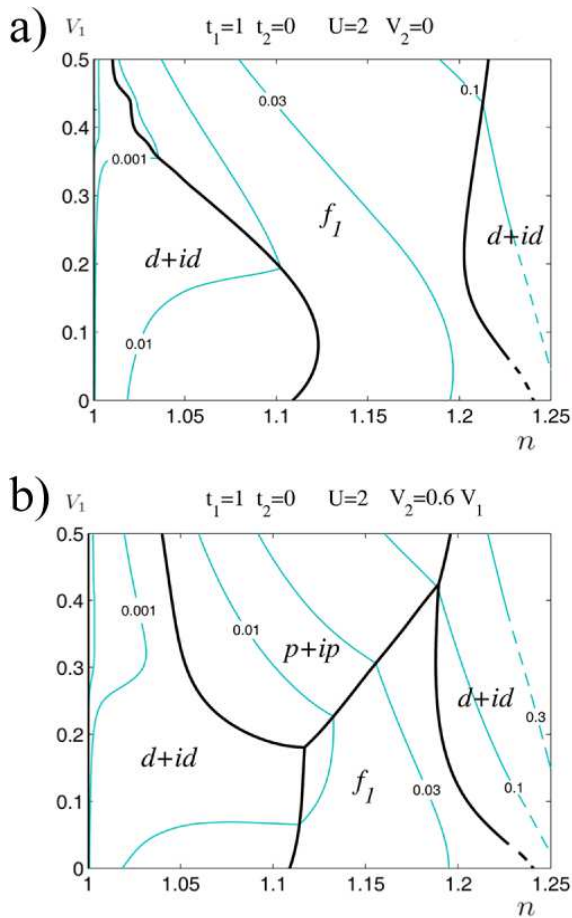}
\caption{Fig.~12. The "$n-V_1$" phase diagram of the
superconducting state of the graphene monolayer at $t_2=0,\,U=2$
for $V_2=0$ (a) and $V_2=0.6V_1$ (b) (all the parameters in the
units of $|t_1|$). The thin lines are lines of constant
$|\lambda|$~\cite{Kagan15a}.}\label{PD_limiting}
\end{center}
\end{figure}

Figure~\ref{PD_limiting}a shows the phase diagram of the
Kohn-Luttinger superconducting state when the energy spectrum of
the monolayer is described by only one hopping parameter
($t_1\neq0,\,t_2=0$), the Hubbard repulsion, $U = 2$ (hereinafter,
all parameters are given in the units of $|t_1|$), as well as the
Coulomb repulsion of electrons located only on the neighboring
carbon atoms ($V_1\neq0,\,V_2=0$) are taken into account.

\begin{figure}[t]
\begin{center}
\includegraphics[width=0.42\textwidth]{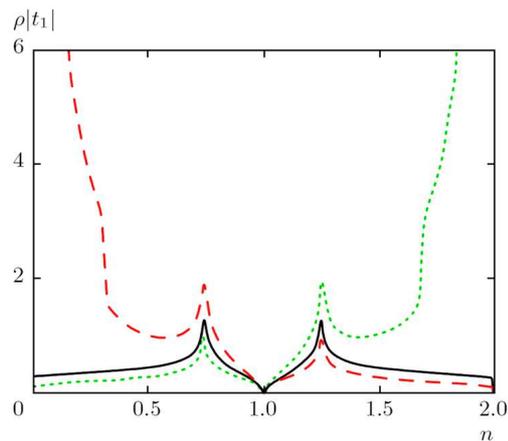}
\caption{Fig.~13. Modification of the electron density of states
for the graphene monolayer upon switching of the hoppings to the
next-nearest atoms for $t_2 =0$ (solid curve), $t_2 = -0.2|t_1|$
(dashed curve), and $t_2 = 0.2|t_1|$ (dotted
curve)~\cite{Kagan14}. }\label{DOS_mono}
\end{center}
\end{figure}

We see from fig.~\ref{PD_limiting}a that the phase diagram
contains three regions. At low electron densities $n$,
superconductivity with the $d + id$-wave symmetry of the order
parameter is realized. At intermediate electron concentrations, a
superconducting $f$-wave pairing takes place. The contribution to
it is determined by the harmonics
$g_{m}^{(f_1)}(\phi)=\displaystyle\frac{1}{\sqrt{\pi}}\,
\textrm{sin}\,(6m+3)\phi$, while a contribution from the harmonics
$g_{m}^{(f_2)}(\phi)=\displaystyle\frac{1}{\sqrt{\pi}}\,
\textrm{cos}\,(6m+3)\phi$ is absent. At larger densities $n$, a
region of the chiral $d + id$-wave pairing appears
again~\cite{Black07,Gonzalez08,Nandkishore12a,Kiesel12}. With an
increase in the parameter of the intersite Coulomb interaction
$V_1$, in the region of small $n$, the $d + id$-wave pairing is
suppressed, and the pairing with $f$-wave symmetry of the order
parameter is realized. The thin curves in Fig.~\ref{PD_limiting}
correspond to the lines of constant $|\lambda|$. It can be seen
that at the approaching the Van Hove filling $n_{VH}$ (solid curve
in Fig.~\ref{DOS_mono}), the values of the effective coupling
constant reach quite reasonable values $|\lambda|=0.1$.

Here, again to avoid the summation of parquet
diagrams~\cite{Dzyaloshinskii88a,Dzyaloshinskii88b,Zheleznyak97},
only the regions of electron concentrations that are not to close
to the Van Hove singularity in the density of electronic states of
graphene are analyzed (see Fig.~\ref{DOS_mono}). For this reason,
the boundaries between the different regions of the realization of
the Kohn-Luttinger superconducting pairing and the lines of the
constant $|\lambda|$ located very close to the Van Hove
singularity are depicted in the phase diagram as dashed curves.

When we take into account the long-range Coulomb interactions of
electrons $V_2$ on the hexagonal lattice of graphene, a
qualitative change occurs in the phase diagram of the
superconducting state~\cite{Kagan15a}. This can be seen from
Fig.~\ref{PD_limiting}b, which was obtained at a fixed relation
between the parameters of the long-range Coulomb interactions
$V_2=0.6V_1$. In this case, a suppression occurs of the wide
region of the superconducting state with the $f$-wave symmetry at
the intermediate electron densities, and a superconducting pairing
with the $p + ip$-wave symmetry is realized. Furthermore, an
account of $V_2$ leads to an increase in the values of the
effective coupling constant to $|\lambda|=0.3$.

Account for the electron hoppings to the next-to-nearest carbon
atoms $t_2$ in the graphene monolayer does not qualitatively
affect the competition between the superconducting phases with
different symmetry types, as can be seen from
Fig.\ref{PD_limiting}b~\cite{Kagan14}. This behavior of the system
is explained by the fact that the hoppings with $t_2>0$ or $t_2<0$
do not cause an essential modification of the density of states of
the monolayer in the ranges of concentrations of charge carriers
between the Dirac point and both Van Hove singularities (see
Fig.~\ref{DOS_mono}). But when we take the hoppings $t_2$ into
account we get an increase in the absolute values of the effective
interaction and, consequently, the realization of the higher
critical temperatures in the idealized graphene
monolayer~\cite{Kagan14}.

\begin{figure}[b]
\begin{center}
\includegraphics[width=0.43\textwidth]{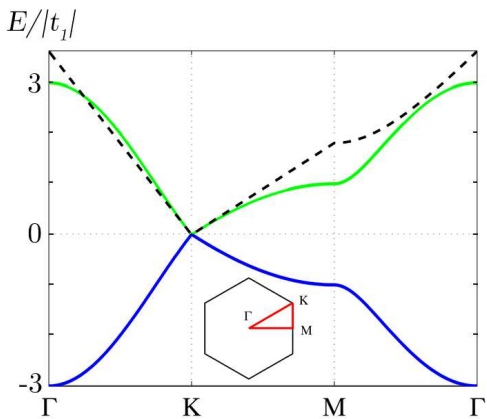}
\caption{Fig.~14. Comparison of the energy spectrum of a graphene
monolayer determined by Eqn (\ref{monolayerspectra}) (blue and
green solid lines) and of the spectrum obtained in the Dirac
approximation (black dashed line). The insert depicts the contour
for moving around the Brillouin zone.} \label{Dirac}
\end{center}
\end{figure}

We note that the Kohn-Luttinger superconductivity (and the
corresponding value of $|\lambda|$) in graphene is never connected
with the Dirac points. Calculations show that in the vicinity of
these points, where a linear approximation to the energy spectrum
of the monolayer (and a parabolic approximation for the spectrum
of graphene bilayer, see Section~\ref{sec_bilayer}~) works well,
the density of states is very low, and the value of the effective
coupling constant $|\lambda|$ is less than $10^{-2}$. Larger
values of $|\lambda|$, which indicate the development of a Cooper
instability, appear at electron densities $n>1.15$. However, at
these concentrations, the energy spectrum of the monolayer along
the $KM$ direction of the Brillouin zone (Fig.~\ref{Dirac})
already differs significantly from the Dirac
approximation~\cite{Kagan15a}.

In Refs~\cite{Zaitsev11,Zaitsev12}, the possibility of Cooper
pairing in graphene was investigated in the opposite limit of
strong coupling, $U\gg t$~\cite{Wehling11}, which is based on the
kinematic mechanism of superconductivity and using the diagram
technique for the Hubbard
operators~\cite{Zaitsev75,Zaitsev76,VVVSGO01}. In particular, a
phase diagram for the superconducting ordering was constructed and
the BCS coupling constant was calculated depending on the filling
of the $\pi$ or $\sigma$ shell.

\section{Enhancement of superconductivity in an idealized graphene bilayer}
\label{sec_bilayer}

\begin{figure}[t]
\begin{center}
\includegraphics[width=0.48\textwidth]{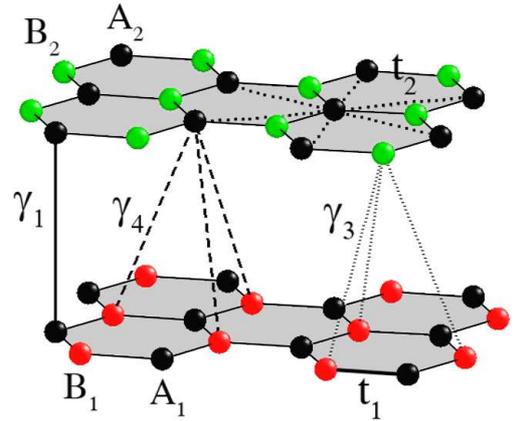}
\caption{Fig.~15. Crystal structure of the graphene bilayer.
Carbon atoms $A_1$ and $B_1$ in the lower layer are shown by red
and black circles; atoms $A_2$ and $B_2$ in the upper layer are
shown by black and green circles. Intralayer electron hoppings are
marked by $t_1$ and $t_2$; $\gamma_1,\,\gamma_3$, and $\gamma_4$
show the interplane
hoppings~\cite{Kagan15a}.}\label{bilayer_structure}
\end{center}
\end{figure}

As far as the electronic properties of graphene depend on the
number of carbon layers~\cite{Guinea06}, we therefore analyze the
possibility of the development of superconducting instability in
an idealized graphene bilayer~\cite{Kagan14c,Kagan15a}, whose
crystalline structure is shown in Fig.~\ref{bilayer_structure}.
The energy band structure of a monolayer is characterized by a
linear dispersion near the Dirac points, while the bilayer has a
quadratic energy spectrum in the low-energy limit
(see~\cite{McCann06,Nilsson06,Partoens06}, and also
reviews~\cite{McCann07,McCann13}).

The authors of~\cite{Nunes12} examined the effect of the
interplanar electron hopping in bilayer graphene and graphite on
the formation of the superconducting phase. Assuming that the
Hubbard interaction is attractive and leads to a superconducting
$s$-wave pairing, the authors of~\cite{Nunes12} showed in the
mean-field approximation that the interplanar hopping increases
the critical temperature $T_c$ of the superconducting transition
at low values of the chemical potential.

The authors of~\cite{Vucicevic12} investigated the possibility of
the realization of the superconducting phase in the mean-field
approximation in the framework of the $t-J$ model for a graphene
bilayer with the direct interlayer hopping $\gamma_1$, (see
Fig.~\ref{bilayer_structure}) and the superexchange interaction
caused by the strong Hubbard repulsion $U$ of electrons. It was
shown that in the bilayer with moderate doping and for the
low-energy scales, the $d+id$-wave pairing is predominant.
In~\cite{Milovanovic12}, using the same model the authors
discussed the coexistence of chiral superconductivity with the
$d+id$-wave symmetry and antiferromagnetism near the half-filling
for the graphene bilayer.

The authors of~\cite{Hosseini12a,Hosseini12b} studied the exotic
superconductivity mechanism based on the interlayer pairing of
chiral electrons in a graphene bilayer, which leads to anomalous
thermodynamic properties. According to the conclusions
in~\cite{Hosseini12a,Hosseini12b}, this mechanism of
superconductivity in a graphene bilayer is quite similar to color
superconductivity~\cite{Alford99} (based on the pairing of quarks
in high-density quark matter at low temperatures) and to gapless
states in nuclear matter~\cite{Sedrakian00}.

The authors of~\cite{Hwang08} used the random phase approximation
(RPA) to calculate the screening of Coulomb interaction in a
graphene bilayer in both doped and undoped regimes. They found
that the static polarization operator of a doped bilayer contains
the Kohn anomaly that is considerably larger than in the case of
graphene monolayer or a 2D electron gas. As it was already noted
in Section~\ref{sec_gas}~, the singular part of the polarization
operator, or the Kohn anomaly, favors an effective attraction of
two particles ensuring a contribution that (for the orbital
angular momenta of the pair $l\neq0$) always exceeds the repulsive
contribution caused by the regular part of the polarization
operator. Thus, the authors of~\cite{Hwang08} concluded that in
the idealized bilayer, it is possible to expect an increase in the
critical temperatures $T_c$ of the superconducting transition in
comparison with $T_c$ in the idealized graphene monolayer.

According to the results in~\cite{Gonzalez13}, the ferromagnetic
instability near the Van Hove singularities prevails over the
Kohn-Luttinger superconductivity. Nevertheless, in~\cite{Vafek14},
the possibility of the superconducting pairing in the repulsive
case on the hexagonal lattice for the graphene bilayer was
investigated within the renormalization group formalism in the
weak-coupling regime far away from the half-filling. Although the
utilization of the renormalization group approach in this regime
can be substantiated only formally~\cite{Vafek14}, the authors
discovered, a chiral $d$-wave type superconductivity besides the
$f$-wave type of the superconducting pairings. Estimating the
critical temperature of the superconducting transition in the
idealized system, the authors of~\cite{Vafek14} obtained
$T_c\sim1\,\textrm{K}$ and noted that the critical temperature can
be lower if we take into account the electron scattering on
impurities.

In~\cite{Kagan14c,Kagan15a}, the authors investigated in details
the influence of the Coulomb interaction of the Dirac fermions on
the formation of the Kohn-Luttinger superconducting state in doped
bilayer graphene neglecting the van der Waals potential of the
substrate and the role of impurities. Within the Shubin-Vonsovsky
model in the weak-coupling Born approximation, taking into account
the Hubbard, interatomic (inside the layer), and interlayer
Coulomb interactions of electrons, a phase diagram of
superconducting state was constructed. It has been shown that the
Kohn-Luttinger polarization contributions on the second order of
the perturbation theory and the long-range intraplane Coulomb
interactions substantially influence the competition between the
superconducting phases with the $f$-, $p + ip$-, and $d + id$-wave
symmetries. It has been demonstrated that the interlayer Coulomb
interaction can lead to an increase in the critical temperature of
the superconducting transition. Now, we discuss the results of the
calculations performed in~\cite{Kagan14c,Kagan15a} in more
details.

We consider an idealized graphene bilayer, assuming that two
monolayers are located according to the $AB$-stacking, i.e., one
layer is turned through 60° relative to the other
one~\cite{McCann06,McCann13}. In this case, we choose an
arrangement of the sublattices in the layers in such a way that
the sites from the different layers located on top of one another
will be refered to the sublattices $A_1$ and $A_2$, and the
remaining sites to the sublattices $B_1$ and $B_2$ (see
Fig.~\ref{bilayer_structure}). In this case, the Hamiltonian of
the Shubin-Vonsovsky model for bilayer graphene taking into
account the electron hoppings between the nearest and
next-to-nearest atoms, the Coulomb repulsion of electrons located
on the same and on different atoms of one layer, as well as with
the interlayer Coulomb interaction, has the following form in the
Wannier representation:
\begin{eqnarray}\label{HamiltonianBilayer}
\hat{H'}&=&\hat{H'}_0+\hat{H}_{\textrm{int}},\\
\hat{H'}_0&=&(\varepsilon-\mu)\Biggl(\sum_{if\sigma}\hat{n}^{A}_{if\sigma}+
\sum_{ig\sigma}\hat{n}^{B}_{ig\sigma}\Biggr)\nonumber\\
&-&t_1\sum_{f\delta\sigma}(a^{\dag}_{1f\sigma}b_{1,f+\delta,\sigma}+
a^{\dag}_{2f\sigma}b_{2,f-\delta,\sigma}+\textrm{h.c.})\nonumber\\
&-&t_2\sum_{i\sigma}\Biggl(\sum_{\langle\langle
fm\rangle\rangle}a^{\dag}_{if\sigma}a_{im\sigma}+\sum_{\langle\langle
gn\rangle\rangle}b^{\dag}_{ig\sigma}b_{in\sigma}+
\textrm{h.c.}\Biggr)\nonumber\\
&-&\gamma_1\sum_{f\sigma}(a^{\dag}_{1f\sigma}a_{2f\sigma}+\textrm{h.c.})\nonumber\\
&-&\gamma_3\sum_{g\delta\sigma}(b^{\dag}_{1g\sigma}b_{2,g+\delta,\sigma}+\textrm{h.c.})\nonumber\\
&-&\gamma_4\sum_{f\delta\sigma}(a^{\dag}_{1f\sigma}b_{2,f-\delta,\sigma}+
a^{\dag}_{2f\sigma}b_{1,f+\delta,\sigma}+\textrm{h.c.}),\label{H0Bilayer}
\end{eqnarray}
\begin{eqnarray}
\hat{H}_{\textrm{int}}&=&U\biggl(\sum_{if}
\hat{n}^{A}_{if\uparrow}\hat{n}^{A}_{if\downarrow}+ \sum_{ig}
\hat{n}^{B}_{ig\uparrow}\hat{n}^{B}_{ig\downarrow}\biggr)\nonumber\\
&+&V_1\sum_{f\delta\sigma\sigma'}
\Bigl(\hat{n}^{A}_{1f\sigma}\hat{n}^{B}_{1,f+\delta,\sigma'}+
\hat{n}^{A}_{2f\sigma}\hat{n}^{B}_{2,f-\delta,\sigma'}\Bigr)\nonumber\\
&+&\frac{V_2}{2}\sum_{i\sigma\sigma'}\Biggl(\sum_{\langle\langle
fm\rangle\rangle}\hat{n}^{A}_{if\sigma}\hat{n}^{A}_{im\sigma'}+\sum_{\langle\langle
gn\rangle\rangle}\hat{n}^{B}_{ig\sigma}\hat{n}^{B}_{in\sigma'}\Biggr)\nonumber\\
&+&G_1\sum_{f\sigma\sigma'}
\hat{n}^{A}_{1f\sigma}\hat{n}^{A}_{2f\sigma'} +
G_3\sum_{g\delta\sigma\sigma'}
\hat{n}^{B}_{1g\sigma}\hat{n}^{B}_{2,g+\delta,\sigma'}\nonumber\\
&+&G_4\sum_{f\delta\sigma\sigma'}
\Bigl(\hat{n}^{A}_{1f\sigma}\hat{n}^{B}_{2,f-\delta,\sigma'}+
\hat{n}^{A}_{2f\sigma}\hat{n}^{B}_{1,f+\delta,\sigma'}\Bigr).\label{HintBilayer}
\end{eqnarray}
The notations used here are similar to ones used for monolayer
Hamiltonian (\ref{grapheneHamiltonian}). In Hamiltonian
(\ref{HamiltonianBilayer}), the index $i=1,2$ stands for the
number of a layer. The vector $\delta (-\delta)$ connects the
nearest-neighbor atoms of the hexagonal lattice of the lower
(upper) layer. It is assumed that the on-site energies are
$\varepsilon_{Ai}=\varepsilon_{Bi}=\varepsilon$. The symbols
$\gamma_1,\,\gamma_3$ and $\gamma_4$ are the parameters of
interlayer electron hoppings (see Fig.~\ref{bilayer_structure}),
and $G_1$, $G_3$ and $G_4$ are the interlayer Coulomb
interactions.

As in the case of a monolayer (see Section~\ref{sec_monolayer}~),
the Hamiltonian $\hat{H'}_0$ is diagonalized by the Bogoliubov
transformation
\begin{eqnarray}\label{uv2}
\alpha_{i\textbf{k}\sigma}&=& w_{i1}(\textbf{k}){a_{1
\textbf{k}\sigma }} + w_{i2}(\textbf{k}){a_{2\textbf{k}\sigma
}}\\
&+&w_{i3}(\textbf{k}){b_{1\textbf{k}\sigma }} +
w_{i4}(\textbf{k}){b_{2\textbf{k}\sigma }},\quad
i=1,2,3,4,\nonumber
\end{eqnarray}
and is reduced to the form
\begin{eqnarray}
\hat H'_0 =\sum\limits_{i=1}^4 \sum\limits_{ \textbf{k}\sigma }
E_{i\textbf{k}}
{\alpha_{i\textbf{k}\sigma}^{\dag}\alpha_{i\textbf{k}\sigma}}.
\end{eqnarray}
Since the results of \emph{ab initio}
calculations~\cite{Dresselhaus02,Brandt88} performed for graphite
indicate a very low value of the parameter of the interlayer
hopping $\gamma_4$, we assume that $\gamma_4=0$. In this case, the
four-band energy spectrum of bilayer graphene is given by
\begin{eqnarray}\label{spectra_bilayer}
&&E_{i\textbf{k}}=\varepsilon\pm\sqrt{A_{\textbf{k}}\pm\sqrt{B_{\textbf{k}}}}-t_2f_{\textbf{k}},\\
&&A_{\textbf{k}}=\frac14\Bigl(2a^2+4|b_{\textbf{k}}|^2+2|d_{\textbf{k}}|^2\Bigr),\nonumber\\
&&B_{\textbf{k}}=\frac14\Bigl(|d_{\textbf{k}}|^2(|d_{\textbf{k}}|^2-2a^2+4|b_{\textbf{k}}|^2)+a^4+4a^2|b_{\textbf{k}}|^2\nonumber\\
&&\qquad+4ab^2_{\textbf{k}}d_{\textbf{k}}+4ab_{\textbf{k}}^{*2}d^*_{\textbf{k}}\Bigr),\nonumber\\
&&a=\gamma_1,\quad b_{\textbf{k}}=t_1u_{\textbf{k}},\quad
d_{\textbf{k}}=\gamma_3u_{\textbf{k}}.\nonumber
\end{eqnarray}

\begin{figure}[t]
\begin{center}
\includegraphics[width=0.40\textwidth]{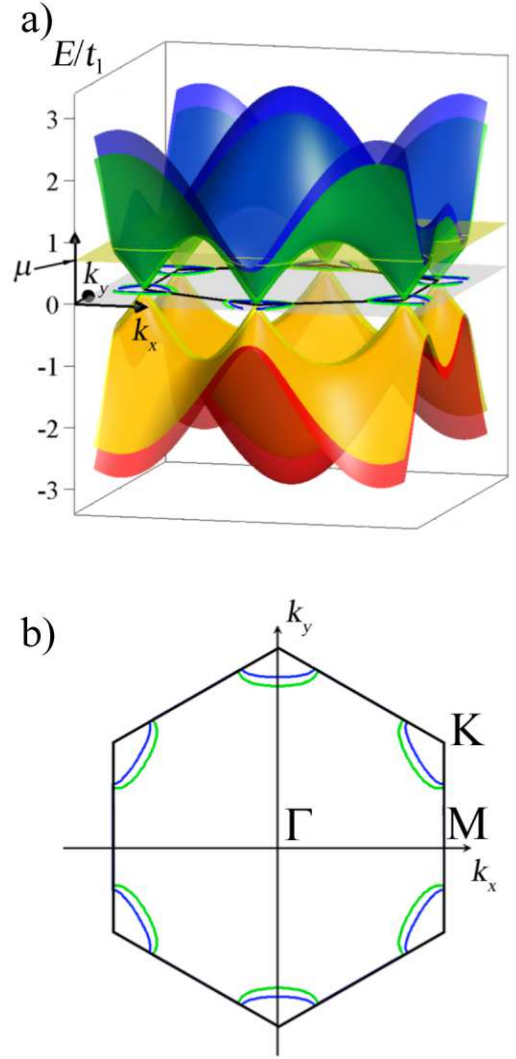}
\caption{Fig.~16. (a) Energy structure of the graphene bilayer
near the Dirac points and (b) the formation of a multisheet Fermi
contour at $t_2=0,\,\gamma_1=0.12,\,\gamma_3=0.1$ and $\mu=0.7$
(all the parameters are in the units of
$|t_1|$)~\cite{Kagan15a}.}\label{two_contours}
\end{center}
\end{figure}

The conditions for the appearance of the Kohn-Luttinger
superconductivity in the graphene bilayer in the Born
weak-coupling approximation are analyzed using the hierarchy of
the model parameters
\begin{equation}\label{hierarchy_bilayer}
W>U>V_1>V_2>G_1>G_3,\,G_4,
\end{equation}
(where $W$ is the bandwidth of the graphene bilayer), according to
the general scheme presented in Section~\ref{sec_monolayer}~. Upon
doping of the bilayer, the chemical potential is assumed to be
located inside the two upper energy bands, $E_{1\textbf{k}}$ and
$E_{2\textbf{k}}$, as it is shown in Fig.~\ref{two_contours}a. In
this case, if $\gamma_1\neq0$ and $\mu>\gamma_1$, in the vicinity
of each Dirac point at the electron densities $1<n<n_{VH}$ (where
$n$ is the electron concentration calculated per number of atoms
of one layer), the Fermi contour consists of two lines, as it is
shown in Fig.~\ref{two_contours}b. The initial and the final
momenta of electrons in the Cooper channel then also belong to the
two upper bands and, consequently, the indices $i$ and $j$ in the
Kohn-Luttinger diagrams in the case of a bilayer (see
Fig.~\ref{diagrams_alpha}) take the values 1 or 2. As a result, we
arrive at expression (\ref{Gamma_wave}) for the effective
interaction of electrons in the Cooper channel, in which the bare
amplitudes for the graphene bilayer are given by
\begin{eqnarray}
\Gamma_{ij;lm}^{||}&&(\textbf{k},\textbf{p}|\textbf{q},\textbf{s})\nonumber\\
&&=
\frac12\Bigl(V_{ij;lm}(\textbf{k},\textbf{p}|\textbf{q},\textbf{s})
+V_{ji;ml}(\textbf{p},\textbf{k}|\textbf{s},\textbf{q})\nonumber\\
&&+G^{(1)}_{ij;lm}(\textbf{k},\textbf{p}|\textbf{q},\textbf{s})
+G^{(1)}_{ji;ml}(\textbf{p},\textbf{k}|\textbf{s},\textbf{q})\nonumber\\
&&+G^{(3)}_{ij;lm}(\textbf{k},\textbf{p}|\textbf{q},\textbf{s})
+G^{(3)}_{ji;ml}(\textbf{p},\textbf{k}|\textbf{s},\textbf{q})\nonumber\\
&&+G^{(4)}_{ij;lm}(\textbf{k},\textbf{p}|\textbf{q},\textbf{s})
+G^{(4)}_{ji;ml}(\textbf{p},\textbf{k}|\textbf{s},\textbf{q})\Bigr),\\
V_{ij;lm}&&(\textbf{k},\textbf{p}|\textbf{q},\textbf{s})=V_1\Bigl(
u_{\textbf{q}-\textbf{p}} w_{i1}(\textbf{k}) w_{j3}(\textbf{p})
w^*_{l3}(\textbf{q}) w^*_{m1}(\textbf{s})\nonumber\\
&&+u_{\textbf{q}-\textbf{p}}^* w_{i2}(\textbf{k})
w_{j4}(\textbf{p})
w^*_{l4}(\textbf{q}) w^*_{m2}(\textbf{s})\Bigr)\nonumber\\
&&+\frac{V_2}{2}\sum_{r=1}^4
f_{\textbf{q}-\textbf{p}}w_{ir}(\textbf{k}) w_{jr}(\textbf{p})
w^*_{lr}(\textbf{q}) w^*_{mr}(\textbf{s}),
\end{eqnarray}
\begin{eqnarray}
G^{(1)}_{ij;lm}&&(\textbf{k},\textbf{p}|\textbf{q},\textbf{s})=G_1
w_{i1}(\textbf{k}) w_{j2}(\textbf{p}) w^*_{l2}(\textbf{q})
w^*_{m1}(\textbf{s}),\\
G^{(3)}_{ij;lm}&&(\textbf{k},\textbf{p}|\textbf{q},\textbf{s})\nonumber\\
&&=G_3 u_{\textbf{q}-\textbf{p}} w_{i3}(\textbf{k})
w_{j4}(\textbf{p})
w^*_{l4}(\textbf{q})w^*_{m3}(\textbf{s}),\\
G^{(4)}_{ij;lm}&&(\textbf{k},\textbf{p}|\textbf{q},\textbf{s})=G_4\Bigl(
u_{\textbf{q}-\textbf{p}}^* w_{i1}(\textbf{k}) w_{j4}(\textbf{p})
w^*_{l4}(\textbf{q}) w^*_{m1}(\textbf{s})\nonumber\\
&&+u_{\textbf{q}-\textbf{p}} w_{i2}(\textbf{k}) w_{j3}(\textbf{p})
w^*_{l3}(\textbf{q}) w^*_{m2}(\textbf{s})\Bigr),
\end{eqnarray}
\begin{eqnarray}
\Gamma_{ij;lm}^{\bot}&&(\textbf{k},\textbf{p}|\textbf{q},\textbf{s})
=U_{ij;lm}(\textbf{k},\textbf{p}|\textbf{q},\textbf{s})
\nonumber\\
&&+V_{ij;lm}(\textbf{k},\textbf{p}|\textbf{q},\textbf{s})+
V_{ji;ml}(\textbf{p},\textbf{k}|\textbf{s},\textbf{q})\nonumber\\
&&+G^{(1)}_{ij;lm}(\textbf{k},\textbf{p}|\textbf{q},\textbf{s})
+G^{(1)}_{ji;ml}(\textbf{p},\textbf{k}|\textbf{s},\textbf{q})\nonumber\\
&&+G^{(3)}_{ij;lm}(\textbf{k},\textbf{p}|\textbf{q},\textbf{s})
+G^{(3)}_{ji;ml}(\textbf{p},\textbf{k}|\textbf{s},\textbf{q})\nonumber\\
&&+G^{(4)}_{ij;lm}(\textbf{k},\textbf{p}|\textbf{q},\textbf{s})
+G^{(4)}_{ji;ml}(\textbf{p},\textbf{k}|\textbf{s},\textbf{q}),\\
U_{ij;lm}&&(\textbf{k},\textbf{p}|\textbf{q},\textbf{s})
\nonumber\\
&&=U\sum_{r=1}^4 w_{ir}(\textbf{k}) w_{jr}(\textbf{p})
w^*_{lr}(\textbf{q}) w^*_{mr}(\textbf{s}).
\end{eqnarray}
Here, the indices $l$ and $m$ can take the values 1,2, 3, or 4.

As it was already noted in the case of the monolayer, there is no
common opinion in the literature concerning the values of the
intraplane and interplane Coulomb interaction parameters in a
graphene bilayer~\cite{Wehling11,Milovanovic12}. In our
calculations, we used the hierarchy of parameters given by
(\ref{hierarchy_bilayer}), which allows the application of the
Born weak-coupling approximation. For the parameters of the
interlayer hopping $\gamma_1$ and $\gamma_3$, values close to
those determined for graphite in
Refs~\cite{Dresselhaus02,Brandt88} are used.
\begin{figure}[t]
\begin{center}
\includegraphics[width=0.49\textwidth]{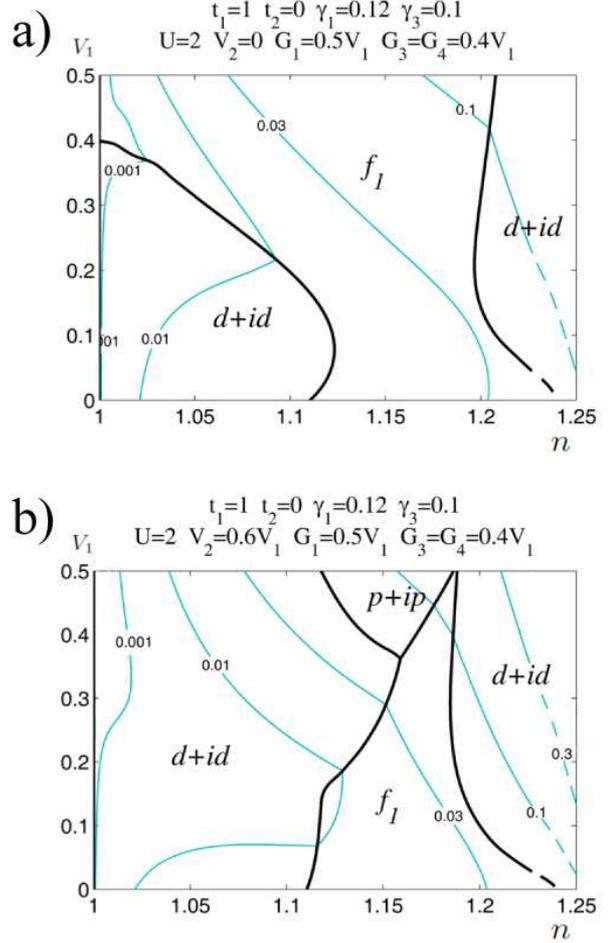}
\caption{Fig.~17. The "$n-V_1$" phase diagram of the
superconducting state of the graphene bilayer at
$t_2=0,\,\gamma_1=0.12,\,\gamma_3=0.1,\,U=2,\,G_1=0.5V_1,\,G_3=G_4=0.4V_1$
at $V_2=0$ (a) and $V_2=0.6V_1$ (b) (all the parameters are in the
units of $|t_1|$). The thin lines are lines of constant
$|\lambda|$~\cite{Kagan15a}.}\label{PD_G3G4}
\end{center}
\end{figure}

Let us consider the superconducting phase diagram of the graphene
bilayer and its modifications by different interactions. First of
all, we note that at $\gamma_1=\gamma_3=\gamma_4=0$ and
$G_1=G_3=G_4=0$, when the graphene bilayer represents two isolated
monolayers, a limiting transition to the results obtained for the
monolayer in Section 8 is checked in the numerical calculations.
Choosing the interlayer electron hopping parameters as
$\gamma_1=0.12,\,\gamma_3=0.1$ (hereinafter, all parameters are
taken in the units of $|t_1|$), with all the other parameters
having the same values as in Fig.~\ref{PD_limiting}, we do not get
considerable changes in the phase diagram of the graphene bilayer.
Including the Coulomb interaction $G_1$, we get only a weak shift
of the boundaries of the $f_1$ and $d+id$-wave pairing in the
phase diagram of Fig.~\ref{PD_limiting}. Moreover, this inclusion
does not affect the absolute values of $\lambda$.

Figure~\ref{PD_G3G4} demonstrates the influence of the interlayer
Coulomb interactions $G_3$ and $G_4$ on the phase diagram of the
graphene bilayer. This diagram was obtained using the set of
parameters $t_2=0,\,\gamma_1=0.12,\,\gamma_3=0.1,\,U=2$ and
$V_2=0$ for the relations between the interlayer and intersite
(intralayer) Coulomb interactions. In Fig.~\ref{PD_G3G4}a, we
chose the set of parameters $G_1=0.5V_1,\,G_3=G_4=0.4V_1$ in
accordance with the hierarchy of the parameters specified by
(\ref{hierarchy_bilayer}). The results of the calculations show
that an increase in $G_3$ and $G_4$ separately leads to a
suppression of the $d + id$-wave pairing and to an expansion of
the region of the $f$-wave pairing at low electron densities. In
this case, a stronger suppression of the superconducting $d +
id$-wave phase can be achieved by an increase in the interlayer
Coulomb interaction parameter $G_4$. Simultaneously, when we take
the interactions $G_3$ and $G_4$ into account, as it is shown in
Fig.~\ref{PD_G3G4}a, we get not only an intensive suppression of
the superconducting $d+id$-wave pairing at low electron densities
and realization of the superconductivity with the $f$-wave
symmetry, but also an increase in the absolute values of the
effective coupling constant $\lambda$.
\begin{figure}[b]
\begin{center}
\includegraphics[width=0.49\textwidth]{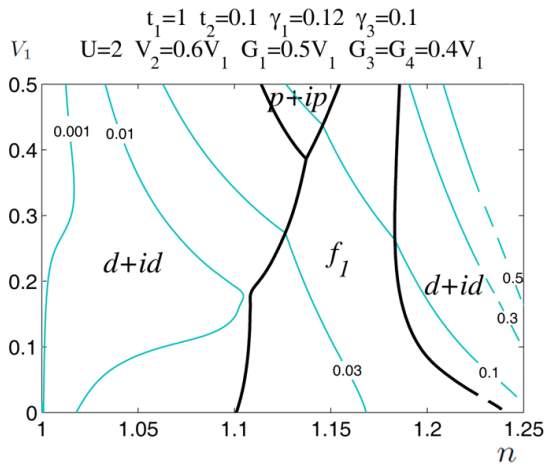}
\caption{Fig.~18. The "$n-V_1$" phase diagram of the
superconducting state of a graphene bilayer at
$t_2=0.1,\,\gamma_1=0.12,\,\gamma_3=0.1,\,U=2,\,G_1=0.5V_1,\,G_3=G_4=0.4V_1$
(all the parameters are in the units of $|t_1|$). The thin lines
are lines of constant $|\lambda|$~\cite{Kagan15a}.}\label{PD_t2}
\end{center}
\end{figure}

Figure~\ref{PD_G3G4}b shows the phase diagram of a graphene
bilayer calculated at the same parameters as in
Fig.~\ref{PD_G3G4}a, but with an account of the long-range Coulomb
repulsion of electrons $V_2$. It can be seen from the comparison
of Fig.~\ref{PD_G3G4}b and Fig.~\ref{PD_limiting}b that the
inclusion of $G_3\neq0$ and $G_4\neq0$ leads to a strong
competition between the $d + id$- and $p + ip$-wave pairing and to
a significant suppression of the latter in the region of the
intermediate electron concentrations. In this case, in the
preserved region of $p + ip$-wave pairing, the value of
$|\lambda_{p+ip}|$ exceeds the $f$-wave coupling constant
$|\lambda_f|$ insignificantly.

\begin{figure}[t]
\begin{center}
\includegraphics[width=0.49\textwidth]{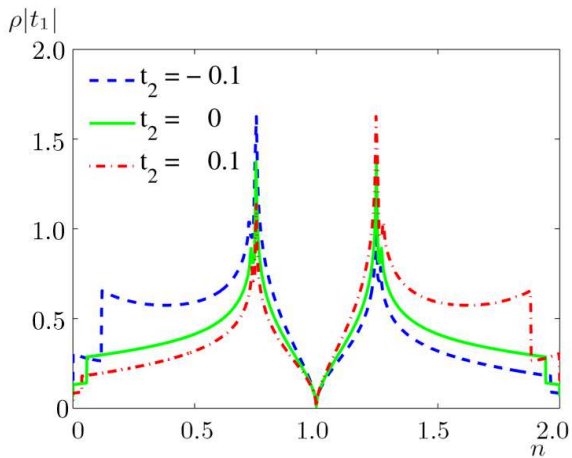}
\caption{Fig.~19. Dependence of the density of states for the
graphene bilayer per unit cell of one layer on the electron
concentration for the set of parameters
$t_2=0,\,\gamma_1=0.12|t_1|,\,\gamma_3=0.1|t_1|$~\cite{Kagan15a}.}\label{DOS_bilayer}
\end{center}
\end{figure}

The inclusion of electron hoppings $t_2$ to the next-to-nearest
carbon atoms does not affect qualitatively the competition between
the superconducting phases shown in Fig.~\ref{PD_G3G4}. This can
be seen from Fig.~\ref{PD_t2}, which shows the phase diagram of a
graphene bilayer obtained with the parameter values
$t_2=0.1,\,\gamma_1=0.12,\,\gamma_3=0.1,\,U=2,\,G_1=0.5V_1$ and
$G_3=G_4=0.4V_1$. This behavior of the system is explained by the
fact that the hopping $t_2 > 0$ or $t_2 < 0$ for the graphene
bilayer, just as in the case of the monolayer, does not lead to an
essential modification of the density of states in the regions of
charge-carrier concentrations between the Dirac point and both
$n_{VH}$ points (Fig.~\ref{DOS_bilayer}). However, it can be seen
from Fig.~\ref{PD_t2} that an account for hoppings $t_2$ leads to
an increase in the absolute values of the effective interaction
and, consequently, to the higher critical temperatures of the
superconducting transition in the idealized graphene bilayer.

Our calculations show that the Kohn-Luttinger mechanism can result
in critical temperatures of the superconducting transition as high
as $T_c\sim 20-40~\textrm{K}$ in the idealized graphene bilayer.
In spite of these rather optimistic estimations, the
superconductivity in real graphene, as it was stressed in
Section~\ref{sec_monolayer}~, has not been discovered yet. Real
graphene is only close to superconductivity.

There are several reasons why the results of the theoretical
calculations presented in this review can be in disagreement with
the experimental data. First of all, in our calculations, we did
not take into account the influence of the van der Waals potential
of the substrate
~\cite{Girifalco02,Hasegawa04,Hasegawa07,Bostrom12,Klimchitskaya13}.
With an increase in the number of layers, the role of this
potential should be weakened apparently. However, even in the case
of a multilayer system, the van der Waals forces can worsen the
conditions for the development of the Cooper instability.

Secondly, as we noted, there is no common opinion in the
literature concerning the values of the parameters of the
intraplane and interplane Coulomb interactions in the graphene
bilayer. In this review, the values close to those obtained from
\textit{ab initio} calculations (performed in~\cite{Wehling11} for
graphite) were used for the intraplane Coulomb interactions. The
values of the interplane Coulomb interactions were taken such that
they would satisfy the hierarchy of the parameters of the Born
weak-coupling approximation.

Thirdly, in our calculations, a clean and structurally ideal
graphene bilayer is considered, whereas a real material contains
various impurities and structural imperfections, including grain
boundaries and twinning planes. For conventional $s$-wave pairing,
the singlet superconducting state is destroyed by magnetic
impurities\cite{Abrikosov60,Anderson59,Tsuneto62,Markowitz63}, but
for anomalous pairing with $f$-, $p + ip$-, and $d+id$-wave
symmetry even nonmagnetic impurities~\cite{Larkin70} and
structural imperfections are known to contribute to the
destruction of the superconducting
state~\cite{Vafek14,Black14d,Posazhennikova96}.

We also emphasize one additional possible reason for the
discrepancy between the results of the theoretical calculations in
graphene and experimental observations. In recent
work~\cite{Kats14}, the role of quantum ($T=0$) fluctuations in
the graphene layers was investigated. These fluctuations were
shown to initiate logarithmic corrections to the elasticity and
bending moduli of the layers. In other words, according
to~\cite{Kats14}, quantum fluctuations of the flexural vibrations
of graphene layers can lead to a situation when the electrons move
along strongly curved string-like trajectories rather than along
the atomically smooth layers. This situation requires examination,
although in this case superconductivity can not be excluded and
can even be enhanced via the exchange by the quanta of the bending
vibrations between the pairing electrons.

\section{New promising systems with anomalous pairing}
\label{sec_perspectives}

We can assume today that there is a number of systems in which
anomalous superconducting pairing can be realized, and in
particular, the pairing based on the Kohn-Luttinger mechanism and
its generalizations. One such a system is strontium titanate
SrTiO$_3$. At room temperature, SrTiO$_3$ has a cubic crystal
structure, while at $T^*=105\,\textrm{K}$, its structure becomes
tetragonal as a result of a phase transition. The electron
structure of SrTiO$_3$ is characterized by the presence of an
energy gap with a width of $3\,\textrm{eV}$, which separates the
filled $2p$ bands of oxygen from the empty $3d$ bands of
titanium~\cite{Cardona65,Shanthi98}.

It was shown in~\cite{vanderMarel11} that in a limited region of
the momentum space near the center of the Brillouin zone, the band
structure of the strontium titanate can be effectively described
by $d_{xy}$, $d_{yz}$, or $d_{zx}$ Bloch waves, each having two
directions with a strong dispersion ($k_x$ and $k_y$ for the
$d_{xy}$-wave orbital, etc.) and one direction with a weak
dispersion, which is orthogonal to the first two
directions~\cite{vanderMarel11}. As a result, three degenerate
energy bands are formed, which with good accuracy can be
approximated by parabolas, and the Fermi surface consists of three
overlapping ellipsoids with the center at the center of the
Brillouin zone, which are oriented along the axes $x$, $y$, and
$z$ of the reciprocal cubic lattice.

Effectively, the strontium titanate is a semiconductor, which in
the case of weak electronic doping demonstrates metallic
properties with a relatively high mobility of charge carriers,
quadratic temperature dependence of the electrical resistivity,
and a strong temperature dependence of the infrared optical
conductivity~\cite{vanderMarel11}. At low temperatures, the
material becomes superconducting~\cite{Schooley64}, with the
maximal critical temperature
$T_c=1.2\,\textrm{K}$~\cite{Bednorz86}, although superconductivity
is usually observed at lower temperatures,
$T_c\leq0.7\,\textrm{K}$, and is characterized by a domelike
$T_c(n)$ dependence~\cite{Koonce67,Binning80}. Superconducting
pairing is also observed at $T_c\sim0.2\,K$ in the 2D electron
gas, which is formed at the interface of the SrTiO$_3$/LaAlO$_3$
heterostructures~\cite{Reyren07,Heber09}. In this case, the
superconducting transition temperature can be increased to
$T_c\sim0.3-0.4\,\textrm{K}$ by the application of an electric
field~\cite{Caviglia08,Bell09}.

We note that at present the nature of superconductivity in
strontium titanate and in related heterostructures remains
unclear. Superconductivity in SrTiO$_3$ was first investigated in
Ref.~\cite{Cohen64} based on the mechanism of electron-electron
attraction due to the exchange of phonons of the same and the
different valleys. This study was stimulated by the earlier
results of the band-structure calculations~\cite{Kahn64}, which
demonstrated the multivalley band structure of SrTiO$_3$.
Subsequently, the mechanism of multivalley superconductivity in
strontium titanate was investigated
in~\cite{Eagles67,Eagles69,Koonce69}. However, later on, when the
data were accumulated indicating that SrTiO$_3$ is a
superconductor with three almost parabolic bands, other mechanisms
of superconductivity, which are not connected with the multivalley
structure were suggested (see,
e.g.,~\cite{Appel69,Zinamon70,Jarlborg00}).

In Ref.~\cite{Fernandes13}, in the framework of the
phenomenological 2D model, were investigated manifestations of
multiband superconductivity in thin films of SrTiO$_3$ and at the
interfaces of the SrTiO$_3$/LaAlO$_3$ heterostructures at various
doping levels. The authors of~\cite{Fernandes13} did not discuss
the nature of the superconducting instability and limited
themselves to an examination of $s$-wave pairing. In their model,
the $d_{xz}$, $d_{yz}$, and $d_{xy}$-orbitals form two electron
bands, as in a weakly doped compound, and only one of them
intersects the Fermi level~\cite{Fernandes13}. With an increase in
the doping level, the chemical potential intersects the second
band, leading to a strong modification of the superconducting
properties of the system. Theoretical results for the calculations
of the dependences of $T_c$ and local density of states on the
doping level~\cite{Fernandes13} have been compared with the
experimental data~\cite{Binning80}. It was shown that the
intraband (intraorbital) effective attraction in SrTiO$_3$
prevails over the interband (interorbital) attraction, whereas in
other multiband superconductors, such as in pnictides, no
predominance of the intraband effective attraction over the
interband one was observed.

Another systems in which the development of Cooper instability is
possible via the Kohn-Luttinger mechanism is a family of the
‘vertical’ heterostructures, which consist of graphene layers
separated by intercalating layers of boron
nitride~\cite{Ponomarenko11,Britnell12a,Britnell12b,Gorbachev12}.
These structures demonstrate a number of interesting properties
connected with the interaction of electrons from different layers.
In these systems, the boron nitride (h-BN), like graphene, has a
hexagonal structure (because of its color and the similarity of
its structure to the structure of graphite, it is frequently
called ‘white graphite’~\cite{Sakai11}). At the same time, at the
sites of the $A$ and $B$ sublattices, it contains atoms of boron
and nitrogen. This causes the appearance of a wide energy gap
(5.2-5.9 eV) in the electronic structure of
h-BN~\cite{Hoffman84,Blase95,Watanabe04,Kubota07,Lee11}, which
underlies the wide usage of h-BN as a high-quality dielectric in
graphene devices~\cite{Dean10}. Note that h-BN is chemically and
thermally stable and is not characterized by the presence of
broken bonds or surface traps for the charge carriers. That is why
graphene structures based on h-BN substrates demonstrate a higher
mobility of charge carriers~\cite{Mayorov11,Gannett11}, smaller
roughness, and twofold-lower fluctuations of the potential than
the similar graphene structures on the substrates of
SiO$_2$~\cite{Xue11,Decker11}.

Graphene structures with h-BN can easily be modified. The
concentration of charge carriers in the graphene layers, the
spacings between these layers, and the nature of the substrate can
be changed independently in a wide range of parameters. These
changes can lead to a strong modification of electron-electron
interaction in this family of heterostructures, which can open the
possibility of realization of superconductivity at relatively
small concentrations of carriers and in the absence of any
specific properties of the density of states~\cite{Guinea12}.

We note that the possibility of Cooper pairing due to
electron-electron interaction was investigated in
Ref.~\cite{Guinea12} in the model of a vertical heterostructure
consisting of two graphene layers with the concentrations of
carriers $n_1$ and $n_2$ and with three dielectric intercalating
layers with different static dielectric constants $\epsilon_1$,
$\epsilon_2$, and $\epsilon_3$. The possibility of the
superconducting instabilities was analyzed in the framework of the
Kohn-Luttinger mechanism and the most probable superconducting
phase was described. In particular, it was shown that the
superconducting state with odd momenta, at which the
superconducting gaps have opposite signs in different Dirac cones,
is the ground state of the system due to the intervalley
scattering at high densities.

Note that in solids, the crystalline structure strictly determines
the effective mass, the velocity of electrons, and the strength of
their interactions. This constraint significantly limits the
development and the verification of different theoretical and
experimental methods of physics of strongly correlated electronic
systems. Another, more flexible method for studying the different
models of strongly correlated electrons is connected with the
systems of ultracold atoms captured by a periodic potential
obtained by the interference of three laser
beams~\cite{Lewenstein07,Bloch08}.

In Ref.~\cite{Zhu07}, an experimental scheme was suggested of
simulation and observation of the Dirac fermions in a system of
the ultracold atoms in a two-dimensional hexagonal optical
lattice. The authors of~\cite{Zhu07,Wunsch08} showed theoretically
that it is possible to control the anisotropy of the optical
lattice by changing the intensity of the trapping laser and
realize both the regimes of massive and massless Dirac fermions,
as well as to observe a phase transition between these two
regimes. In fact, the authors of~\cite{Zhu07,Wunsch08} predicted a
topological semimetal-dielectric Lifshitz transition with the gap
opening in the fermionic spectrum and with a change in the
temperature behavior of the electronic heat capacity. It was noted
in~\cite{Zhu07} that Bragg spectroscopy~\cite{Stamper99} and
different methods of determining the atomic-density profile in
magnetic or optical traps~\cite{Anglin02,Zwierlein06,Shin06} can
be used for the experimental detection of both gapped and gapless
regimes and the phase transitions between them. The physical
picture observed in this case, according to the authors
of~\cite{Zhu07}, must be very close to the picture of the ensemble
of the Dirac fermions in the graphene monolayer.

Recently, the researchers of Eslinger’s group~\cite{Tarruell12} in
Zurich experimentally realized the Dirac points with well
controlled properties using the ultracold fermionic atoms of
$^{40}$K in a hexagonal optical lattice. The presence of the Dirac
points in the band structure was detected by the authors
of~\cite{Tarruell12} by the observation of a minimal gap in the
Brillouin zone. The authors of~\cite{Tarruell12} used the unique
experimental technique of the optical tuning of the lattice
potential for controlling the effective mass of the Dirac
fermions. Moreover, the change in the lattice anisotropy allowed
the authors of~\cite{Tarruell12} to change the position of the
Dirac points in the Brillouin zone. It turns out that if the
anisotropy exceeds a certain critical value, then the two Dirac
points merge and annihilate. This phenomenon generated a broad
theoretical
interest~\cite{Hasegawa06,Wunsch08,Montambaux09,Lee09}, but at the
same time the difficulties concerning the possibility of its
experimental observation in solids became obvious to the
community~\cite{Pereira09}.

The experimental realization of the Dirac points in a system of
the ultracold atoms in the hexagonal optical
lattices~\cite{Tarruell12} also opens the great prospects for the
experimental and theoretical study of a broad class of the
physical phenomena caused by the complex topology of the lattice,
including the anomalous superconducting pairing and the appearance
of the various chiral phases.

\section{Conclusions}

The Kohn-Luttinger mechanism and its generalizations (which assume
the appearance of the anomalous pairing in the systems with a
purely repulsive interaction) represent a universal pairing
mechanism for many superconductive systems. We demonstrated the
instability of the Fermi gas with repulsion towards the
superconducting transition in a triplet $p$-wave state. The
initial conclusion about the possibility of the Cooper instability
for the Fermi-gas with a short-range (Hubbard) repulsion and a
quadratic dispersion law was generalized for the electrons in the
real crystalline solids in the tight-binding approximation. It
turns out that the character of the energy spectrum of electrons
(which is determined by the hopping parameters) greatly affects
the symmetry of the superconducting order parameter and the phase
diagram of superconductivity.

Nevertheless, in a fundamental sense, the conclusion about the
development of the Cooper instability via the Kohn-Luttinger
mechanism and its generalizations remains valid even in the
presents of the lattice. Moreover, in many cases this mechanism
can lead to a substantial increase in the superconducting
transition temperature already at low density of charge carriers
in particular in a spin-polarized case and in a two-band
situation.

It has been demonstrated that the universality of the
Kohn-Luttinger mechanism is preserved when we take into account
the finite screening radius in electron systems with repulsion. In
the framework of the Shubin-Vonsovsky model, we showed that the
Coulomb repulsion of electrons located at different sites of the
crystalline lattice could substantially modify the superconducting
phase diagram and can lead to an increase of the critical
temperature under appropriate conditions. In particular, at the
electron concentrations close to the Van Hove singularity in the
density of states the critical temperatures can reach the values
of the order of 100 K (realistic for cuprates) at the moderate
ratios of $U/W$ between the Hubbard repulsion $U$ and the
bandwidth $W$.

We have also shown in this review that the Kohn-Luttinger
mechanism of the superconducting pairing can be realized in the
systems with a linear (Dirac-like) dispersion relation. This
possibility was demonstrated for an idealized graphene monolayer
whose atoms form a hexagonal lattice. It was shown that in this
system, the polarization effects also lead to the effective
attraction of electrons in the Cooper channel. The results
obtained for the graphene monolayer were generalized on the case
of a graphene bilayer, which consists of two layers that interact
by means of interlayer Coulomb repulsion. It was shown that the
examination of the idealized two-layer system of graphene leads to
an increase in the critical temperature of the superconducting
transition in the framework of the Kohn-Luttinger mechanism.

Along with the analysis of the superconducting state, we also
analyzed the structure of the normal state of the basic models
with the Hubbard repulsion and found nontrivial corrections to the
Fermi gas Galitskii-Bloom expansion caused by the presence of the
upper Hubbard band in the lattice models or by the presence of a
singularity due to the Landau $f$-function of quasiparticles
interaction at low electron densities. However, these corrections
do not destroy the Landau Fermi-liquid picture in
three-dimensional and two-dimensional systems, and also preserve
all the results concerning the realization of the superconducting
pairing in them.

In the review, the significant attention has been paid to the
description of the systems and materials important for the
development of the microelectronics, such as the vertical
heterostructures of graphene/boron-nitride/graphene, strontium
titanate, and the related heterostructures. We have analyzed in
details both an anomalous superconductivity and the possibility of
fermionic superfluidity in 3D and 2D solutions of $^3$He in
$^4$He, and also in a system of ultracold $^6$Li and $^{40}$K
atoms in the magnetic traps and the optical lattices. Thus, we
built a bridge connecting the interests of the solid-state and the
low-temperature scientific communities.

To conclude, we emphasize once again the universal nature of the
Kohn-Luttinger mechanism and its generalizations for the formation
of the Cooper instability in repulsive Fermi systems and its
importance for the realization of an anomalous superconducting and
superfluid pairing with a nonzero value of the orbital angular
momentum ($l\neq0$).

\section*{Acknowledgments}

We are grateful to M A Baranov, A V Chubukov, D V Efremov, M V
Feigel'man, V V Kabanov, K I Kugel', M S Marienko, N M Plakida, N
V Prokof'ev, A Ya Tzalenchuk, and V V Val’kov for the fruitful
discussions and constant attention to our work. The work was
supported by the Russian Foundation for Basic Research (project
nos. 14-02-00058 and 14-02-31237). M~Yu~K thanks the Program of
Basic Research of the National Research University Higher School
of Economics for support. The work of M~M~K was supported by grant
of the President of the Russian Federation (SP-1361.2015.1) and
the Dinasty Foundation.

\end{document}